\documentclass{aa}  

\newcommand{\HII}{H\,{\scriptsize II}}

\newcommand{\CII}{[C\,{\scriptsize II}]}

\usepackage{graphicx}
\usepackage{txfonts}
\usepackage{xcolor}
\usepackage[draft]{hyperref}
\usepackage{amssymb}
\usepackage{units}
\begin{document}

   \title{A new phase of massive star formation? \\
   A luminous outflow cavity centred on an infrared quiet core}


   \author{L. Bonne
          \inst{1,2,3}
          \and
          N. Peretto \inst{1}
          \and
          A. Duarte-Cabral \inst{1}
          \and
          A. Schmiedeke \inst{4}
          \and
          N. Schneider \inst{5}
          \and
          S. Bontemps \inst{2}
          \and
          A. Whitworth \inst{1}
          }

   \institute{School of Physics \& Astronomy, Cardiff University, Queens Buildings, The parade, Cardiff CF24 3AA, UK\\
              \email{lbonne@usra.edu}
         \and
             Laboratoire d'Astrophysique de Bordeaux, Universit\'e de Bordeaux, CNRS, B18N,
              all\'ee Geoffrey Saint-Hilaire, 33615 Pessac, France
        \and
             SOFIA Science Center, USRA, NASA Ames Research Center, Moffett Field, CA 94045, USA
        \and
             Max-Planck-Institut f\"ur extraterrestrische Physik, Gie{\ss}enbachstra{\ss}e 1, 85748 Garching, Germany
        \and
             I. Physik. Institut, University of Cologne, Z\"ulpicher Str. 77, D-50937 Cologne, Germany
             }

   \date{Received ...; -}

 
  \abstract
   {Due to the sparsity and rapid evolution of high-mass stars, a detailed picture of the evolutionary sequence of massive protostellar objects still remains to be drawn. Some of the early phases of their formation are so short that only a handful of objects throughout the Milky Way currently find themselves spending time in those phases. }
   {Star-forming regions going through the shortest stages of massive star formation will present different observational characteristics than most regions. By studying the dust continuum and line emission of such unusual clouds, one might be able to set strong constraints on the evolution of massive protostellar objects.}
   {We present a detailed analysis of the G345.88-1.10 hub filament system, a newly discovered star-forming cloud that hosts an unusually bright bipolar infrared nebulosity at its centre. We used archival continuum observations from \textit{Herschel}, WISE, {\it Spitzer}, 2MASS, and SUMSS in order to fully characterise the morphology and spectral energy distribution of the region. We further made use of APEX $^{12}$CO(2-1), $^{13}$CO(2-1), C$^{18}$O(2-1) and H30$\alpha$ observations to investigate the presence of outflows and map the  kinematics of the cloud. Finally, we performed  RADMC-3D radiative transfer calculations to constrain the physical origin of the central nebulosity.}
   {At a distance of 2.26$^{+0.30}_{-0.21}$ kpc, G345.88-1.10 exhibits a network of parsec-long converging filaments. At the junction of these filaments lie four infrared-quiet fragments. H1 is the densest fragment (with M=210~M$_{\odot}$, R$_{\rm{eff}}=0.14$~pc) and sits right at the centre of a wide (opening angle of $\sim$ 90$\pm$15$^{o}$) bipolar nebulosity where the column density reaches local minima. The $^{12}$CO(2-1) observations of the region show that these infrared-bright cavities are spatially associated with a powerful molecular outflow that is centred on the H1 fragment. Negligible radio continuum and no H30$\alpha$ emission is detected towards the cavities, seemingly excluding that ionising radiation drives the evolution of the cavities. 
   Radiative transfer calculations of an embedded source surrounded by a disc and/or a dense core are unable to reproduce the observed combination of a low-luminosity  ($\lesssim$ 500 L$_{\odot}$) central source and a surrounding high-luminosity ($\sim 4000$~L$_{\odot}$) mid-infrared-bright bipolar cavity. This suggests that radiative heating from a central protostar  cannot be responsible for the illumination of the outflow cavities.
   
   }
   {This is, to our knowledge, the first reported object of this type. The rarity of objects like G345.88-1.10 is likely related to a very short phase in the massive star and/or cluster formation process that was so far unidentified. 
   We discuss whether mechanical energy deposition by one episode or successive episodes of powerful mass accretion in a collapsing hub might explain the observations. While promising in some aspects, a fully coherent scenario that explains the presence of a luminous bipolar cavity centred on an infrared-dark fragment remains elusive at this point.
   }


   \maketitle
%

\section{Introduction}
The series of processes that lead to the transport of ambient molecular gas down to the surface of young accreting protostars is  not fully understood yet \citep{Li2014}. Questions related to the nature of the initial instability, the star formation timescale and the morphology and size of the mass reservoir 
are examples of areas one needs to make progress on. The study of nearby (d~$<500$~pc) star-forming regions presents a picture in which prestellar cores, i.e. the progenitors of individual stars or small stellar systems \citep{WardThompson2007,DiFrancesco2007}, form as the result of the fragmentation of gravitationally unstable 
filaments \citep{Andre2014}. These cores then collapse to form a protostar at their centres. Conservation of angular momentum means that a circumstellar disc is formed alongside the protostar. The size and time at which such discs first appear, along with the exact role they play in the accretion process, are still a matter of active debate \citep[e.g.][]{Li2014,Hartmann2016}. 
As far as low-mass star formation is concerned, most of the stellar mass is thought to be accreted during the first (i.e. Class 0) stages \citep{Andre2000}. 
Note that there is now plenty of evidence that accretion is not a steady process but rather episodic \citep[e.g.][]{Hartmann2016}. 
A consequence of the mass accretion activity is the formation of high-velocity collimated jets that evacuate a part of the angular momentum from the core during protostellar accretion. This jet entrains with it ambient molecular gas that is seen as a molecular protostellar outflow \citep{Bachiller1996,Bally2016}. The presence of a protostellar outflow thus provides a probe of protostellar mass accretion.
These outflows might also contribute to the self-regulation of star formation within molecular clouds by injecting turbulence and slowing down star formation activity \citep[e.g.][]{Arce2011,DuarteCabral2012}.

The results presented above have mostly been obtained by studying low-mass star-forming regions. It is however debated whether this applies to the formation of the rarer and faster evolving high-mass stars \citep{Zinnecker2007,Motte2017}. For instance, despite extensive searches in the Galactic plane, no significant population of massive ($M>100$~M$_{\odot}$, $R\le0.05$~pc) prestellar cores has been found \citep{Motte2007,Motte2010,Russeil2010,Louvet2019,Sanhueza2019}. Instead, a growing body of evidence is showing that the mass of massive  protostellar sources increases, at least initially, with time \citep{Tige2017,Peretto2020,Rigby2021}, most likely as the result of the global collapse of their parent clumps \citep{Peretto2006,Peretto2013,Peretto2014,Schneider2010,Schneider2015b,Kirk2013,Csengeri2017a,Csengeri2017b,Watkins2019,Jackson2019}. The presence of hub filament systems, i.e. networks of converging filaments, might be a tracer of such collapse. In fact, it has been shown that the centre of such systems represent ideal locations for the formation of massive cores \citep[e.g.][]{Schneider2012,Williams2018}, with hubs being more efficient at concentrating a large fraction of their mass within the most massive cores \citep{Anderson2021}. The large-scale kinematics of the parent cloud around massive cores is therefore key to better understand high-mass star formation. A direct link between the mass infall rate of the prototypical hub SDC335 and its mass outflow rate even suggests a continuity of mass transfer from parsec-scales down to core scale \citep{Avison2021}. 

Another major difference between the formation of low-mass and high-mass stars is their relative Kelvin-Helmoltz time scale, i.e. the time it takes for pre-main sequence stars to radiate away its thermal energy. For low-mass stars, this time is larger than the lifetime of pre-main sequence stars, allowing low-mass stars to start the ignition of hydrogen burning after it finished accreting. This is not the case for high-mass stars (>10 M$_{\odot}$) which have shorter Kelvin-Helmoltz timescales than the accretion timescale. As a result, these stars start burning hydrogen while still deeply embedded within their parent molecular cloud. This used to be a major hurdle for our understanding of high-mass star formation since the large radiation pressure from massive stars should halt accretion \citep{Larson1971,Wolfire1987}. However, this is resolved when considering non-spherically symmetric accretion flows, such as through an accretion disc. Accretion discs provide a low-density channel for protostellar feedback to escape, the so-called flashlight effect, preventing the reversal of the mass inflow \citep[e.g.][]{Yorke1999,Yorke2002,Kuiper2010}. 
With the advent of (sub)millimeter high-angular resolution observatories such as ALMA and NOEMA, an increasing number of massive discs/torus have been found \citep{Johnston2015,Beltran2016,Ilee2016,Cesaroni2017,Csengeri2018}. However, these discs are often fragmented, a possible consequence of the high density of these structures \citep{Beuther2018,Ilee2018,Ahmadi2018}. Like in the low-mass case, it has been proposed that high-mass protostars undergo episodic accretion bursts \citep[e.g.][]{CarrascoGonzalez2015,Stecklum2021,Hunter2021}. 

\begin{figure*}
\includegraphics[width=0.5\hsize]{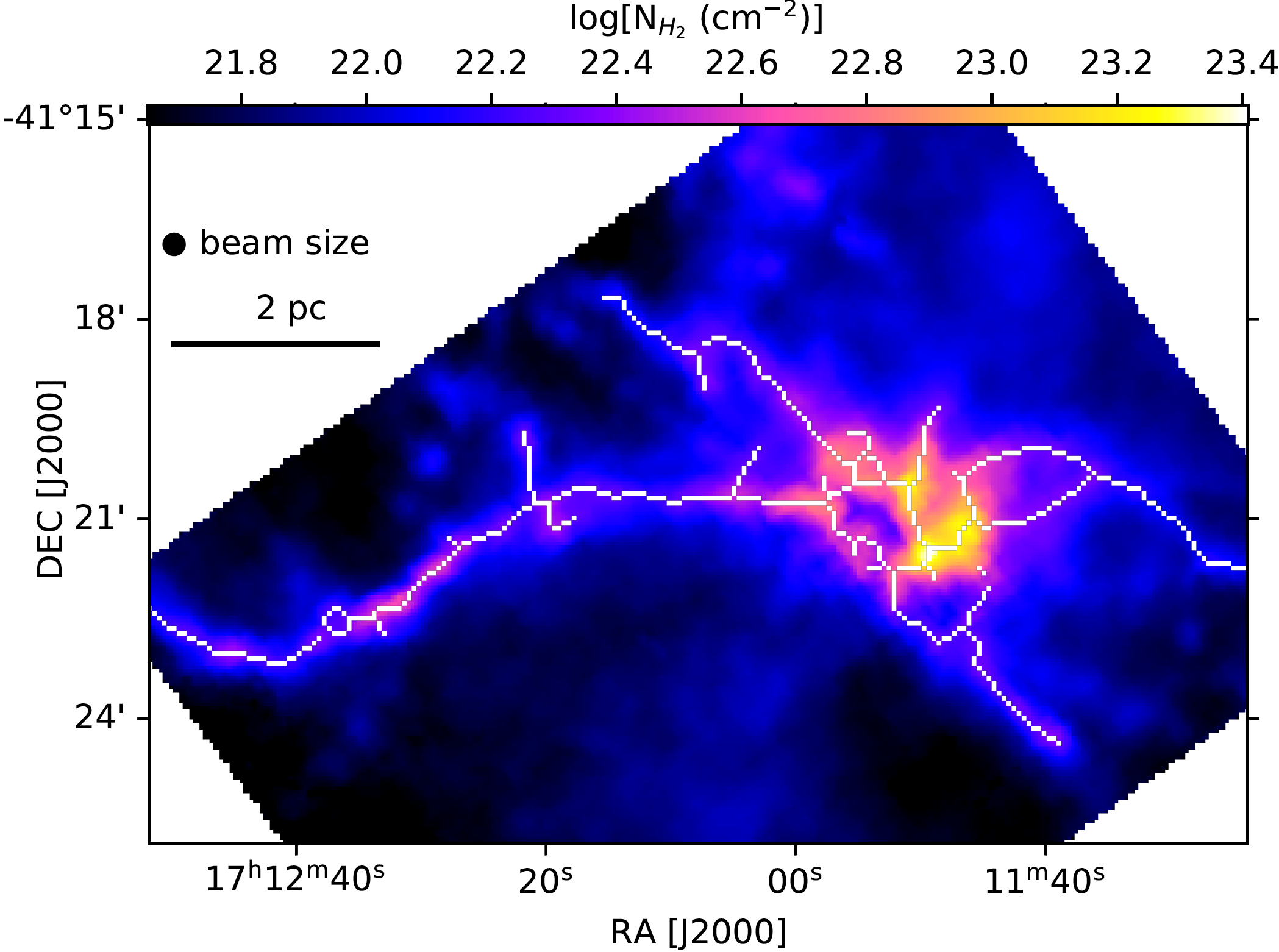}
\includegraphics[width=0.49\hsize]{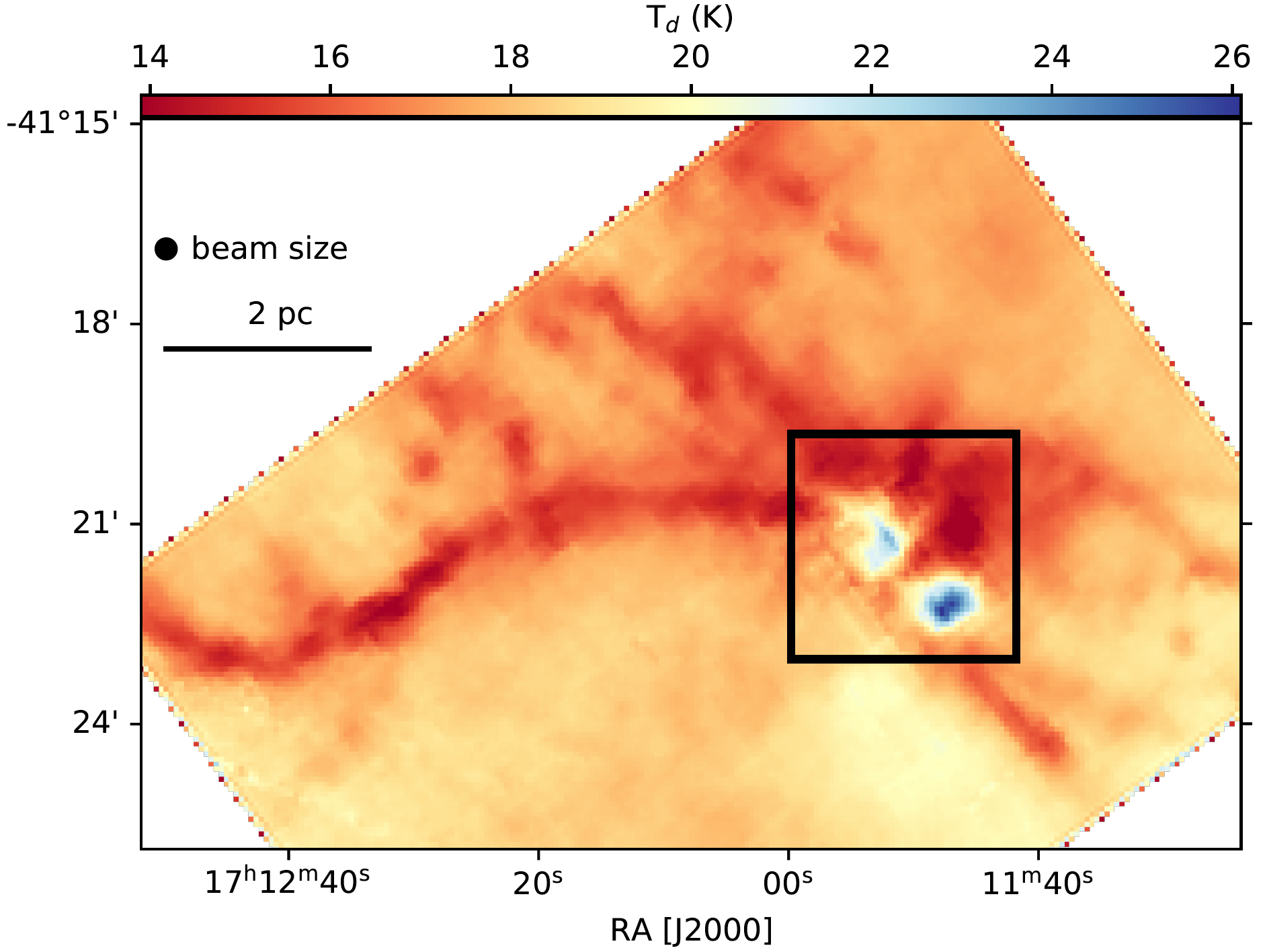}
\caption{ (left): H$_2$ column density image of G345.88-1.10 on a log scale at a spatial resolution of 18$''$ derived from {\it Herschel} data at 160 and 250 $\mu$m. The hub morphology has been highlighted by the skeleton of the filament network present in the region. 
(right): The corresponding {\it Herschel} dust temperature map. The black box indicates the region of interest for this paper which is presented in more detail in Fig. \ref{fluxDensCont}. 
}
\label{filamentsFig}
\end{figure*}

In order to make further progress in our understanding of the earliest stages of massive star formation, two approaches can be considered. The first one consists in studying a large sample of sources in order to get statistics on a set of properties. The second one consists in performing a detailed analysis of an individual source that, for one reason or another, is of particular interest. It is this second approach that we are taking in this paper.  We present here the first analysis of the G345.88-1.10 star-forming cloud (see Fig.~\ref{filamentsFig}), a hub filament system with a bright infrared bipolar nebulosity at the centre. G345.88-1.10 has been serendipitously discovered, and, as of today,  no literature exists on it. As discussed in this paper, we believe that the study of G345.88-1.10 will shed new light on the  mechanisms linked to the very first phases of high-mass star formation. In section 2 we describe the observational data we used in the present study. Section 3 presents the hub characteristics. Section 4 describes the morphology and spectral energy distribution (SED) of the bipolar nebulosity. In section 5 we identify and analyse the properties of the central molecular outflow, while in section 6 we investigate the possibility whether the bipolar nebulosity could be the result of ionizing radiation. In section 7 the radiative transfer (RT) calculations are presented and compared with the observed \textit{Herschel} continuum emission. In Section 8, we bring all the results together, and discuss their implications for massive star/cluster formation. Finally, Section 9 presents a summary of the paper and future prospects.

\section{Observational data}

As a result of the relatively large galactic latitude of G345.88-1.10 ($|$b$| >$ 1) a few key Galactic plane surveys performed in the past 20 years, such as MIPSGAL \citep{Carey2009} or more recently SEDIGISM \citep{Schuller2017,Schuller2021}, have not observed it. This could explain why it did not attract attention yet. In this paper, we study G345.88-1.10 across a number of wavelengths, using  dust continuum data from \textit{Herschel}, Spitzer, WISE, and new spectral line observations obtained with the Atacama Pathfinder Experiment telescope (APEX).

\subsection{\textit{Herschel}}

The \textit{Herschel} data of G345.88-1.10 is provided by the Hi-GAL survey which is dedicated to the far-infrared dust continuum imaging of the Galactic plane, and was obtained from the HiGAL archive\footnote{\text{http://vialactea.iaps.inaf.it/}} \citep{Molinari2010,Traficante2011,Molinari2016}. The photometric maps at 70 and 160 $\mu$m, obtained with the PACS instrument \citep{Poglitsch2010}, have an angular resolution of $8.4''$ and $13.5''$, respectively. The maps at 250, 350 and 500 $\mu$m, obtained with the SPIRE instrument \citep{Griffin2010}, have a resolution of $18.2''$, $24.9''$ and $36.3''$, respectively.

\subsection{WISE, {\it Spitzer} and 2MASS IR surveys}

The WISE all-sky survey provides data at 3.4 $\mu$m, 4.6 $\mu$m, 12 $\mu$m and 22 $\mu$m for G345.88-1.10 with an angular resolution of $6.1''$, $6.4''$, $6.5''$ and $12.0''$, respectively \citep{Wright2010}.\\
Although G345.88-1.10 has not been observed with the MIPS instrument on the {\it Spitzer} space telescope, the source has been covered by the IRAC instrument within the GLIMPSE survey \citep{Benjamin2003}, providing images at 3.6 $\mu$m, 4.5 $\mu$m, 5.8 $\mu$m and 8 $\mu$m at an angular resolution of $1.66''$, $1.72''$, $1.88''$ and $1.98''$, respectively \citep{Fazio2004}.\\
Lastly, the 2MASS survey provides a map in the J, H and K-band for G345.88-1.10 with an angular resolution of $\sim$ 2$''$ \citep{Skrutskie2006}.

\subsection{APEX}

In march 2016, the $^{12}$CO(2-1), $^{13}$CO(2-1) and C$^{18}$O(2-1) lines were mapped around G345.88-1.10 using the APEX-1 receiver which is installed on the APEX telescope \citep{Guesten2006}. The frequencies (and wavelengths) of these rotational transitions are 230.54 GHz (1.301 mm), 220.40 GHz (1.361 mm) and 219.56 GHz (1.365 mm), respectively. The setup that covers $^{12}$CO(2-1) simultaneously also covers the H30$\alpha$ hydrogen recombination line at 231.90 GHz, which is a tracer of \HII\ regions. This data has an  angular resolution of $\sim 28''$. The $^{13}$CO(2-1) and C$^{18}$O(2-1) lines were mapped simultaneously by tuning the receiver to an intermediate frequency. All lines were observed with a precipitable water vapor (pwv) between 0.8 and 1.3 mm and the data was converted to the main beam temperature (T$_{mb}$) using a main beam efficiency $\eta_{mb}$ = 0.81 \citep{Vassilev2008}. The final maps were produced using the GILDAS CLASS software \footnote{\text{https://www.iram.fr/IRAMFR/GILDAS/doc/html/class-html/class.html}} with a channel width of 0.3 km s$^{-1}$ for all lines. The noise, estimated as the line standard deviation within signal-free channels, is $\sim$ 0.2 K for $^{12}$CO(2-1) and H30$\alpha$. For $^{13}$CO(2-1) and C$^{18}$O(2-1), the typical noise over the map is $\sim$ 0.15 K. The observations were carried out using the on-the-fly mode centered on $\alpha_{\text{(2000)}}$ = 17$^{h}$11$^{m}$43.9$^{s}$, $\delta_{\text{(2000)}}$ = -41$^{\circ}$18$^{\prime}$25.5$^{\prime\prime}$. The $^{12}$CO(2-1) and H30$\alpha$ map covers an area on the sky of 4$'$x4$'$ while the $^{13}$CO(2-1) and C$^{18}$O(2-1) maps cover a smaller area of 2$'$x2$'$.

\section{G345.88-1.10: A nearby hub filament system}

\begin{figure}
 \centering
\includegraphics[width=\hsize]{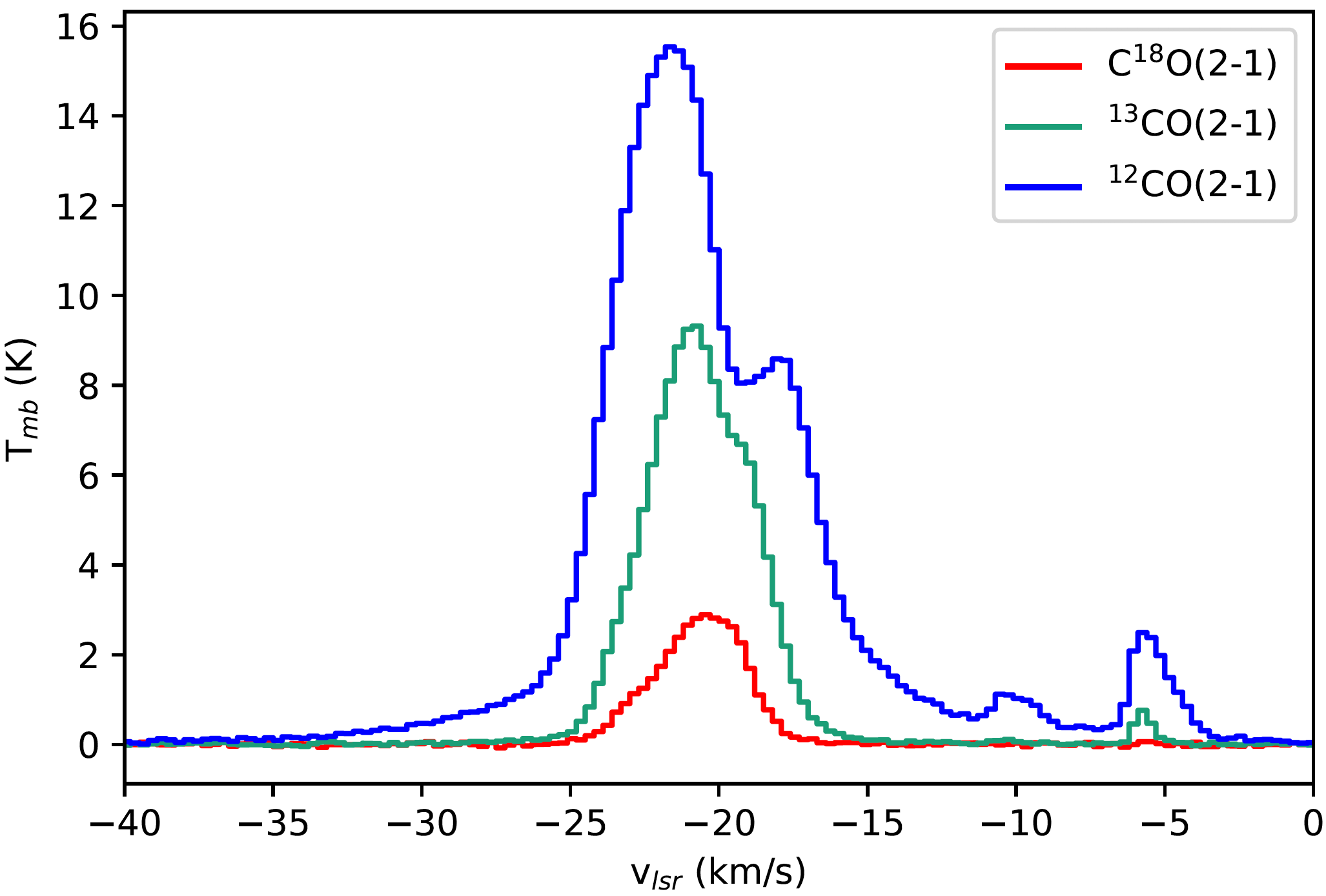}
\caption{$^{12}$CO(2-1), $^{13}$CO(2-1), and C$^{18}$O(2-1) emission averaged over the extent of the hub. 
}
   \label{fig:invPCygni}
\end{figure}

\subsection{The distance of G345.88-1.10}
As this is the first paper on G345.88-1.10, no published distance exists. To estimate it, we use the galactic rotation model from \citet{Reid2014}. Our APEX data show (see Fig.~\ref{fig:invPCygni}) that the velocity of G345.88-1.10 is centred at -21 km s$^{-1}$. This results in a heliocentric kinematic near distance of 2.26$^{+0.30}_{-0.21}$ kpc, where the uncertainties are given by the Monte Carlo calculations\footnote{http://www.treywenger.com/kd/index.php} presented in \citet{Wenger2018}. This distance estimate places G345.88-1.10 in the Scrutum galactic arm \citep[e.g.][]{Pineda2013}, at a galactocentric distance of 6.1 kpc. The far distance gives 13.96$^{+0.24}_{-0.48}$ pc. However, this far distance seems unlikely to us for a couple of reasons. First, it would imply that the bolometric luminosity of the cavity exceeds 10$^{5}$ L$_{\odot}$ (see Sec. \ref{sec:lumSect} for the L$_{bol}$ estimate). Galactic sources that bright are generally associated with bright radio continuum emission ($>$ 100 mJy) which we do not observe in G345.88-1.10 (see Sec. \ref{sec:hiisection}). We note that a couple of luminous (L$_{bol} >$ 10$^{5}$ L$_{\odot}$) sources were found with a radio continuum flux $<$ 100 mJy \citep{Lumsden2013}, but only a small fraction of luminous sources satisfies this condition. Second, if placed at the far distance, then the two clouds at -10 km s$^{-1}$ and -6 km s$^{-1}$ that we identify in Fig. \ref{fig:invPCygni} would be foreground clouds. These two clouds overlap in velocity with the observed CO high-velocity wings in Fig. \ref{fig:invPCygni}. Given that outflows in dense molecular regions typically have higher excitation temperatures than the bulk of diffuse molecular gas, at least a part of these clouds would likely be seen in absorption rather than emission. This argument is particularly valid for G345.88-1.10 as we will show in Sec. \ref{sec:outflowmap} that the outflow of G345.88-1.10 is located in the warm nebulosities shown in Fig. \ref{filamentsFig}. Lastly, we note that assuming the far distance for the region implies that the region is located 270 pc above the Galactic plane. This is at the edge of the thin disk in the Milky Way. As the region is forming high-mass stars, see later on, this is a fairly high distance from the Galactic plane. For the remainder of this paper, we therefore select the close distance solution for G345.88-1.10. 

\begin{table*}
\begin{center}
\begin{tabular}{cccccc}
\hline
\hline
Source & Right ascension (J2000) & Declination (J2000) & Gas mass  & R$_{\rm{eff}}$ & N$_{\rm{H_{2},av}}$  \\
    & (hh:mm:ss) & (dd:mm:ss) & (10$^{2}$ M$_{\odot}$) & (pc) & (10$^{23}$ cm$^{-2}$) \\
\hline
H1 & 17:11:43.70 & -41:18:24.32 & 2.1$^{+1.2}_{-1.1}$ & 0.14$^{+0.03}_{-0.02}$ & 2.0 $\pm$ 1.0 \\
H2 & 17:11:40.76 & -41:18:03.14 & 1.7$^{+1.0}_{-0.9}$ & 0.14$^{+0.03}_{-0.02}$ & 1.7 $\pm$ 0.9 \\
H3 & 17:11:45.38 & -41:17:20.79 & 2.8$^{+1.6}_{-1.5}$ & 0.20$^{+0.05}_{-0.03}$ & 1.2 $\pm$ 0.6 \\
H4 & 17:11:51.74 & -41:17:45.21 & 1.4$^{+0.8}_{-0.7}$ & 0.21$^{+0.05}_{-0.03}$ & 0.5 $\pm$ 0.3 \\
\hline
\end{tabular}
\caption{Identified fragment properties. Column 1: Fragment name; Column 2: Right ascension; Column 3: Declination; Column 4: Mass; Column 5: Effective radius; Column 6: Averaged H$_2$ column density. Errors on the size are estimated based on the distance uncertainty. For the column density an error of 50\% is assumed to take into account uncertainties related to the dust opacity \citep{Ossenkopf1994}, data calibration \citep{Griffin2013,Bendo2013} and dust to gas conversion. The error on the mass includes both the 50\% as well as the distance uncertainty.}
\label{sourceTable}
\end{center}
\end{table*}

\begin{figure*}
\begin{center}
\includegraphics[width=0.51\hsize]{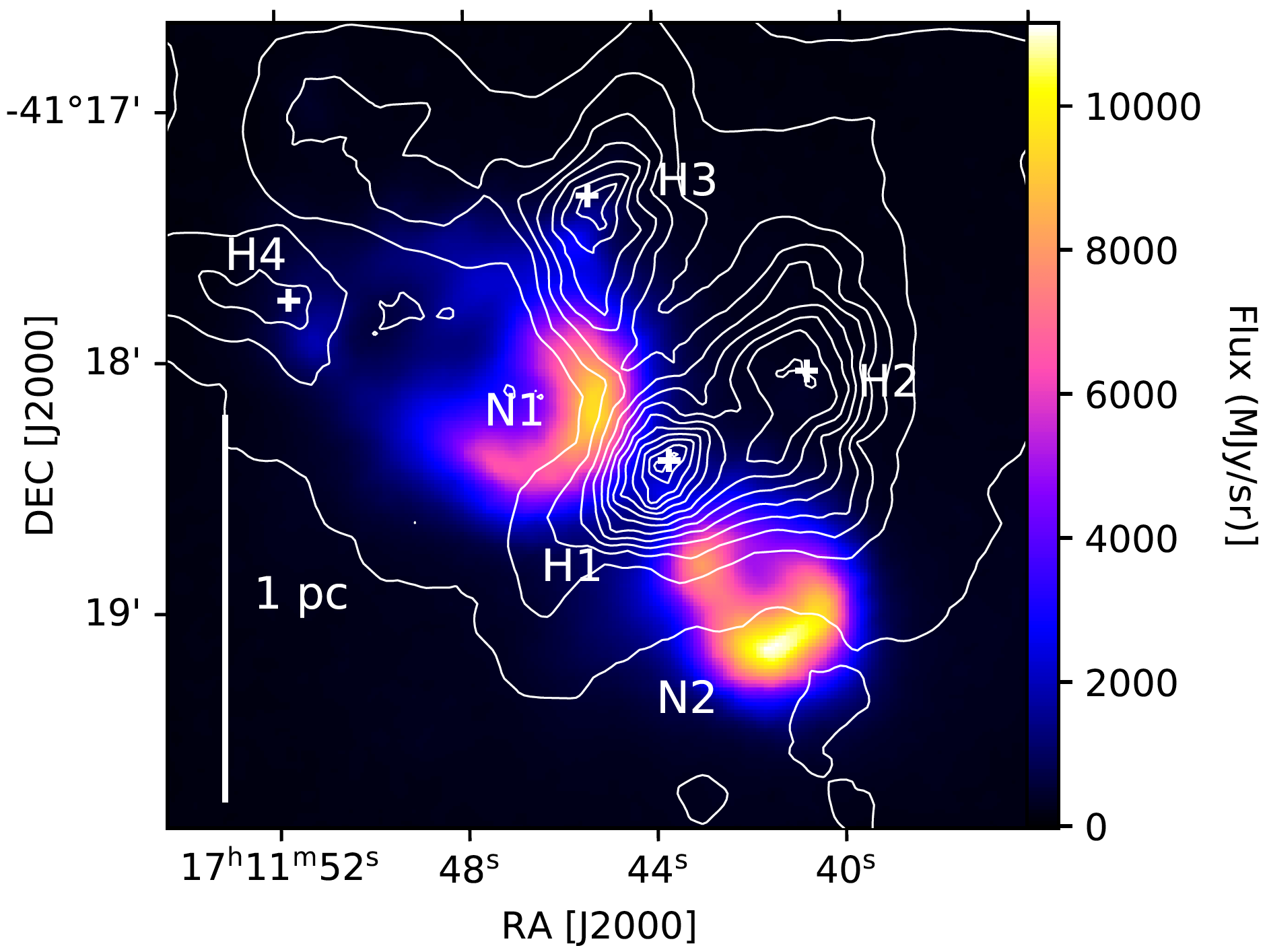}
\includegraphics[width=0.48\hsize]{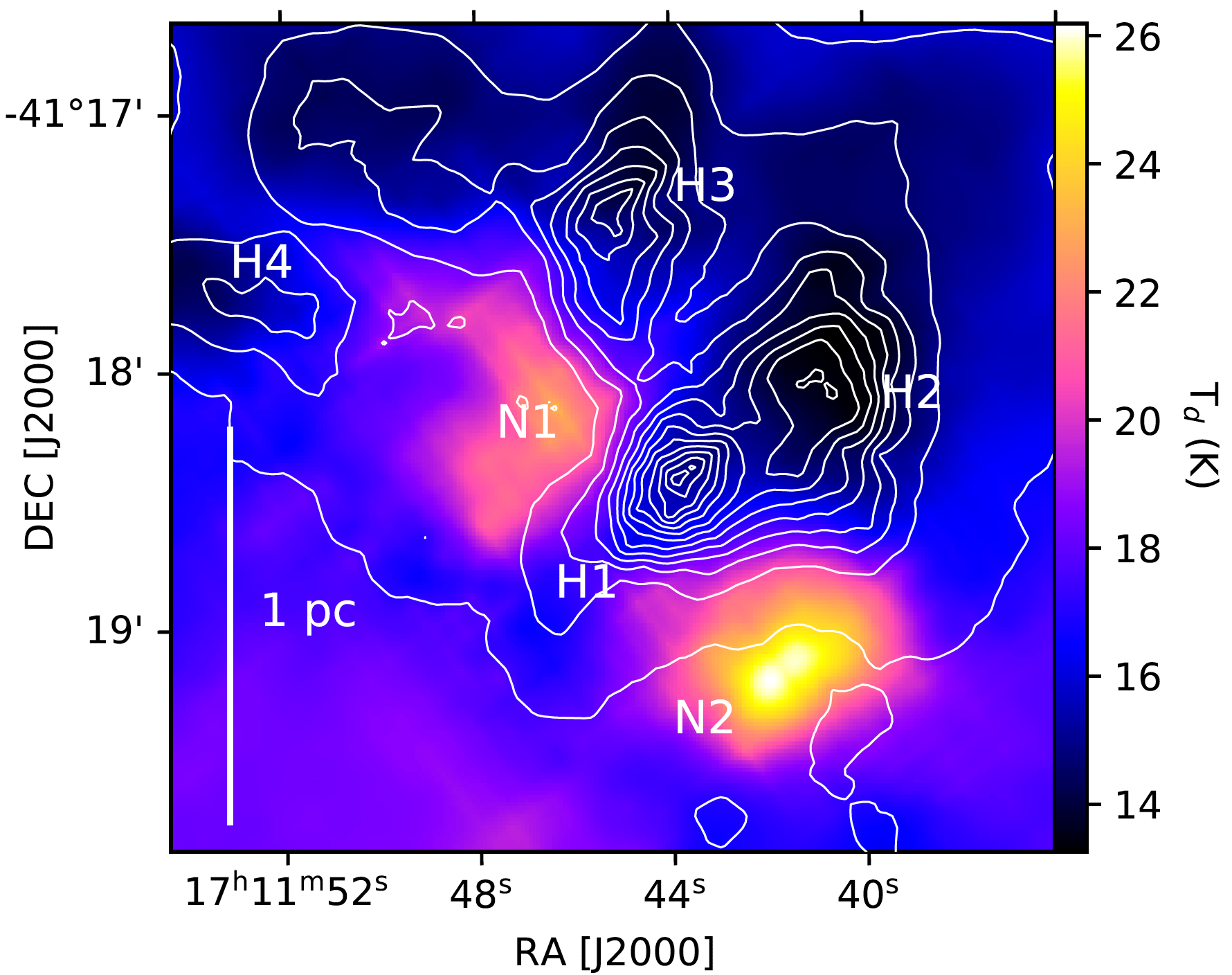}
\caption{(left) \textit{Herschel} 70 $\mu$m image of the central region/hub of G345.88-1.10, showing the presence of a bipolar nebulosity N1 and N2. The H$_{2}$ column density contours are overplotted in white starting at N$_{H_{2}}$ = 2$\times$10$^{22}$ cm$^{-2}$ with increments of 2$\times$10$^{22}$ cm$^{-2}$. The 4 fragments that were extracted from the column density map with the dendrogram technique are indicated by the white crosses together with the two nebulosities. (Right) Image of the {\it Herschel} dust temperature centred on the infrared nebulosity. All symbols and notations are identical to those presented in the left-hand-side panel.
}
\label{fluxDensCont}
\end{center}
\end{figure*}

\subsection{H$_2$ column density and dust temperature mapping}
\label{sec:colDensMap}

Figure \ref{filamentsFig} presents the H$_2$ column density and dust temperature images of G345.88-1.10 at 18$''$ resolution. These have been obtained using the method described in \citet{Peretto2016}. This method only uses the 160 and 250 $\mu$m data from \textit{\textit{Herschel}}. First, we convolved 
the 160 $\mu$m image to the resolution of the 250 $\mu$m image using a Gaussian kernel of FWHM = $\sqrt{\theta_{250}^{2} - \theta_{160}^{2}}$ = $13.4''$. Then, the temperature at every pixel was determined from the ratio of the 160 $\mu$m and 250 $\mu$m flux density using
\begin{equation}
\label{eq:veryHighResTemp}
R_{160/250} = \frac{S_{160}}{S_{250}} = \frac{B_{\nu 160} (T_{d})}{B_{\nu250} (T_{d})}\left( \frac{250}{160}\right)^{\beta}
\end{equation}
where S$_{\lambda}$ is the flux density, B$_{\nu}$ the Planck function, $\beta$ the index of the dust opacity law and T$_{d}$ the dust temperature. For $\beta$ a value of 2 was used \citep{Sadavoy2013}. Since equation ~\ref{eq:veryHighResTemp} has no analytical solution, Brent's method\footnote{\footnotesize{https://docs.scipy.org/doc/scipy/reference/generated/scipy.optimize.brentq.html}} was used to find a zero point within the 10 - 100 K temperature range. For all pixels, a solution was found in the 13 - 30 K range. From the resulting dust temperature map, the H$_2$ column density image can then be obtained via the following equation:
\begin{equation}
\label{eq:veryHighResDens}
N_{H_{2}} = S_{\nu_{250}}/ \big[ B_{\nu_{250}}(T_{d}) \kappa_{\nu_{250}}\mu m_{H} \big]
\end{equation}
where $\kappa_{\nu_{250}}$ = 0.012 cm$^{2}$ g$^{-1}$ is the specific dust opacity at 250 $\mu$m that already incorporates a dust to gas mass ratio of 1\% \citep{Hildebrand1983}, $\mu$ = 2.33 is the average molecular weight and m$_{H}$ is the mass of atomic hydrogen.

As a safety-check, we also produced the temperature and column density maps by performing a pixel-by-pixel SED fit to the 160 $\mu$m, 250 $\mu$m, 350 $\mu$m and 500 $\mu$m maps, at an angular resolution of 36$''$ \citep[e.g.][]{Peretto2010,Battersby2011,Hill2011}. The resulting maps were very similar to those presented in Fig.~\ref{filamentsFig}, but at a lower angular resolution. Furthermore, the calculation of the total gas mass within the central parsec of G345.88-1.10 using both column density maps provided similar values within 10~\% of each other, i.e. 2900 M$_{\odot}$ for the ratio method and  2600 M$_{\odot}$ for the SED method. In the remainder of this paper, we will consider the column density and temperature map produced by the ratio method of \cite{Peretto2016}.


\subsection{Identifying filaments and fragments}\label{Sec:IDfils}

Figure \ref{filamentsFig} shows two very interesting features. First, there is a clear hub filament system with a network of a few parsec long filaments converging towards a high column density region. In order to highlight this structure, we have performed a skeleton extraction using the second derivative method presented in, e.g., \citet{Schisano2014,Williams2018,Watkins2019,Orkisz2019}. The second striking feature is the presence of two warm regions right at the centre of G345.88-1.10, on either side of the highest column density peak.

In order to identify sub-structures within the column density map we used the {\it astrodendro} package\footnote{http://www.dendrograms.org/} \citep{Rosolowsky2008}. Astrodendro segments the input image by constructing a dendrogram tree. The most compact  non-fragmented structures of this tree are referred to as {\it leaves} in the dendrogram terminology, and {\it fragments} in the remainder of this paper. To be identified as a fragment that is spatially resolved, the leaf has to cover at least two times the beamsize at 250 $\mu$m. This limits the resolution of the dendrogram to an effective radius of $\sim$ 0.1 pc. The tree starts at N$_{H_{2}}$ = 10$^{22}$ cm$^{-2}$, to focus on the dense gas only, and is constructed using steps of N$_{H_{2}}$ = 2.5$\times$10$^{21}$ cm$^{-2}$. With these parameters, Astrodendro identifies 4 fragments in the hub. These are labeled as H1, H2, H3 and H4 in Fig. \ref{fluxDensCont}.\\
The fragment with the highest column density (H1) has a mass of 210$^{+50}_{-30}$ M$_{\odot}$ within a radius of 0.14 pc. The coordinates, mass, effective radius and average column density of the identified fragments are given in Table \ref{sourceTable}. The other identified fragments (H2-H4) also have masses in the 100 to 300 M$_{\odot}$ range. While we are expecting these sources to be sub-fragmented on smaller scales, their masses suggest that G345.88-1.10 has the potential to form a rich stellar cluster.\\

\begin{figure}
    \centering
    \includegraphics[width=\hsize]{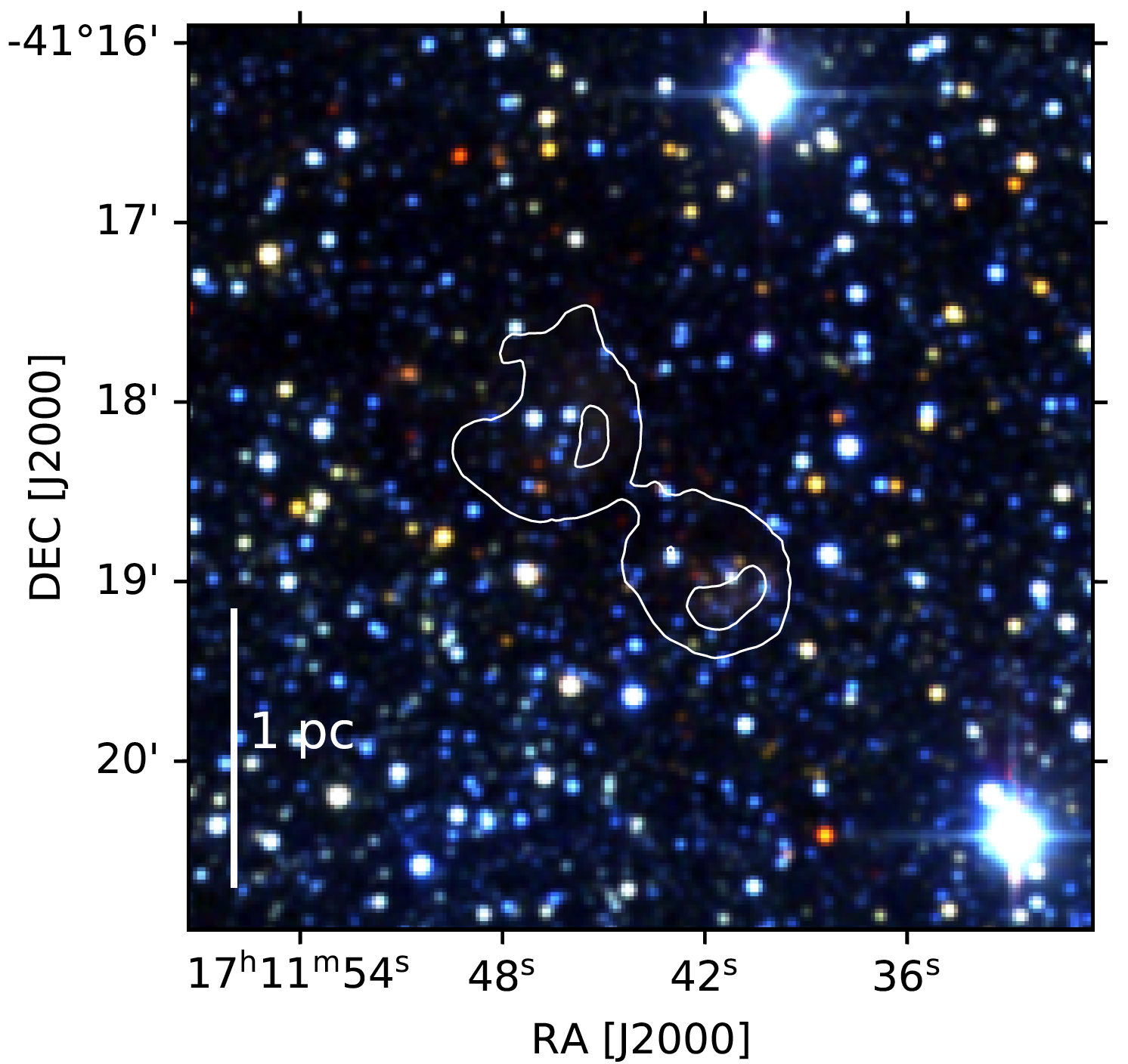}
    \caption{RGB image of the 2MASS J-band (in blue), H-band (in green) and K-band (in red). The white contours indicate the 70 $\mu$m continuum emission at 2000 and 8000 MJy sr$^{-1}$, to highlight the location of the full 70 $\mu$m nebulosities and the brightest regions of the 70 $\mu$m nebulosities, respectively. The 2MASS data shows no evident stellar overdensity in the cavities which would point to the presence of clusters.}
    \label{fig:2MASSrgb}
\end{figure}

\begin{figure*}
\begin{center}
\includegraphics[width=0.32\hsize]{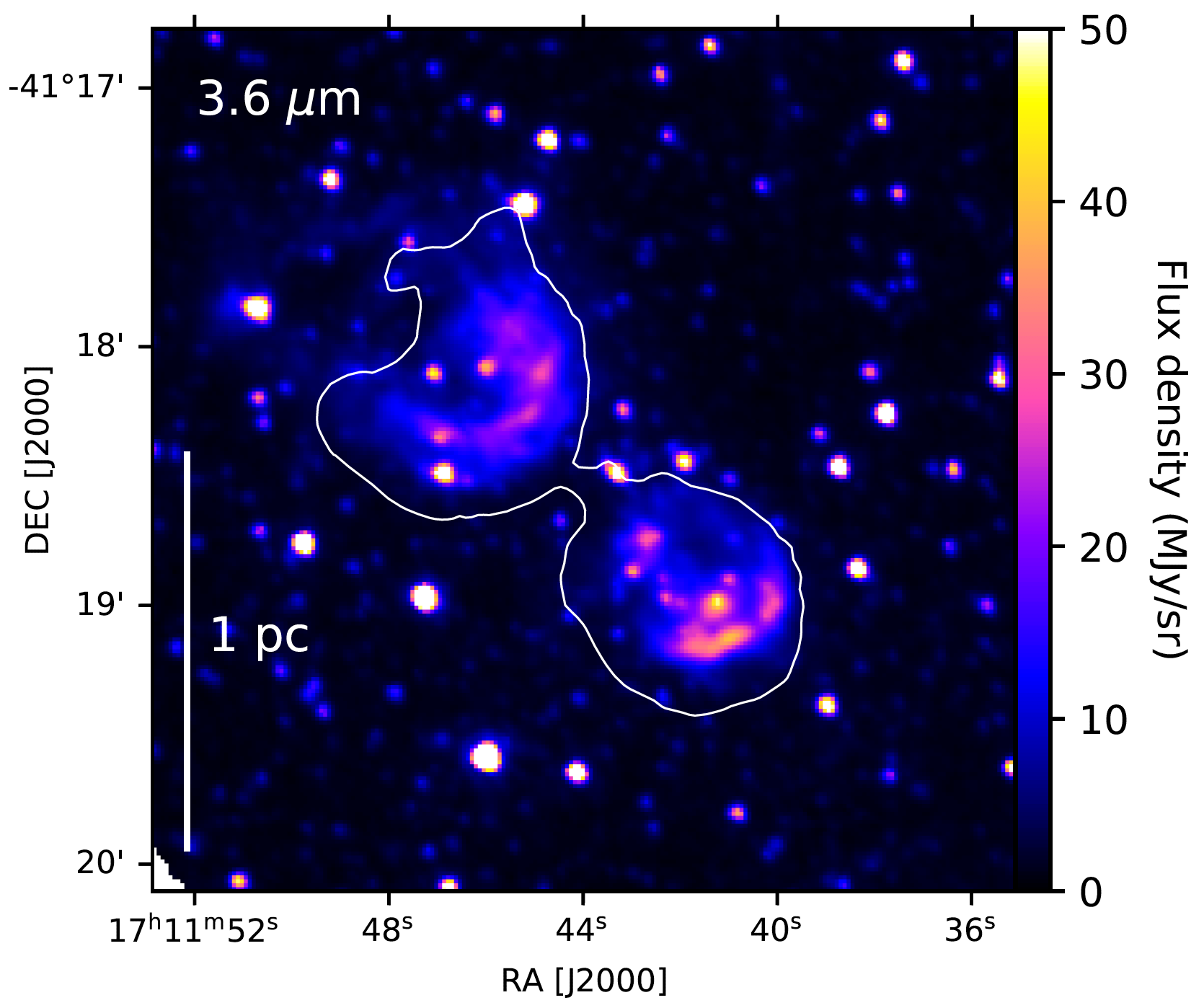}
\includegraphics[width=0.32\hsize]{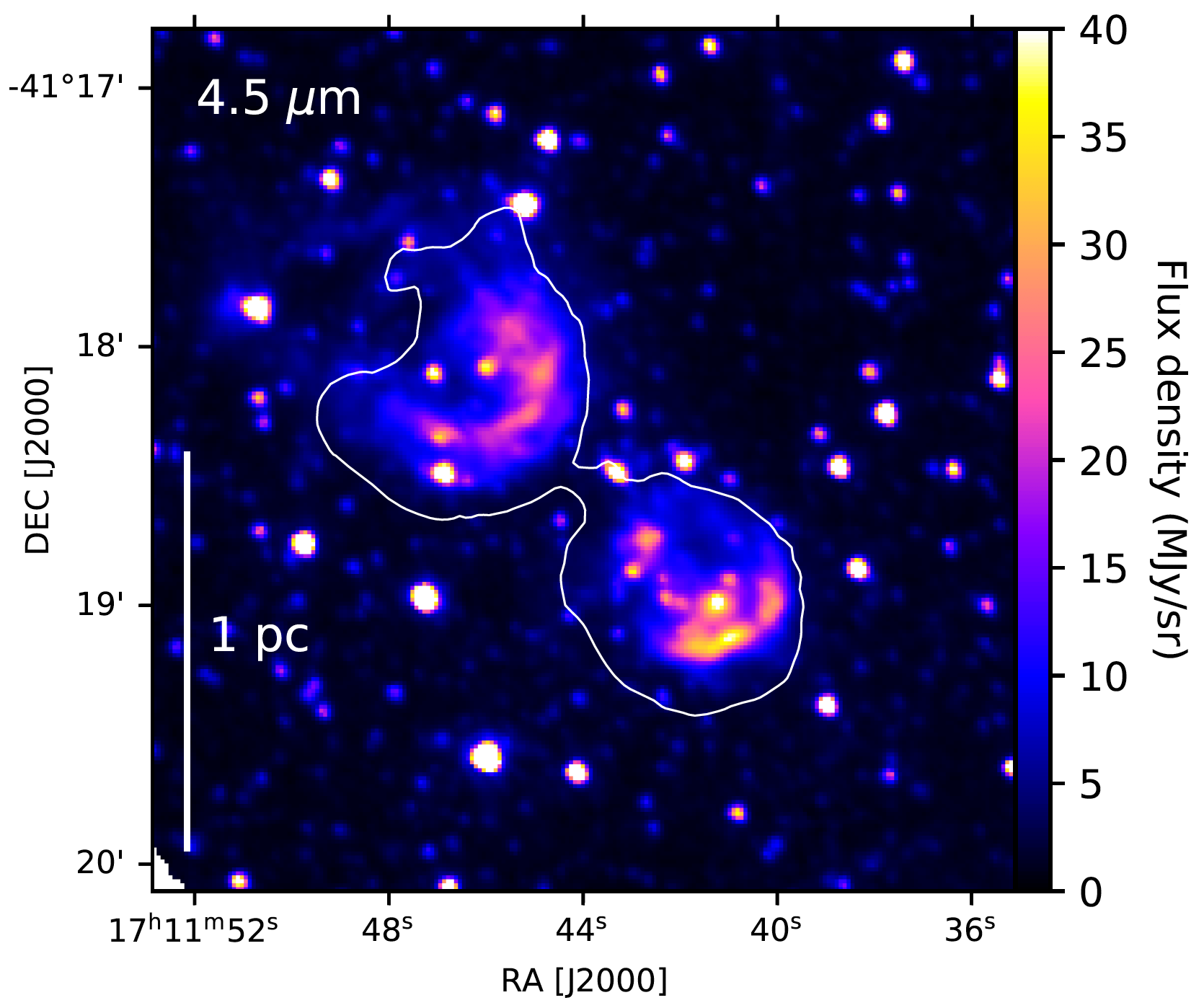}
\includegraphics[width=0.33\hsize]{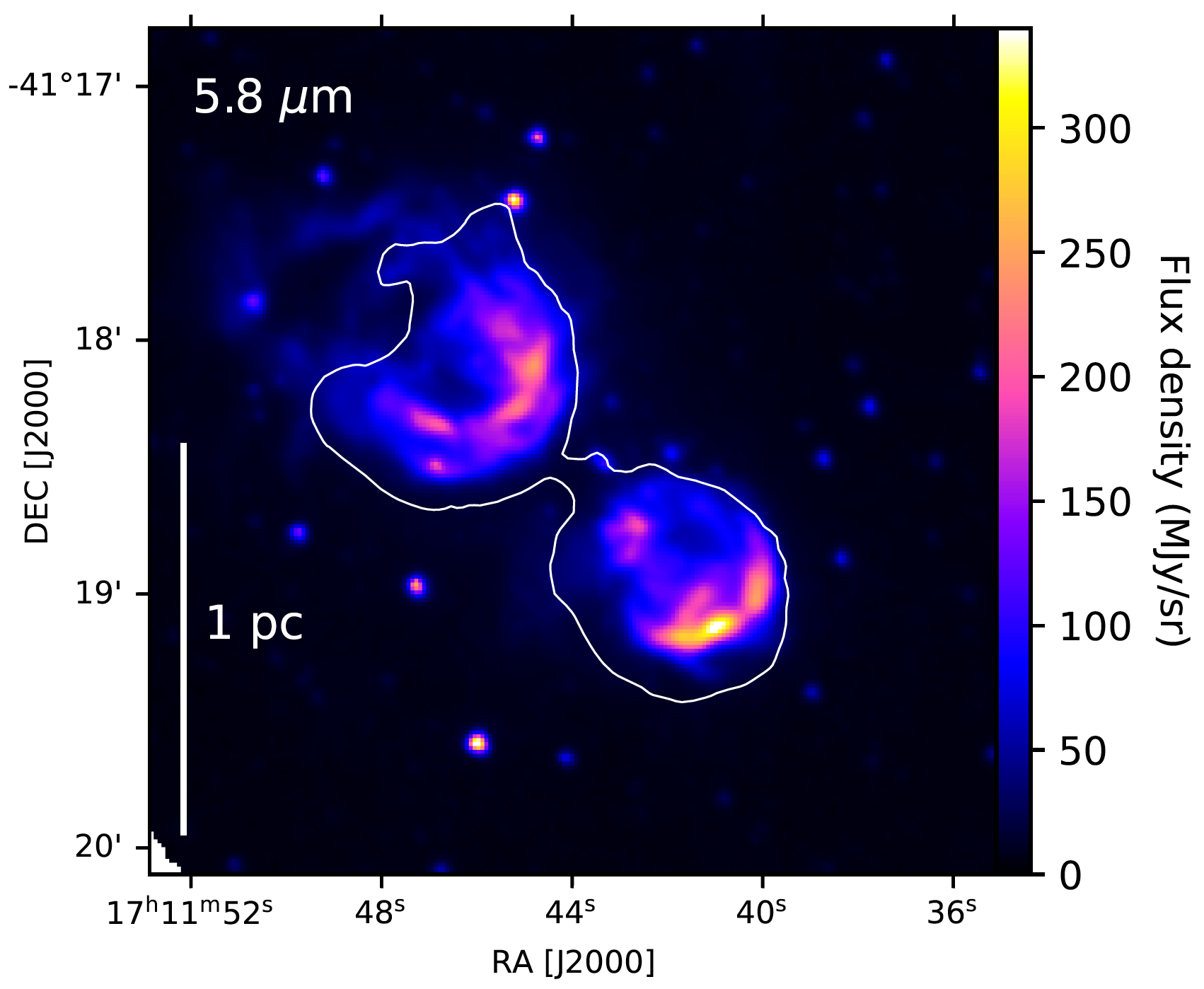}
\includegraphics[width=0.32\hsize]{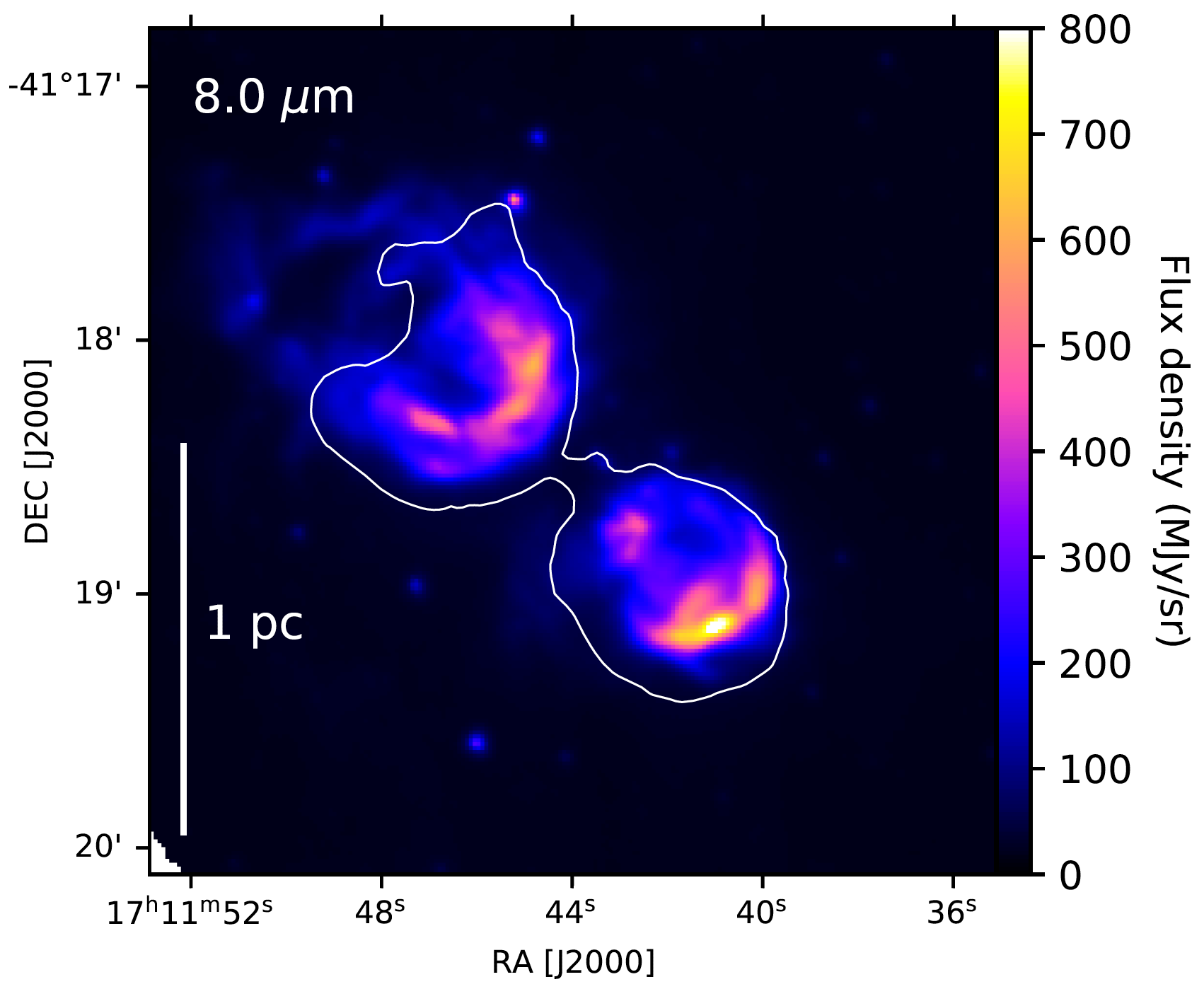}
\includegraphics[width=0.32\hsize]{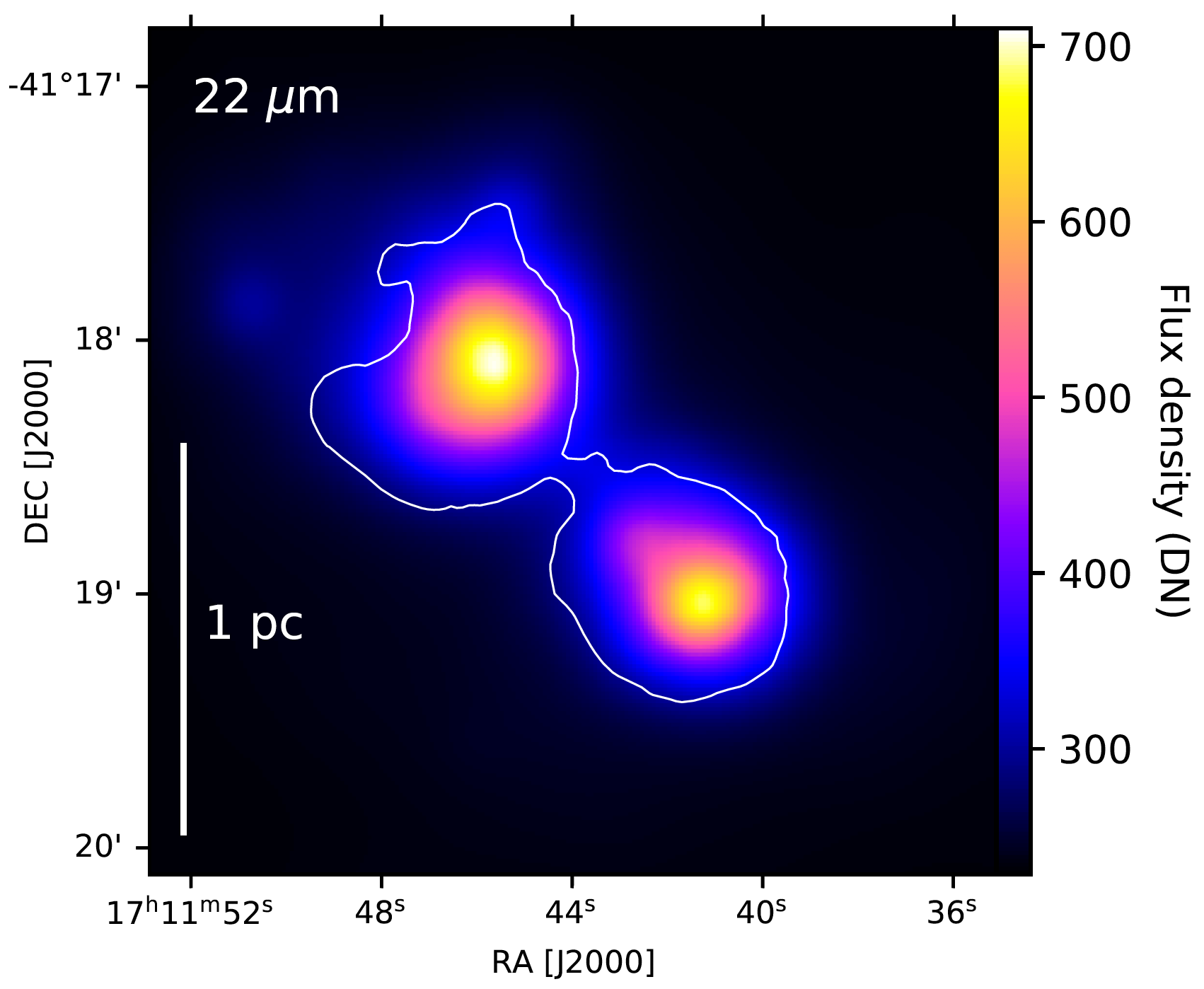}
\includegraphics[width=0.33\hsize]{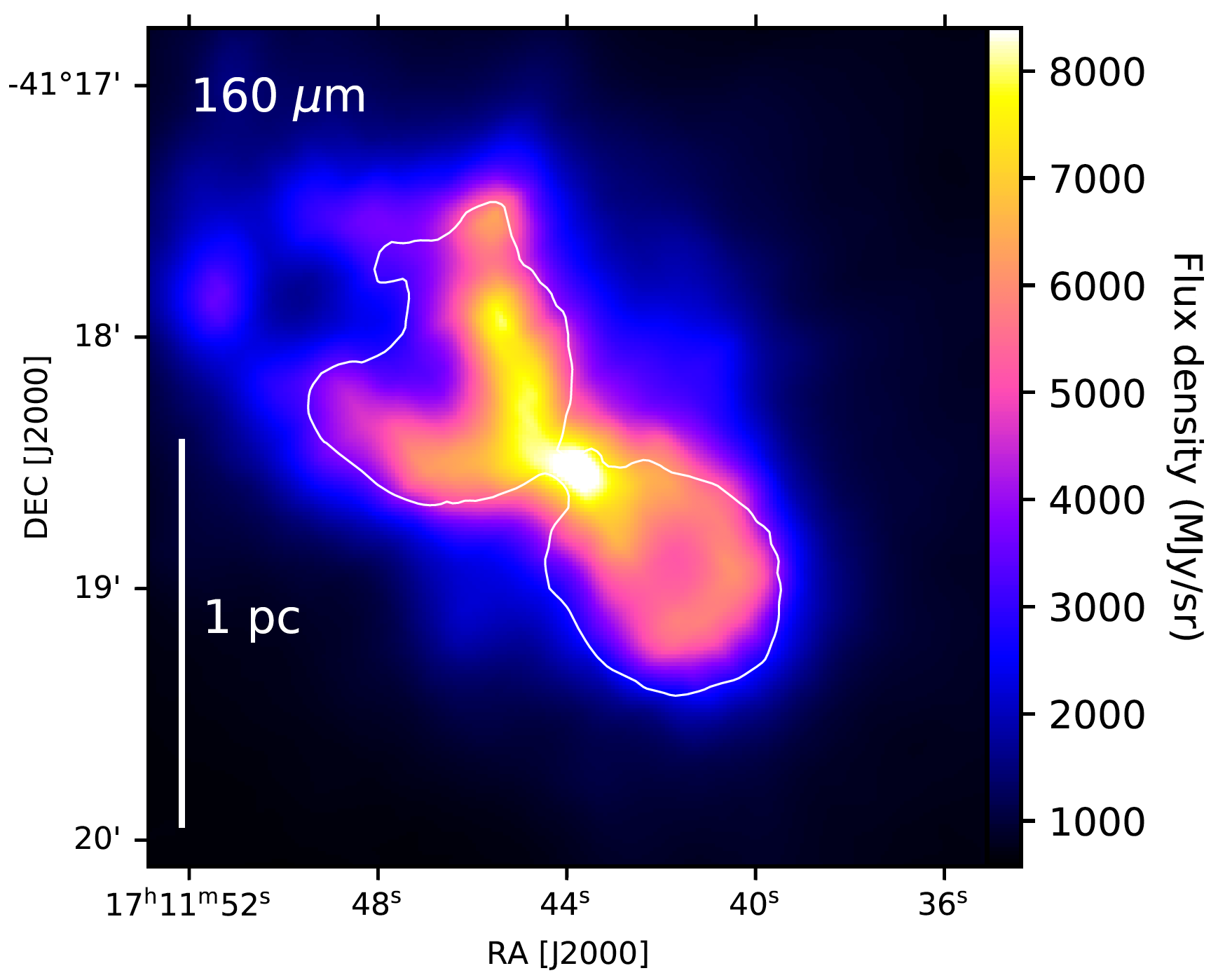}
\includegraphics[width=0.33\hsize]{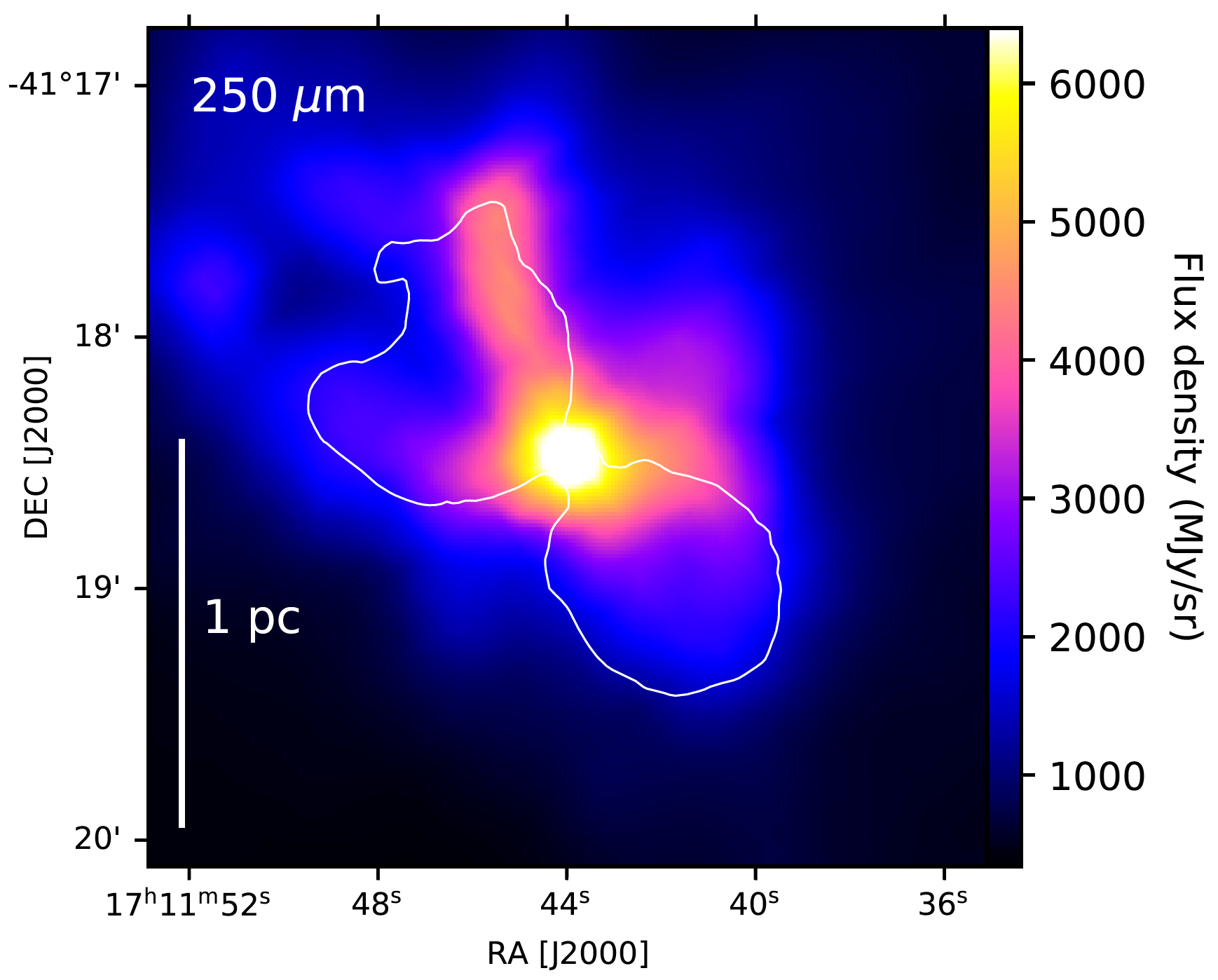}
\includegraphics[width=0.33\hsize]{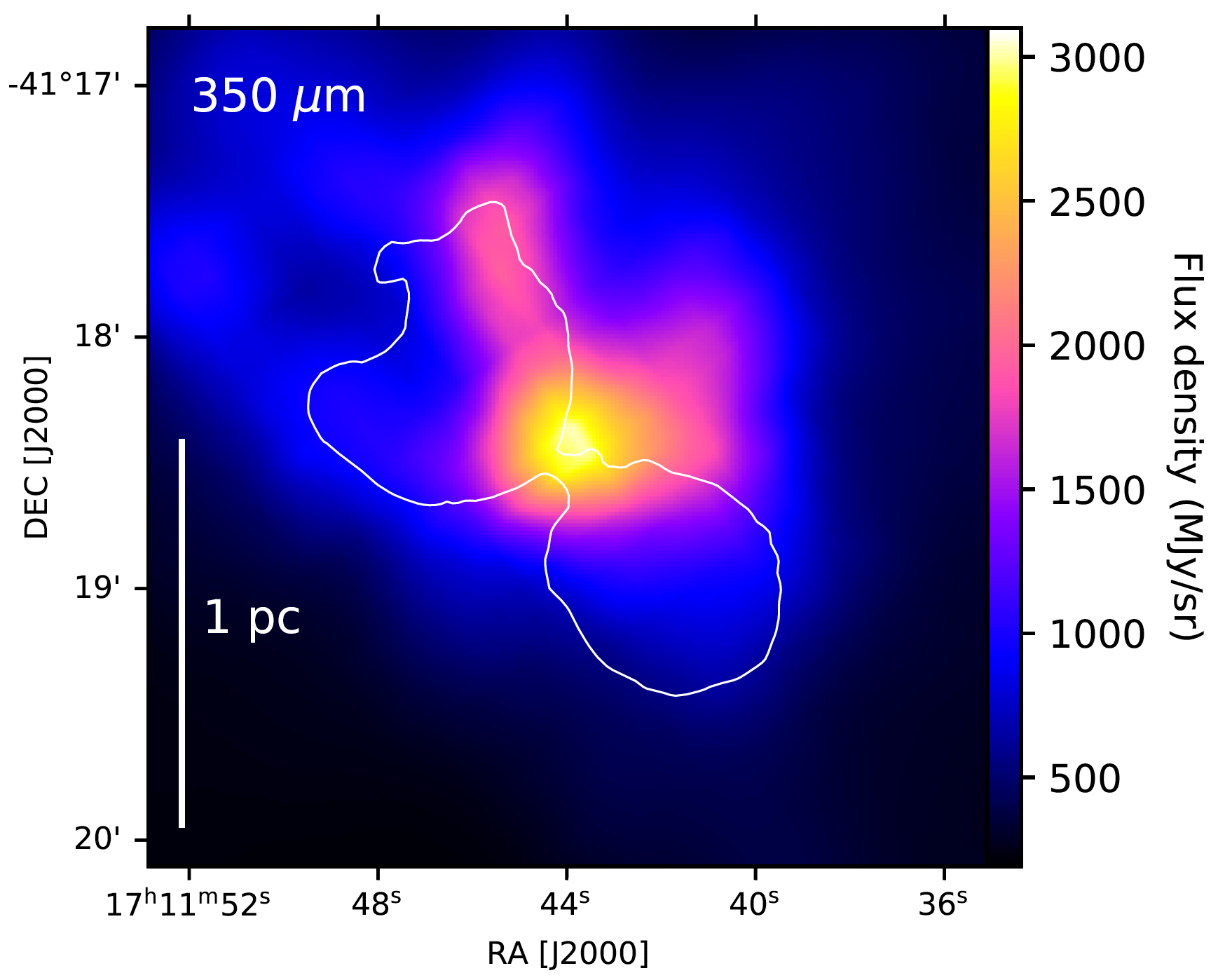}
\includegraphics[width=0.32\hsize]{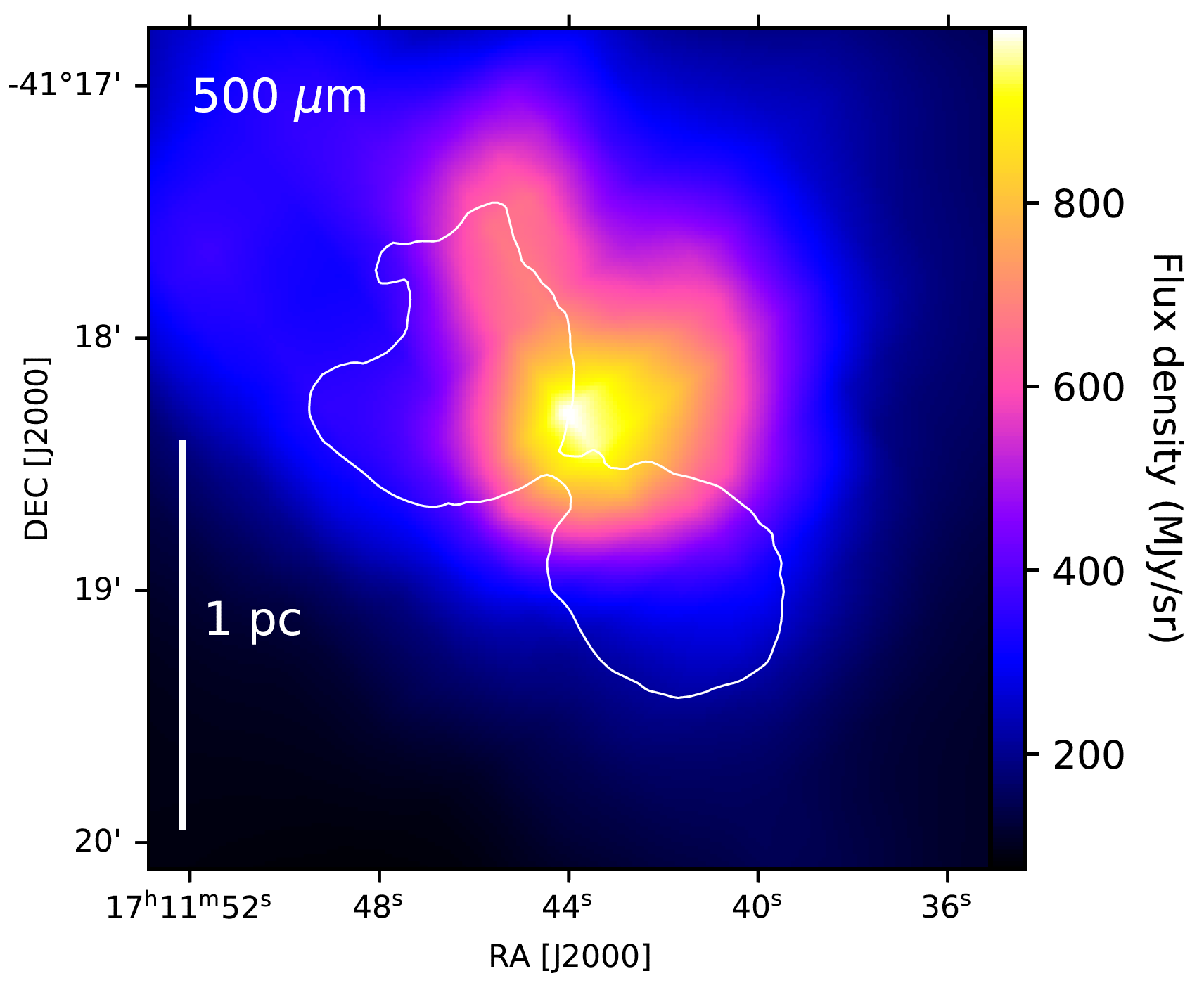}
\caption{Mid-infrared images from IRAC (3.6, 4.5, 5.8 and 8 $\mu$m), WISE (22 $\mu$m) and \textit{Herschel} (160, 250, 350, 500 $\mu$m) of the G345.88-1.10 hub with the overlaid contour indicating the area of the 70 $\mu$m nebulosities (here defined by being brighter than 2000 MJy sr$^{-1}$).
}
\label{photMaps}
\end{center}
\end{figure*}

\section{A luminous bipolar nebulosity embedded within G345.88-1.10}
\subsection{Morphology of the mid-infrared emission}
As mentioned in the introduction, the striking feature of G345.88-1.10 is its central mid-infrared bipolar nebulosity (see Fig.~ \ref{fluxDensCont}). The nebulosity is made of two lobes, N1 and N2, that are located on either side of the H1 fragment. Figure~\ref{fluxDensCont} also shows that, in addition to be warmer, the dust within the nebulosity reaches local H$_2$ column density minima. 
Some 70 $\mu$m emission was recently found in the bipolar cavities of S106 \citep{Adams2015} and RCW 36 \citep{Minier2013} which both contain one or two young O stars at their centre \citep{Ellerbroek2013,Comeron2018,Schneider2018}. However, what is striking about the 70 $\mu$m  emission towards G345.88-1.10 is the dark H1 fragment in which no clear 70 $\mu$m source is observed. If a (proto)star, or a group of (proto)stars, is responsible for powering the mid-infrared nebulosities then it must be deeply embedded within H1. This 70 $\mu$m darkness of the H1 fragment is in stark contrast to S106 and RCW 36, where the 70 $\mu$m emission at the location of the embedded source is an order of magnitude brighter than the cavities \citep{Minier2013,Adams2015}. 

In order to check whether or not the 70 $\mu$m nebulosities might be powered by two independent young clusters, we studied the literature and inspected images at shorter wavelengths down to the J-band. After eye inspection of the VISTA images, \citet{Borissova2011} proposed that G345.88-1.10 hosts two small clusters that are spatially coincident with the mid-infrared nebulosities. For each cluster, they identified 10 potential members which is exactly the lower limit they set in their study to identify a cluster candidate. However, inspecting the 2MASS RGB image of the J-band, H-band and K-band in Fig. \ref{fig:2MASSrgb} we find no convincing indication of a stellar over-density compared to the field stars. In fact, in the systematic 2MASS search for star clusters by \citet{Froebrich2007} no cluster candidates were detected within the G345.88-1.10 region. Furthermore, to our knowledge, no cluster candidates were found towards G345.88-1.10 in the GAIA database either. At longer wavelengths (see Fig. \ref{photMaps}) we do see some compact sources at 3.6 $\mu$m within the boundary of the nebulosities, however, once again, the compact source density is not larger than that of the background. 
Another argument against the presence of two star clusters is the resemblance of the two nebulosities, both in terms of size and brightness. This is visible by eye (see Fig. \ref{photMaps}), but will be quantified further in Sect. \ref{sec:lumSect}. Note that it would also be a remarkable coincidence that both clusters would appear to form a bipolar structure around the fragment with the highest density within the hub. Overall, it seems unlikely that the two infrared nebulosities observed towards G345.88-1.10 are powered by stellar clusters.\\

Finally, looking at Fig. \ref{fluxDensCont} in more detail, two relatively weak compact 70 $\mu$m sources can be spotted, one nearly coincident with the dust column density peak of H3, and the other one just at the boundary of H4. These two sources have emission counterparts at all wavelengths presented in Fig.~\ref{photMaps}, demonstrating that these are protostellar sources. In fact, of the four fragments identified, only the H2 fragment shows absolutely no association with mid-infrared emission. 

\subsection{The spectral energy distribution and luminosity}\label{sec:lumSect}

\begin{figure}
\includegraphics[width=\hsize]{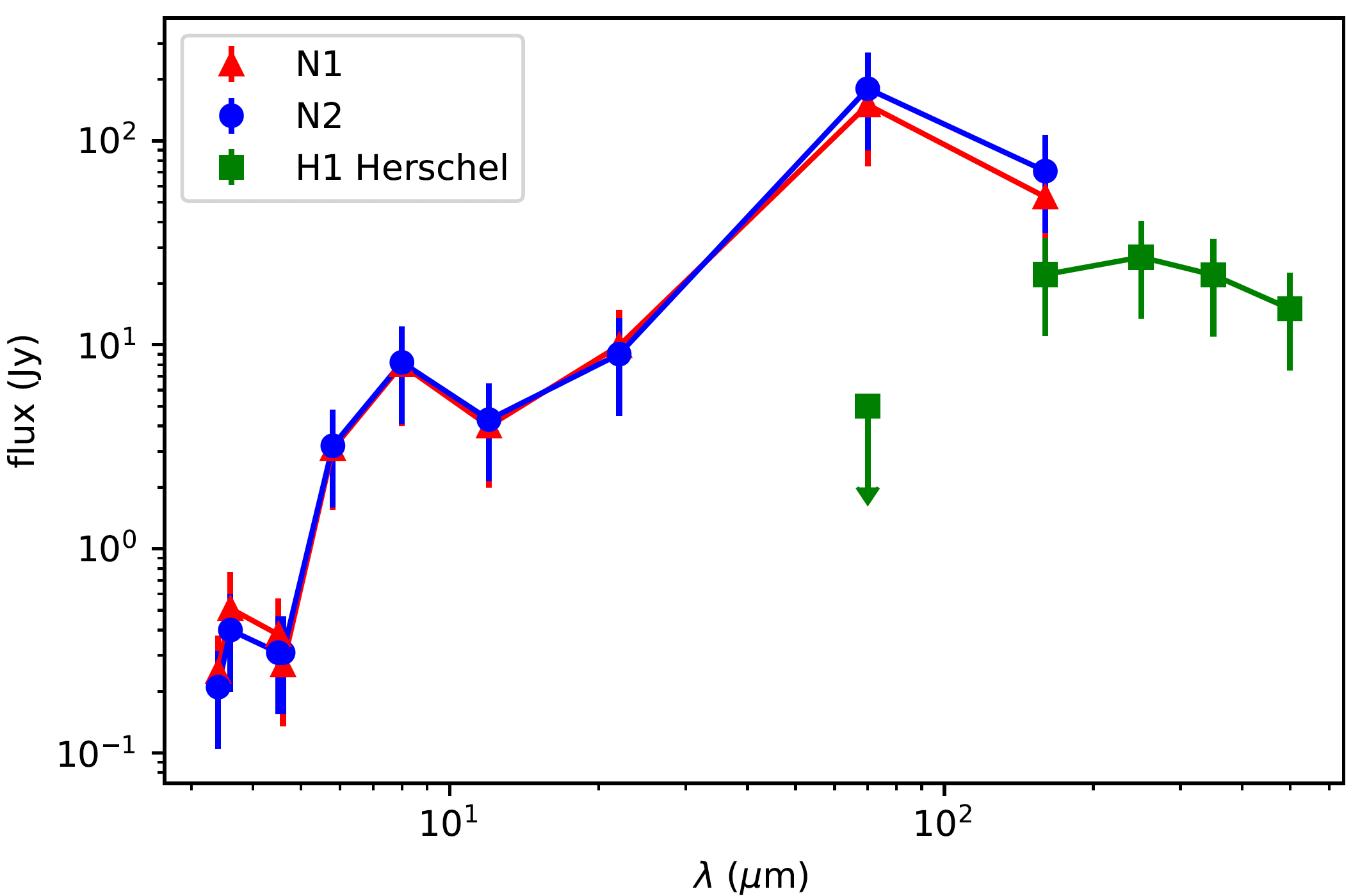}
\caption{Spectral energy distributions obtained from \textit{Herschel}, WISE and IRAC images, for the two nebulosities N1 and N2, and the H1 fragment.}
\label{sedH1}
\end{figure}

In order to determine the respective luminosities of the nebulosities and the H1 fragment we construct the SEDs of N1, N2, and H1 separately. For that purpose, fluxes were obtained from the \textit{Herschel}, WISE and IRAC photometric maps using aperture photometry (see Appendix \ref{app: fluxCalc}). For the H1 fragment, no fluxes were determined at wavelengths shorter than 160 $\mu$m since it is not possible to disentangle the fragment from the nebulosities at these shorter wavelengths. For the nebulosities no fluxes are calculated at wavelengths above 250 $\mu$m as they are probably associated with the ambient cloud. 
The corresponding SEDs are presented in Figure \ref{sedH1}. The first feature one notices is the strong similarity between the shape and magnitude of the N1 and N2 SEDs. Only at wavelengths covered by \textit{Herschel}, the southern nebulosity (N2) becomes slightly brighter. The luminosity of each source is determined by integrating over the entire wavelength range, from 3.6 $\mu$m to 160 $\mu$m for N1 and N2, and from 160 $\mu$m to 500 $\mu$m for H1. For the two nebulosities, we obtain luminosities of $2.1^{+1.2}_{-1.1}\times10^{3}$ L$_{\odot}$ and $2.3^{+1.3}_{-1.2}\times10^{3}$ L$_{\odot}$, respectively. 
The luminosity of the H1 fragment was found to be 51$^{+29}_{-27}$ L$_{\odot}$ only. The luminosity of the entire region (N1 + N2 + H1) is 4.4$\times$10$^{3}$ L$_{\odot}$, and thus dominated by the nebulosities. When considering the upper limit of 5 Jy at 70 $\mu$m for the H1 core, an upper limit of 100 L$_{\odot}$ is obtained for the H1 fragment, which is still a fraction (i.e. $\sim2\%$) of the total luminosity. For a standard protostellar core, the sum of the core and cavity luminosities is generally considered to be representative of the protostellar luminosity, but we will show later that most of the flux from the nebulosities cannot be related to photon emission originating from an embedded object in H1.\\
\begin{table*}
\tiny
\begin{center}
\begin{tabular}{cccccccccccccc}
\hline
\hline
 & \multicolumn{13}{c}{Flux (Jy)}\\
  & 3.4 $\mu$m & 3.6 $\mu$m & 4.5 $\mu$m & 4.6 $\mu$m & 5.8 $\mu$m & 8 $\mu$m & 12  $\mu$m & 22 $\mu$m & 70 $\mu$m & 160 $\mu$m & 250 $\mu$m & 350 $\mu$m & 500 $\mu$m\\
\hline
H1 &  &  &  &  &  & & & & < 5.0 & 22$\pm$4 & 27$\pm$5 & 22$\pm$4 & 15$\pm$3\\
N1 & 0.25$\pm$0.05 & 0.51$\pm$0.10 & 0.38$\pm$0.08 & 0.27$\pm$0.05 & 3.1$\pm$0.6 & 8.0$\pm$1.6 & 4.0$\pm$0.8 & 9.9$\pm$2.0 & (1.5$\pm$0.3)$\cdot$10$^{2}$ & 53$\pm$11 &  &  & \\
N2 & 0.21$\pm$0.04 & 0.40$\pm$0.08 & 0.34$\pm$0.07 & 0.31$\pm$0.06 & 3.2$\pm$0.6 & 8.2$\pm$1.6 & 4.3$\pm$0.9 & 9.0$\pm$1.8 & (1.8$\pm$0.4)$\cdot$10$^{2}$ & 71$\pm$14 &  &  & \\
\hline
\end{tabular}
\caption{Extracted fluxes in Jy for the H1 fragment and the two nebulosities from the \textit{Herschel}, WISE and IRAC photometric maps. An error of 20\% is considered for all fluxes, to take into account uncertainty on the calibration and aperture.}
\label{SEDtable}
\end{center}
\end{table*}

\begin{figure}
\includegraphics[width=\hsize]{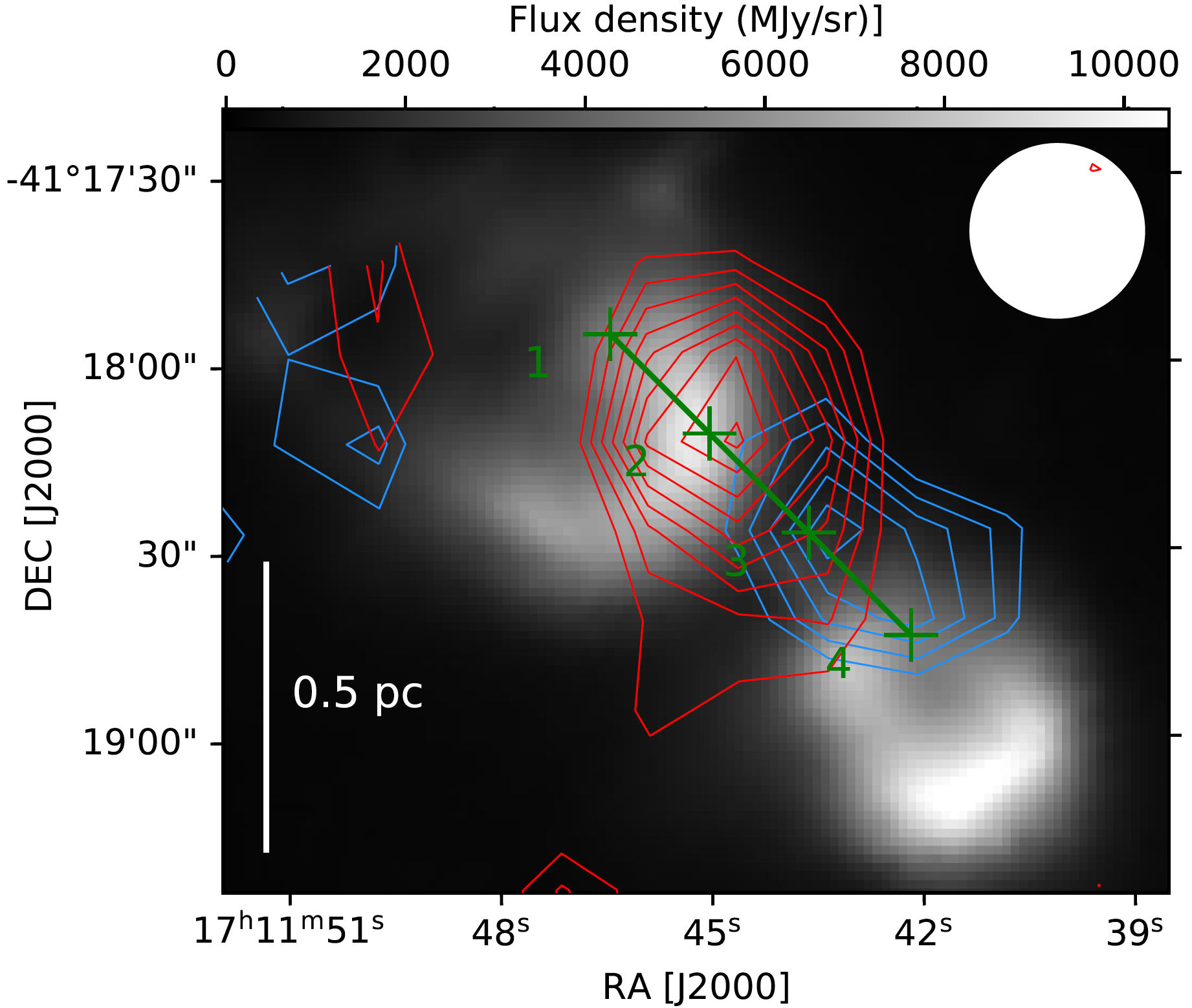}
\includegraphics[width=\hsize]{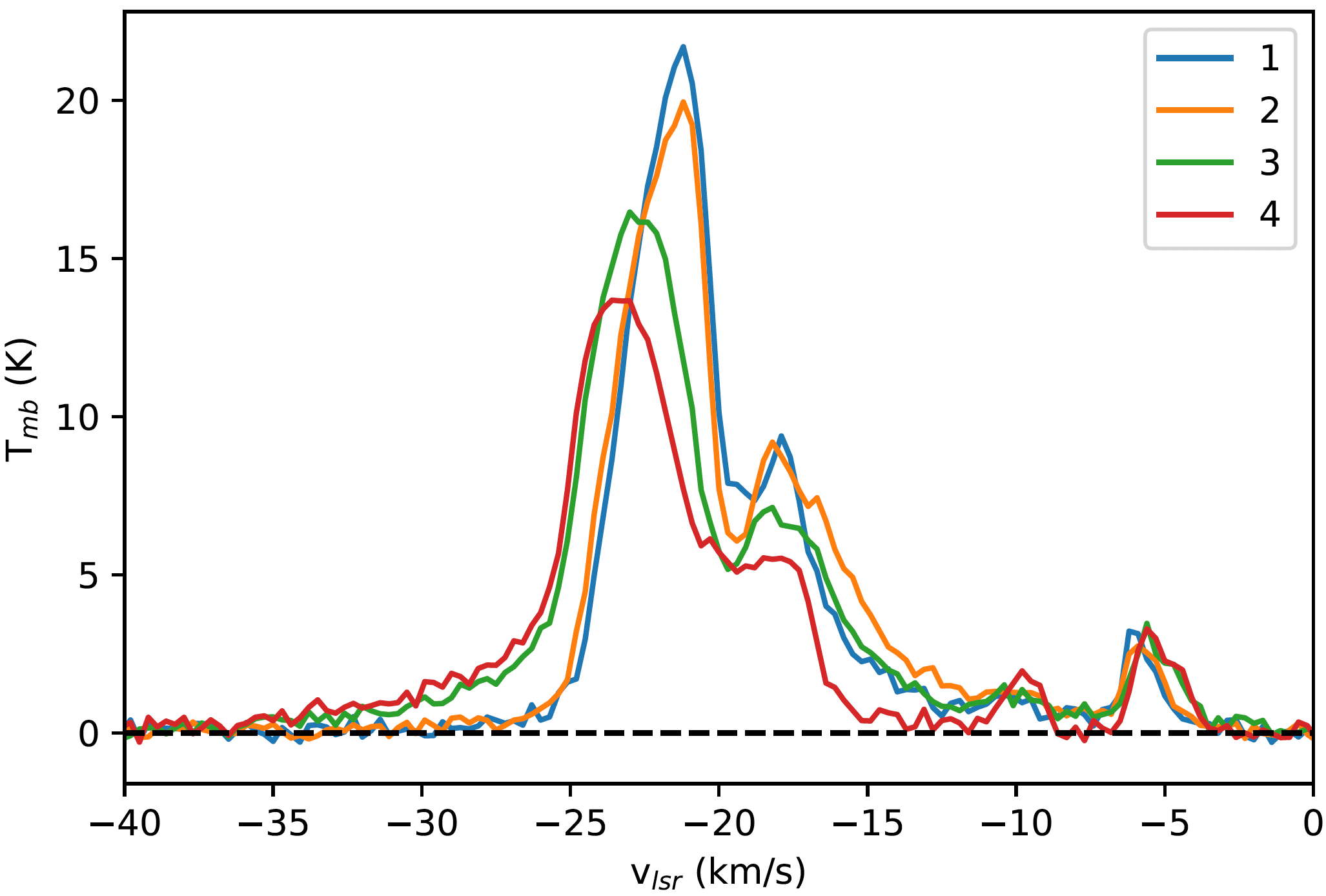}
\caption{(Top) Integrated intensity contours of the redshifted (in red) and blueshifted (in blue) $^{12}$CO(2-1) high-velocity wings overlaid on the 70 $\mu$m map. The contours start at 10 K$\cdot$km s$^{-1}$ with increment of 2 K$\cdot$km s$^{-1}$. The green line and crosses indicate the locations of a spectra presented in the bottom panel with position 3 closely coinciding with the location of the H1 fragment. The white circle at the top indicates the beam size of the $^{12}$CO(2-1) data. (Bottom) $^{12}$CO(2-1) spectra at the indicated locations in the top panel. 
}
\label{outflowDensTemp}
\end{figure}

\section{Associated protostellar outflows}\label{sec:outflowmap}

\subsection{The location of high-velocity wings}
Being centered on the H1 fragment, the bipolar nebulosity at the centre of G345.88-1.10 is reminiscent of outflow cavities. The presence of outflows can be verified by the inspection of the $^{12}$CO(2-1) data. In fact, in Fig.~\ref{fig:invPCygni}, one can already notice the presence of high-velocity wings in the average $^{12}$CO(2-1) spectrum of the hub, a clear signature of outflowing gas. By integrating these $^{12}$CO(2-1) high velocity wings, the location of molecular outflows within G345.88-1.10 can be established. While one could use fixed velocity intervals for the integration of the $^{12}$CO(2-1) wing emission, the presence of velocity gradients within the bulk of the cloud gas would typically lead to large errors. 
In order to minimise such a source of error, the C$^{18}$O(2-1) centroid velocity (v$_{\rm{c^{18}o}}$) and velocity dispersion ($\sigma_{\rm{c^{18}o}}$) were first derived for each pixel (See appendix \ref{sec:coEmissionApp}). After visual inspection, we defined the $^{12}$CO outflow velocity ranges as v$_{\rm{c^{18}o}}+3.5\sigma_{\rm{c^{18}o}} \le$ v$_{\rm{red}} \le$ 0  and -40 $\le$ v$_{\rm{blue}} \le$ v$_{\rm{c^{18}o}}-3.5\sigma_{\rm{c^{18}o}}$. Finally, as can be seen in Fig.~\ref{fig:invPCygni}, contaminating emission is present around v$_{lsr}$ = -5 km s$^{-1}$ km s$^{-1}$ and v$_{lsr}$ = -10 km s$^{-1}$. To avoid contribution from these background clouds, the emission between -6.5 km s$^{-1}$ to -4 km s$^{-1}$ and -11 km s$^{-1}$ to -9 km s$^{-1}$ is left out of the integration of the wings. The resulting integrated intensity maps are represented as contours in Fig. \ref{outflowDensTemp}(top) and show the presence of two unresolved bipolar outflows. The kinematics of these outflows are presented in more detail in App \ref{sec:chanMapSec}, showing no clear signs of further substructure. The brightest of the two bipolar outflows is centered on the H1 fragment, along an axis that is oriented in the same direction as the bipolar nebulosities. The outflow is the brightest towards the region where the nebulosities' edges exhibit the largest H$_2$ column density. Since molecular outflows are considered to be ambient gas that is being entrained by the release of momentum along the jet axis \citep{Gueth1998,Frank2014}, this slight misalignment is to be expected. This is also in agreement with high angular resolution observations of protostellar molecular outflows which show that the low-velocity $^{12}$CO outflow is best detected towards the edges of the outflow cavity \citep[e.g.][]{Tafalla2017,Tabone2017,Louvet2018,DeValon2020}. Altogether, the association between the identified outflow and the bipolar nebulosity confirms the outflow cavity nature of the latter, and from this point forward, we use the terms nebulosity and outflow cavity interchangeably.\\
The two cavities have almost exactly the same luminosity (see above) and similar sizes which suggests that the outflow is close to being in the plane of the sky (i.e. $\theta_{incl} \gtrsim$ 75-80$^{\circ}$, with $\theta_{incl}$ defined from the line of sight) \citep[e.g.][]{Zhang2018}. The N1 cavity is mostly associated with red-shifted outflowing gas and is therefore pointing slightly away from us. On the other hand, N2 is mostly blue-shifted and is thus pointing slightly towards us. As the opening angle of the cavities is relatively large, see Sec. \ref{sec:openingAngleSection} for more details, it would be expected that both lobes contain a blue- and redshifted wing due to projection effects \citep{Cabrit1986,Avison2021}. Although weak, Fig. 7 shows hints of potential blue and red-shifted wings in both lobes even more than a beamsize away from the H1 fragment. Finally, the length of the $^{12}$CO(2-1) outflow, measured starting from the H1 cores, is 0.66~pc for the red-shifted lobe and 0.55 pc for the blue-shifted lobe. These values are not corrected for inclination, but given the large inclination angle of the outflow these values are likely to be close to the actual lengths. 

Although being much weaker than the one associated to the H1 fragment, a second identified bipolar outflow is located around the weak 70 $\mu$m source associated to H4 in the north-east of the map (see Fig. \ref{outflowDensTemp}). No outflow has been detected towards the H2 and H3 fragments with the $^{12}$CO(2-1) data.

\begin{table*}
\begin{center}
\begin{tabular}{cccccc}
\hline
\hline
source & F$_{\text{CO,blue}}$ (56.7$^{\circ}$) & F$_{\text{CO,red}}$ (56.7$^{\circ}$) & F$_{\text{CO,tot}}$ (56.7$^{\circ}$) & F$_{\text{CO,tot}}$ (70$^{\circ}$) & F$_{\text{CO,tot}}$ (80$^{\circ}$)\\
 & 10$^{-5}$ M$_{\odot}$ km s$^{-1}$ yr$^{-1}$ & 10$^{-5}$ M$_{\odot}$ km s$^{-1}$ yr$^{-1}$ & 10$^{-5}$ M$_{\odot}$ km s$^{-1}$ yr$^{-1}$ & 10$^{-5}$ M$_{\odot}$ km s$^{-1}$ yr$^{-1}$ & 10$^{-5}$ M$_{\odot}$ km s$^{-1}$ yr$^{-1}$ \\
\hline
H1 & 27$^{+8}_{-5}$ & 48$^{+16}_{-10}$ & 2.8$\times$10$^{2}$ & 7.7$\times$10$^{2}$ & 3.1$\times$10$^{3}$ \\
H4 & 4.2$^{+1.2}_{-0.7}$ & 24$^{+7}_{-4}$ & 1.4$\times$10$^{2}$ & / & / \\
\hline
\end{tabular}
\caption{Outflow momentum rate obtained for the blue- and redshifted parts of the two outflows. The last three columns display the total outflow momentum rate, assuming symmetric outflow, corrected for the CO opacity at an outflow inclination of 57.6$^{\circ}$, 70$^{\circ}$ and 80$^{\circ}$. The last two columns are only given for the bipolar outflow originating in the H1 fragment as it is expected that it might have such a large inclination angle with the line of sight.}
\label{momentumFluxTable}
\end{center}
\end{table*}

\subsection{The outflow momentum rate}
\label{sec:sectionMomentumFlux}

Integrating the $^{12}$CO(2-1) wing emission over the entire extent of the red- and blue-shifted outflows, one can estimate their CO outflow momentum rate assuming the driving source is located at a velocity of -21 km s$^{-1}$. To calculate this, we used the technique described in \citet{Bontemps1996} and \citet{DuarteCabral2013} (see Appendix \ref{app:calculationMomentumFlux} for more details). 
The results from the calculations are listed in Table~\ref{momentumFluxTable} for both the blue- and red-shifted components of the two bipolar protostellar outflows. To estimate the total outflow momentum rate, it is assumed that both sides eject an equal amount of momentum. The best estimate for the total outflow momentum rate is then obtained by taking two times the maximum value of the two sides \citep{DuarteCabral2013}. Assuming a mean inclination angle of 57.6$^{\circ}$ \citep{CabritBertout1992,Bontemps1996}, the resulting momentum rate of the protostellar outflow originating in H1 is 2.8$\times$10$^{-3}$ M$_{\odot}$ km s$^{-1}$ yr$^{-1}$, which is similar to other CO estimates of the momentum rate at early stages of massive star formation \citep[e.g.][]{DuarteCabral2013,Tan2016}.\\
Using 3$\sigma_{\rm{C^{18}O}}$ or 4$\sigma_{\rm{C^{18}O}}$ (instead of 3.5$\sigma_{\rm{C^{18}O}}$) when defining the outflow wing velocity ranges only changes the outflow momentum rate by $+4\%$ and $-6\%$, respectively. This error is small compared to other sources of uncertainty in the determination of the momentum rate, such as: the inclination angle, the opacity of the CO wings, the symmetry assumption, and the distance. 
As the axis of the protostellar outflow from H1 is expected to be located close to the plane of the sky, the assumed mean inclination could significantly underestimate the momentum rate. Working with a larger inclination of 70$^{\circ}$ or 80$^{\circ}$ increases the outflow momentum rate to 
7.7$\times$10$^{-3}$ and 3.1$\times$10$^{-2}$ M$_{\odot}$ km s$^{-1}$ yr$^{-1}$
, respectively (see Tab. \ref{momentumFluxTable}). 

Finally, even though we have detected only one bipolar outflow centered on the H1 fragment, it is possible that multiple, unresolved, outflows are responsible for the observed outflow cavity. Only higher angular resolution observations of G345.88-1.10 will allow us to conclude on this particular question of outflow multiplicity.

\subsection{Outflow opening angle}
\label{sec:openingAngleSection}

The opening angle of an outflow is often considered an indicator of how evolved the protostellar system is \citep[e.g][]{Arce2007,Kuiper2016}, even though recent Hubble observations have questioned this idea \citep{Habel2021}. 
To determine the opening angle of the outflow emanating from the H1 core, we used both the $^{12}$CO(2-1) emission and the cavities seen with the 70 $\mu$m emission. The estimate of the opening angle was performed by creating a triangle connecting the outer parts of the outflow, which is demonstrated in Fig. \ref{outflowOpeningAngle} of App. \ref{app:appendixOutflowAngle}. This method assumes that the cavity is axi-symmetric. Measurements based on the $^{12}$CO(2-1) outflow give an estimated opening angle of 105$^{\circ}$ for the red-shifted outflow (N1) and 72$^{\circ}$ for the blue-shifted outflow (N2).  Measurements based on the 70 $\mu$m emission give an estimated opening angle of 98$^{\circ}$ for N1 and 76$^{\circ}$ for N2. Despite the limited angular resolution of our $^{12}$CO(2-1) observations, the agreement between the two sets of measurements gives us confidence that the opening angle of the H1 outflow is about 90$^{o}\pm$15$^{o}$. 
These values are significantly larger than  what is traditionally considered to be the upper limit of outflow opening angles for low-mass class 0 (i.e.$\sim 55^{\circ}$) and class I (i.e. $\sim 75^{\circ}$) protostellar outflows \citep{Arce2006}. However, recent work by \citet{Habel2021} did present class 0 protostellar outflows with opening angles up to $\sim$80$^{o}$.

\section{\HII\ emission in the cavities}
\label{sec:hiisection}

\begin{figure}
\includegraphics[width=\hsize]{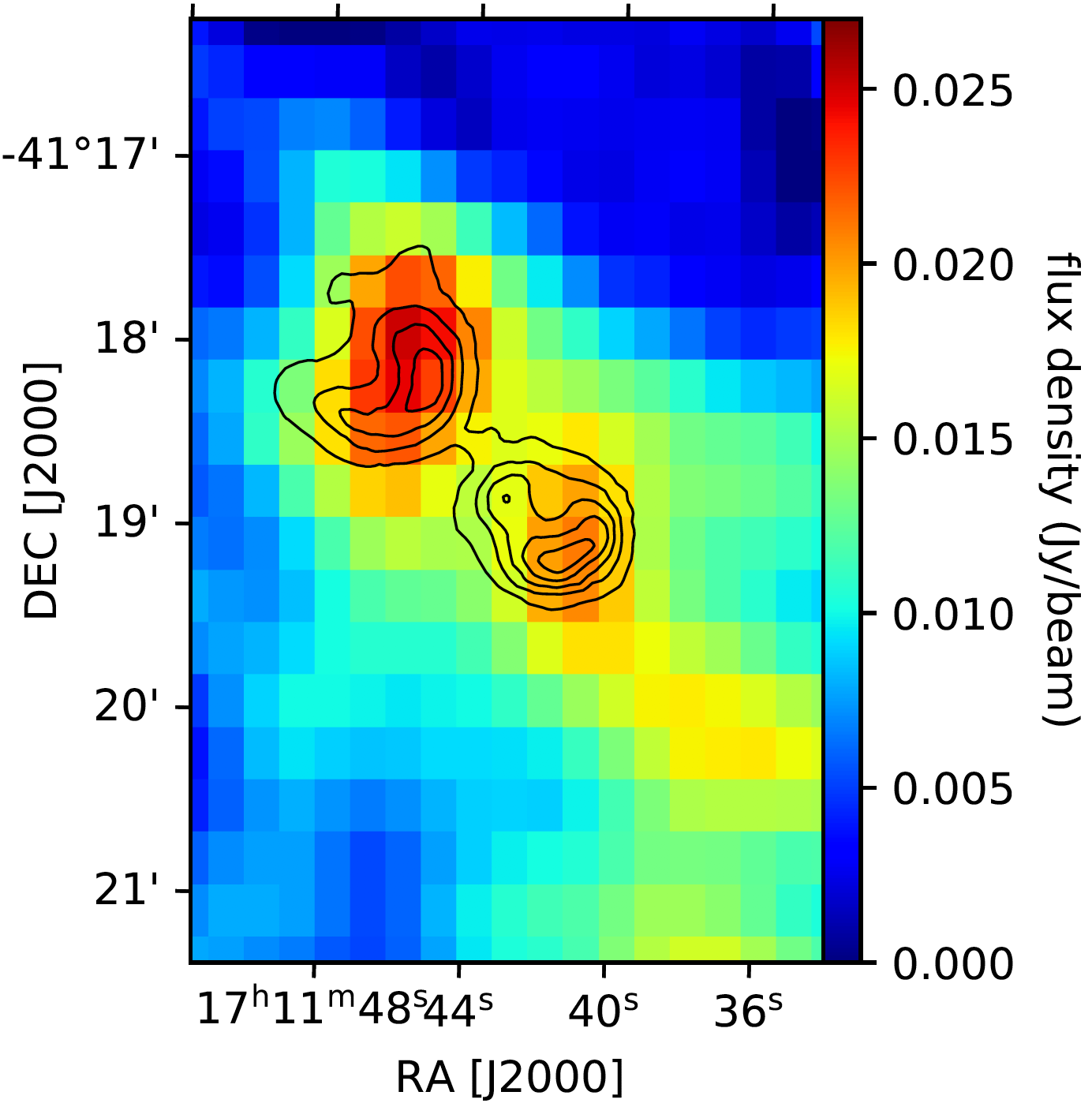}
\caption{Radio continuum image of G345.88-1.10 at 843 MHz from the SUMSS survey \citep{Bock1999}. The 70 $\mu$m emission observed towards the bipolar cavity is highlighted with black contours (2000, 4000, 6000, 8000 and 10 000 MJy sr$^{-1}$).
}
\label{SUMMSplot}
\end{figure}

In the previous section we showed that the presence of mid-infrared bright cavities at the centre of the G345.88-1.10 hubs could be the consequence of protostellar outflow activity. In this section we investigate whether the G345.88-1.10 cavities could also be explained by the presence of ionising radiation from an embedded or undetected massive star (cluster). For that purpose, tracers of \HII\ regions were used. Only two recent radio continuum surveys cover G345.88-1.10. The Parkes-MIT-NRAO (PMN) radio continuum survey \citep{Griffith1993} provides radio continuum data at 4.85 GHz (6.19 cm) at an angular resolution of 5$'$ \citep{Condon1993}. In this survey, G345.88-1.10 remains undetected with an upper limit of 100 mJy. This upper limit can be compared with the flux in the PMN survey for the 16 identified bipolar \HII\ regions near the Galactic plane between galactic longitudes $\pm$ 60$^{\circ}$ \citep{Samal2018,Deharveng2015}. All these bipolar \HII\ regions but one have fluxes that are larger than the flux upper limit for G345.88-1.10 by at least one order of magnitude. Only G051.61-0.36 from \citet{Samal2018} has a flux of $\sim$ 0.2 Jy, which is still a factor two above the upper limit on G345.88-1.10. Having a closer look at the 70 $\mu$m emission of G051.61-0.36 in the HiGAL maps \citep{Molinari2010}, this source looks nothing like G345.88-1.10 as it is a 70 $\mu$m point source without 70 $\mu$m emission in the bipolar cavities. A comparison with this set of bipolar \HII\ regions thus suggests that G345.88-1.10 is not a classical bipolar \HII\ region. 
The other radio continuum survey that covers G345.88-1.10 is the Sydney University Molonglo Sky Survey (SUMSS) at 843 MHz with an angular resolution of 45$''$\citep{Bock1999}. In this survey some very weak radio continuum emission is detected towards G345.88-1.10. (see Fig. \ref{SUMMSplot}). This emission is not centered on the massive H1 fragment but is observed towards both cavities. This emission is very weak, around three times the noise level, which explains why G345.88-1.10 is missing in the SUMSS source catalogue \citep{Mauch2003}. Measuring the SUMSS flux of both cavities we obtain 26 mJy for N1 and 21 mJy for N2. However, it should be noted that G345.88-1.10 is located $\sim$ 30$^{\prime}$ away from a very bright SUMSS source. As a result, a complex low-intensity background of $\sim$10 mJy beam$^{-1}$ can be seen in the vicinity of G345.88-1.10, contributing to an overestimate of the cavity fluxes presented above.\\ 

Lastly, we inspect the H30$\alpha$ recombination line covered by the APEX setup. In order to improve the signal-to-noise ratio we resampled the data to  2~km s$^{-1}$ velocity channels. However, no emission is detected anywhere in the region (see App. \ref{app:h30alpha}). Averaging the spectra over both cavities gives a 3$\sigma$ upper-limit for the line of 63 mK and 110 mK for the N1 and N2 lobes, respectively. Assuming a  typical electron temperature of 7500 K at the galactocentric radius of G345.88-1.10 \citep{Quireza2006}, the equations presented in App. \ref{app:h30alpha} allow us to calculate an upper limit for the cavities' electron density. For N1 and N2 we obtain n$_{e} <$ 3.4$\times$10$^{2}$ cm$^{-3}$ and n$_{e} <$ 4.6$\times$10$^{2}$ cm$^{-3}$, respectively. Combined with the assumed electron temperature, these electron density upper limits translate into thermal pressure upper limits for the ionised gas of 5.0$\times$10$^{6}$ and 6.9$\times$10$^{6}$ K cm$^{-3}$ in N1 and N2, respectively. 
 Compared to the thermal and ram pressure of the molecular gas, estimated to be $\sim 2.6\times10^{5}$ K cm$^{-3}$ (assuming T$_{K}$ = 20 K) and 1.2$\times$10$^{7}$ K cm$^{-3}$ (assuming v = 2 km s$^{1}$) respectively, the thermal pressure upper limit of the ionised gas is below the ram pressure in the hub. This indicates that the clearing of gas and dust at the location of the nebulosities cannot be due to the ionising radiation of an unresolved stellar cluster. This hints at the idea that the protostellar outflow might play a role in the formation of the observed bipolar cavity centered on H1. Using the upper limit at 5 GHz, which typically is optically thin, with the equation from \citet{Schmiedeke2016} gives a more stringent n$_{e} <$ 2.5$\times$10$^{2}$ cm$^{-3}$ and resulting thermal HII pressure $<$ 3.7$\times$10$^{6}$ K cm$^{-3}$.\\
The H30$\alpha$ line and 5 GHz radio continuum thus only provide an upper limit on the pressure from ionised gas, while the 843 MHz continuum emission cannot be used to estimate the electron density, even though it is detected, because the emission can be optically thick at this wavelength. Furthermore, at these long wavelengths the emission might even be the result of synchrotron radiation from a powerful jet associated with the outflow \citep[e.g.][]{CarrascoGonzalez2010,RodriguezKamenetzky2017}. Dedicated radio continuum observations are required to resolve this question.

\section{Radiative transfer calculations}

The mid-infrared images of G345.88-1.10, presented in  Fig.~\ref{fluxDensCont} and \ref{photMaps}, show at its centre a large bright bipolar nebulosity which is closely associated to a bipolar outflow emanating from the H1 fragment. 
Infrared bright outflow cavities are commonly observed in the near-infrared, typically at a wavelength of a couple of microns, where photons are efficiently scattered by dust grains located on the surface of the cavity. In the case of G345.88-1.10, the cavity is the brightest at 70 $\mu$m, a wavelength at which scattering is negligible. Therefore, the mid-infrared dust emission towards the outflow cavities of G345.88-1.10 must be thermal dust emission. The obvious mechanism to heat up dust grains on the walls of an outflow cavity is by direct illumination of a central protostar. The most probable location of that protostar is embedded within the infrared dark H1 fragment. 

To investigate whether radiative heating from a deeply embedded source could explain the observed mid- to far-infrared emission of G345.88-1.10, we performed a series of dust continuum radiative transfer calculations. 

\subsection{Pandora setup}

In this paper we use the python-based radiative transfer framework Pandora \citep{Schmiedeke2016}. Here we briefly summarize the components embedded in Pandora, that we employ in our study. We use RADMC-3D \citep{Dullemond2012} for a self-consistent determination of the dust temperature and calculation of synthetic continuum maps, and MIRIAD \citep{Sault1995} for the post-processing. We explore the parameter space by hand.

\subsubsection{Mesh refinement of RADMC-3D}

For the radiative transfer calculations, we use a square box of 6 pc in size containing a dust distribution and heating sources. In order to resolve dust over-densities present within the box volume, adaptive mesh refinement (AMR) is used \citep{Berger1984,Berger1989} by RADMC-3D. Here, the AMR grid starts with 11 cells along every major axis and is refined into 8 subcells at each refinement level. A cell is refined if the density difference within a cell exceeds 10\%. Then, a minimum cell size is specified, which, in our case, is set to 100 AU to properly resolve all dust distributions that will be studied. For each grid the dust temperature, and therefore the thermal dust emission, is then self-consistently computed using the Monte Carlo method of \cite{Bjorkman2001}.

\subsubsection{Dust continuum images}

When the photon packages leave the volume grid, they produce an image at different wavelengths in units of Jy pixel$^{-1}$. For each of the simulations 10$^{7}$ photon packages were used \citep{Schmiedeke2016}. To compare these synthetic images with the observations of G345.88-1.10, Miriad is employed \citep{Sault1995} in Pandora. Using the beam size at every wavelength, it convolves the artificial image with a Gaussian beam. This returns the convolved image in Jy beam$^{-1}$ which can be easily converted to MJy sr$^{-1}$ units and compared with the observations.

\subsection{Model setup}

\subsubsection{Heating sources}
The central heating source (i.e. the embedded protostar) is modelled as a blackbody with a specific luminosity. Three values were used for this: 500, 2000 or 4000 L$_{\odot}$. In addition to this internal source of heating, the RADMC-3D calculations include an interstellar radiation field (ISRF) which consists of the cosmic microwave background (CMB), the far infrared emission from dust grains \citep{Draine2007}, the starlight background from \citet{Draine2011}, and a UV background from \citet{Mathis1983}.   

\subsubsection{Dust density distribution}
\label{densityDistributions}

Because of the uncertainties related to the unresolved morphology of the H1 fragment, we here used combinations of three different types of density distributions. 

First, we used a modified Plummer model as described in \cite{Qin2011}
\begin{equation}
\label{eq:DensProfProjCode}
n(\bold r) = \frac{n_{c}}{(1+ \mid\bold r\mid^{2})^{\eta/2}}
\end{equation}
where  n$_{c}$ is the central density. The power law index $\eta$ was taken to be 2 \citep{Schmiedeke2016}, and $\mid\bold r\mid$ is given by
\begin{equation}
\label{eq:Requation}
\mid \bold r\mid\hspace{0.1 cm} = \sqrt{\left(\frac{x}{r_{0,x}} \right)^{2} + \left(\frac{y}{r_{0,y}} \right)^{2} + \left(\frac{z}{r_{0,z}} \right)^{2}}
\end{equation}
where r$_{0,x}$, r$_{0,y}$, r$_{0,z}$ are the characteristic sizes of the Plummer model in the direction of the x, y and z axes. The Plummer model is specifically used to model the core or cloud embedding the protostar.\\\\
The second density distribution is a passive flared disk (PFD) which follows the description in \cite{Pineda2011}
\begin{equation}
\label{eq:passiveFlaredDisk}
n(R,z) = n_{0}\left( 1 + \frac{r_{0}}{R} \right)^{\beta - q} \text{exp}\left( -0.5\left(\frac{z}{h(R)}\right)^{2} \right)
\end{equation}
with:
\begin{equation}
\label{eq:scaleHeight}
h(R) = h_{0}\left( \frac{R}{r_{0}} \right)^{\beta}
\end{equation}
In these equations, R is the radial distance in the equatorial plane, n$_{0}$ is the characteristic density, r$_{0}$ the characteristic radius, $\beta$ the disk flaring power, q the surface density radial exponent, and h$_{0}$ the scale height at the characteristic radius. This PFD starts at an inner radius r$_{i}$.

Lastly we also take into account the presence of outflow cavities, which are characterised by a length, an opening angle and a constant density. All models use a cavity opening angle of 80$^{\circ}$ and a length of 10$^{5}$ AU ($\sim$ 0.5 pc).\\\\
Combinations of these three density distributions were used to produce the complete density distribution of the radiative transfer models. The models will be specified in more detail in the next section. When combining the different density distributions in a single model for G345.88-1.10, the total density in each cell $'$j$'$ was determined by
\begin{equation}
\label{eq:totalDensity}
n_{j} = \sum_{i=1}^{N} n_{i,j}(\bold r)
\end{equation}
where N is the number of dust distributions used for the model, and n$_{i,j}$ the density of each density distribution $'$i$'$ in the cell $'$j$'$. This expression is not valid for the outflow cavity regions, where only the constant density of the cavity contributes.

\begin{figure*}
    \centering
    \includegraphics[width=\hsize]{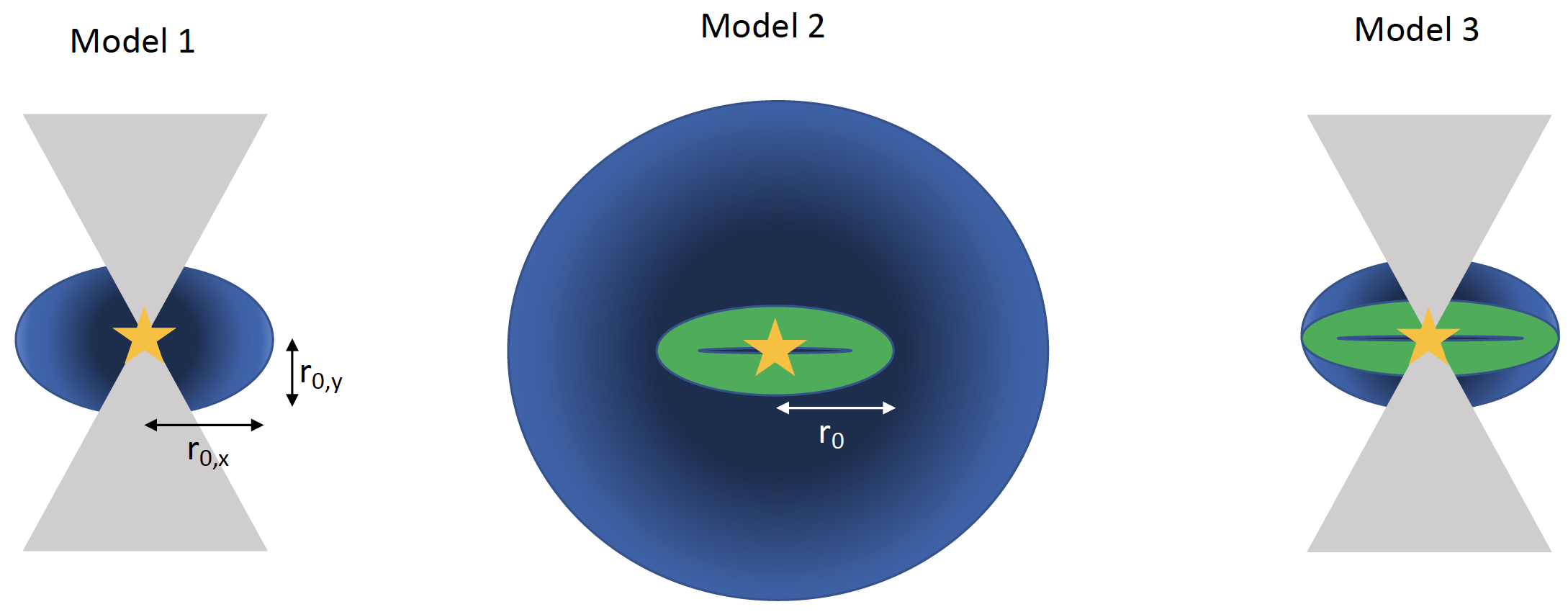}
    \caption{Schematic representation of the three density models used for the radiative transfer calculations. The blue ellipse/circle represents the Plummer (flattened) dense core in Model 1 \& 3, and the Plummer cloud for Model 2. The grey triangle represent the cavities, while the green ring represents the dense central disk. Finally, the central yellow star symbol represents the embedded heating source.}
    \label{fig:sketchModels}
\end{figure*}

\begin{table}
\begin{center}
\begin{tabular}{ccccc}
\hline
\hline
         & Core & Cavity & Disc & Cloud  \\
\hline
 Model 1 & \checkmark & \checkmark&   &  \\
 Model 2 &  & & \checkmark  & \checkmark \\
 Model 3 &\checkmark  &\checkmark & \checkmark   &  \\
\hline
\end{tabular}
\caption{Density distribution elements for all three models}
\label{models}
\end{center}
\end{table}

\subsection{Model grid and parameter space}
To study the observed features in the \textit{Herschel} continuum images of G345.88-1.10 around the H1 fragment, a set of potentially representative radiative transfer models were run. As mentioned before, this set of models uses a central heating source that is embedded in a dust distribution at the origin of the x, y and z axes. The dust distributions for this set of models can be divided into 3 categories, and are from here on referred to as Model 1, 2 and 3. Their composition is presented in Table \ref{models} and Fig. \ref{fig:sketchModels}.\\

\begin{table*}[]
\small
    \begin{center}
    \begin{tabular}{cccc|ccc|cc}
    
    \hline
    \hline
    model & \multicolumn{3}{c}{Plummer core} & \multicolumn{3}{c}{Passive flared disc (PFD)} & \multicolumn{2}{c}{Cavities}\\
    \hline
    & n$_{c,H_{2}}$ (cm$^{-3}$) & r$_{0,x}$ (AU) & $\frac{\text{r}_{0,x}}{\text{r}_{0,y}}$ & n$_{0}$ (cm$^{-3}$) & r$_{0}$ (AU) & h$_{0}$ & size (AU) & n$_{H_{2}}$ (cm$^{-3}$)\\
    \hline
    Model 1& [1, 3, 6]$\times10^{6}$ & [1.5, 4.0, 10]$\times$10$^{3}$ & 0.2,0.6, 1 &  &  &  & 10$^{5}$ & [10$^{0}$,10$^{2}$,10$^{4}$]\\
    Model 2& [10$^{2}$,10$^{4}$] & 5$\times$10$^{4}$ & 1 & [0.2, 0.6, 1]$\times$10$^{7}$ & [0.1, 0.4, 1]$\times$10$^{4}$ & [0.1, 0.3] & & \\ 
    Model 3& 3$\times10^{6}$ & [0.1, 0.4, 1]$\times$10$^{4}$ & 0.7 & [0.2, 0.6, 1]$\times$10$^{7}$ & [0.1, 0.4, 1]$\times$10$^{4}$ & [0.1, 0.3] & 10$^{5}$ & [10$^{0}$,10$^{2}$,10$^{4}$]
    \end{tabular}
    \end{center}
    \caption{Input parameters that were used for the grid of radiative transfer simulations \citep{Bontemps2010,Palau2013,Beuther2015,Beltran2016}, subdivided based on the three different model setups. A combination of all of these input parameters were run. In Model 3, r$_{0,x}$ of the Plummer core and r$_{0}$ of the PFD are fixed to be the same.}
    \label{tab:RTvals}
\end{table*}
For each of the three models, we probed a range of values for most of the parameters in Eqs. (3) to (6). These are summarised in Table \ref{tab:RTvals}. We performed three simulations for each parameter combination, one for a central luminosity of 500 L$_{\odot}$, 2000 L$_{\odot}$ and 4000 L$_{\odot}$ each, leading to a total of 513 different RADMC-3D radiative transfer calculations: 243 setups were run for Model 1, 108 setups for Model 2, and 162 setups for Model 3. The radiative transfer calculations thus cover a wide range of possible configurations. As a consequence of the expected  high inclination angle for the G345.88-1.10 outflow cavity with respect to the line of sight, an inclination of 90$^{o}$ was given to all the models. When running models where the cavities have an inclination of 80$^{\circ}$, it is found, as expected \citep{Johnston2011,Zhang2018}, that there is a substantial difference in brightness for the two cavities. This again indicates that the cavities in G345.88-1.10 have a very small angle with respect to the plane of the sky. 
Note that we limited ourselves to a maximum  luminosity of 4000 L$_{\odot}$ for several reasons: The measured bolometric luminosity of the entire system (cavities and H1 fragment) is about 4000 L$_{\odot}$ (see section \ref{sec:lumSect}); above a luminosity of 4000 L$_{\odot}$ increasingly bright central \HII\ regions are expected to form \citep{Kalcheva2018,Csengeri2018} while we do not detect any in G345.88-1.10; models with L = 4000 L$_{\odot}$ are found to be already too luminous in order to reproduce some key features (see next section).\\
All calculations make use of the thin ice mantle dust grain model as there evidently is some heating in this region \citep{Ossenkopf1994}. But it is likely that, given the range of densities and temperatures within the system, one dust grain model is not enough. This is one of the main limitations of the work we present here.

\subsection{Results}


\begin{figure}
\begin{center}
\includegraphics[width=\hsize]{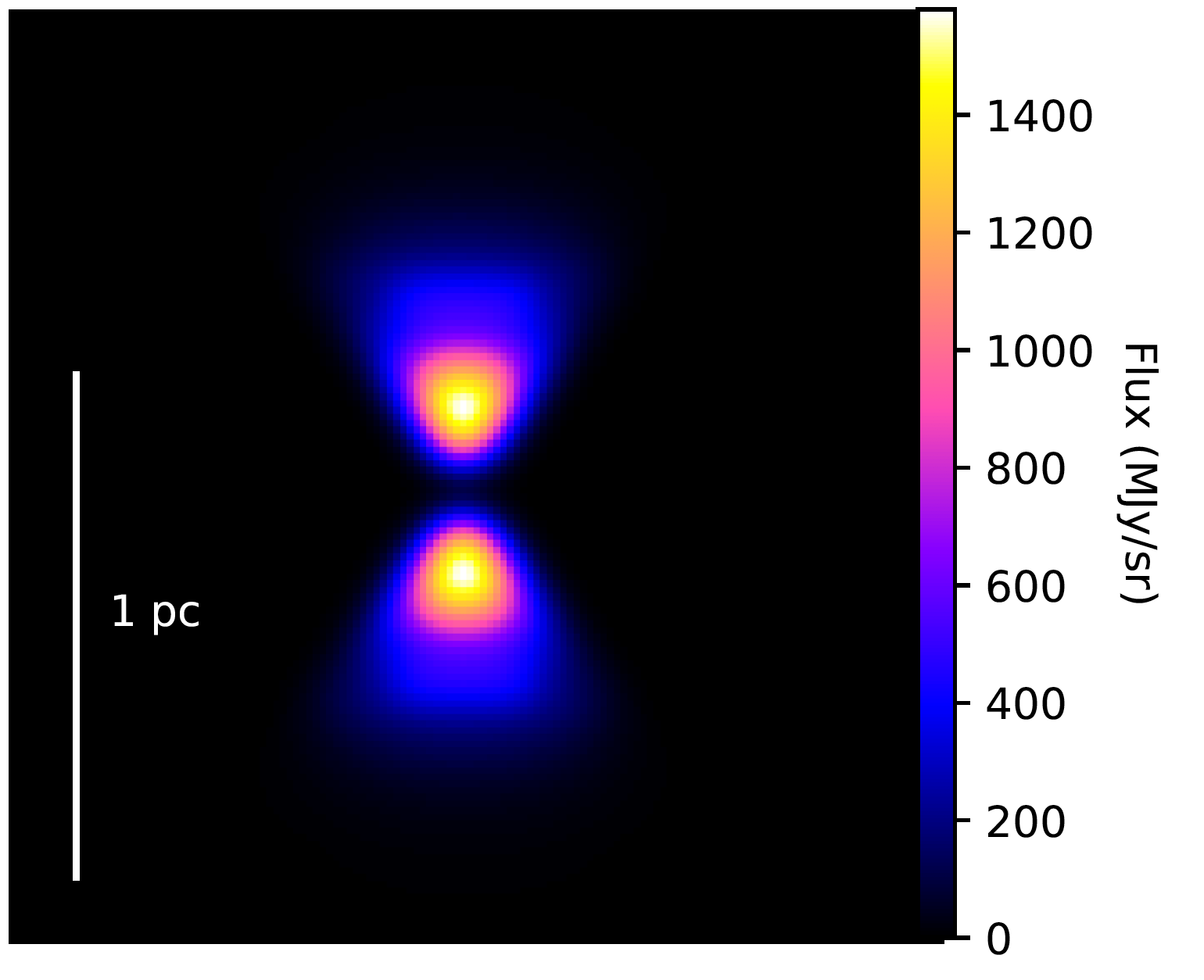}
\caption{70 $\mu$m map of the radiative transfer simulation that produced the brightest nebulosities. These cavities are produced by $'$Model 3$'$ with a central object luminosity of 4000 L$_{\odot}$, a characteristic radius of 10$^{4}$ AU, a characteristic density of 10$^{7}$ cm$^{-3}$ and a cavity density of 10$^{4}$ cm$^{-3}$. This is to be directly compared to Fig.~\ref{fluxDensCont}(left). Note the order of magnitude difference in flux density between the two images.}
\label{nebulositiesRT}
\end{center}
\end{figure}

\subsubsection{The 70 $\mu$m  bright cavities}
\label{RT70micronNebs}
Since the unusually bright 70 $\mu$m cavities is one of the most peculiar aspects of G345.88-1.10, we first focus on whether or not it is possible to produce a bipolar nebulosity that is brighter at 70 $\mu$m than the central core itself. 
Across all 513 radiative transfer calculations we made, about 12\% have their 70 $\mu$m  brightness peaking within the cavities, highlighting the fact that only a small subset of radiative transfer calculations manage to reproduce this unusual feature. The number of runs that do manage to produce a dominant 70 $\mu$m nebulosity is non-uniform across the different models. Model 2 and 3 are more successful at it (see Tab. \ref{RTtabel}) which shows that a dense compact object, a flared disc in our case, is required to produce 70 $\mu$m nebulosities through radiative heating. It can also be inferred from Table \ref{RTtabel} that, within the studied range, the occurrence of nebulosities has little dependence on the source luminosity.\\
Analysing the runs that do produce a 70 $\mu$m-bright nebulosity in more detail, we realise that none of them manages to match the observed brightness of the G345.88-1.10 nebulosity. The run with the brightest nebulosity only recovers about 15\% of the observed 70 $\mu$m intensity for G345.88-1.10. Furthermore, more than 80\% of the nebulosities in the RT calculations produce less 2.5 \% of the observed flux in G345.88-1.10. These low fractions suggest that the observed nebulosities around H1 cannot be solely the result of radiative heating. \\
The 70 $\mu$m dust continuum image with the brightest modelled nebulosity is shown in Fig. \ref{nebulositiesRT}, and was achieved by a run with Model 3. This image can be directly compared to Fig. \ref{fluxDensCont}(left), the observed 70 $\mu$m image of the G345.88-1.10 nebulosity. From this comparison we first notice that there is an order of magnitude difference in brightness, G345.88-1.10 being the brighter of the two. We also find that, while the overall sizes are in relatively good agreement, the nebulosity morphologies are different. Most noticeably, the modelled nebulosities have the highest flux density at the opening of the outflow cavity where dust heating is the most efficient, while the observed nebulosities in G345.88-1.10 have a more complex distribution. This is another indication that radiative heating might not be the driving mechanism behind the observed cavities in G345.88-1.10, although the more complex cavity morphology of the region, compared to the models, likely also plays a role in this.\\
Finally, in G345.88-1.10 the cavity is still bright at 160 $\mu$m, see Fig. \ref{photMaps}. The resulting 70 $\mu$m to 160 $\mu$m flux ratios of the G345.88-1.10 cavities are found to be 2.8 and 2.5 for N1 and N2, respectively (see Tab. \ref{SEDtable}). However, in all radiative transfer calculations, the modelled nebulosity is always brighter at 160 $\mu$m, with maximum 70 $\mu$m to 160 $\mu$m flux ratios of 0.45, 0.58, and 0.65 for the 500 L$_{\odot}$, 2000 L$_{\odot}$ and 4000 L$_{\odot}$ runs, respectively. 

All these results show that there are important differences between the observed nebulosities in G345.88-1.10 and those obtained within our radiative transfer calculations. This implies that even though the flashlight effect, giving rise to an anisotropic thermal radiation field \citep{Yorke2002,Kuiper2010}, could give rise to weak 70 $\mu$m nebulosities around young protostars, it cannot explain the far/mid-infrared properties of the outflow cavities that are observed towards G345.88-1.10 in the performed simulations.

\begin{table*}
\begin{center}
\begin{tabular}{cccc|ccc|ccc}
\hline
\hline
 & \multicolumn{3}{c|}{Model 1} & \multicolumn{3}{|c|}{Model 2} & \multicolumn{3}{|c}{Model 3}\\
 & 500 L$_{\odot}$ & 2000 L$_{\odot}$ & 4000 L$_{\odot}$ & 500 L$_{\odot}$ & 2000 L$_{\odot}$ & 4000 L$_{\odot}$ & 500 L$_{\odot}$ & 2000 L$_{\odot}$ & 4000 L$_{\odot}$\\
\hline
\% of nebulosities & 2.4 \% & 2.4 \% & 0.0 \% & 20 \% & 22 \% & 22 \% & 19 \% & 20 \% & 20 \% \\
\end{tabular}
\caption{Fraction of radiative transfer 70 $\mu$m nebulosities appearing in a perfectly edge-on configuration (in \%) as a function of luminosity and density distribution.}
\label{RTtabel}
\end{center}
\end{table*}

\begin{figure}
\includegraphics[width=\hsize]{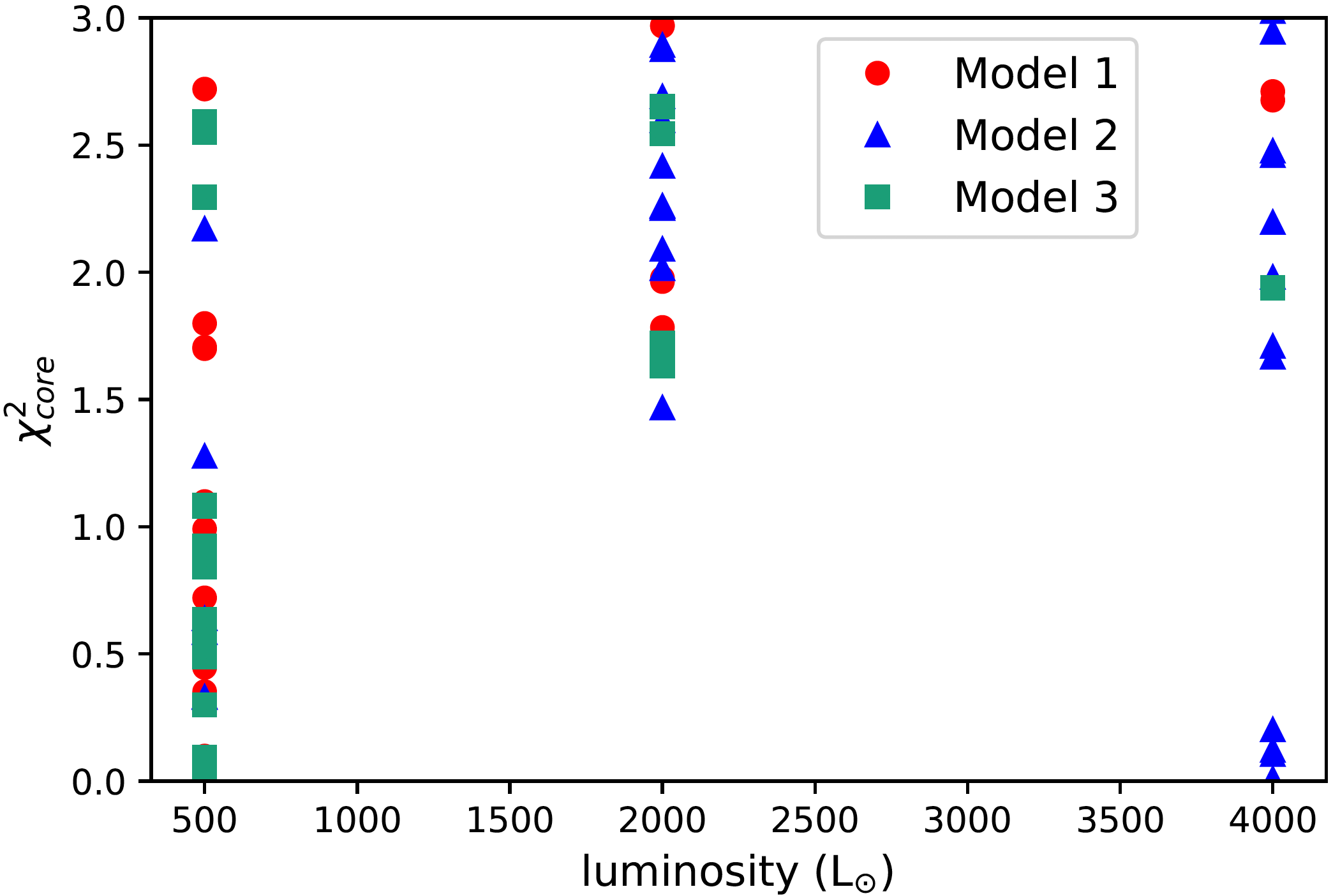}
\caption{$\chi^{2}_{\rm{core}}$ as a function of the central source luminosity for all models with $\chi^{2}_{\rm{core}} < 3$. Model 1: red circle; Model 2: blue triangle; Model 3: green square. 
}
\label{chi2RT}
\end{figure}

\subsubsection{The 70 $\mu$m quiet fragment}\label{sec:70quietRT}

In Fig.~\ref{photMaps}, the H1 fragment is best identified from 250 $\mu$m upwards. At shorter wavelengths, dust continuum emission traces either both the cavities and the core (160 $\mu$m) or the cavities only (from 70 $\mu$m downwards). In order to test the ability of the radiative transfer calculations to reproduce the emission properties of the H1 fragment, we thus focus on the wavelength range between 250 $\mu$m and 500 $\mu$m which also includes the \textit{Herschel} 350 $\mu$m data. The goodness of a specific run is estimated via a reduced $\chi^{2}$ calculation:
\begin{equation}
\chi^{2}_{\rm{core}} = \frac{1}{N}\sum_{\lambda}\frac{\left(S_{\lambda}^{\rm{core}} - O_{\lambda}^{\rm{core}}\right)^{2}}{\sigma_{\lambda}^{2}}
\end{equation}
where S$_{\lambda}^{\rm{core}}$ are the core fluxes estimated from the synthetic images at the  wavelength  $\lambda$ (i.e. 250 $\mu$m, 350 $\mu$m and 500 $\mu$m, respectively). 
$O_{\lambda}^{\rm{core}}$ are the observed fluxes towards the H1 fragment (see Table \ref{SEDtable}). N is the number of wavelengths (i.e. 3), and $\sigma_{\lambda}$ is the uncertainty on the observed flux at the wavelength $\lambda$ (i.e. 20\% of the observed flux), to take into account the distance and calibration uncertainties.

Figure \ref{chi2RT} shows the corresponding $\chi^{2}_{\rm{core}}$ values as a function of the central source luminosity. Only models achieving $\chi^{2}_{\rm{core}} < 3$ are displayed. At first sight, there appears to be no clearly preferred luminosity, both 500 L$_{\odot}$ and 4000 L$_{\odot}$ runs achieve low $\chi^{2}_{\rm{core}}$ values. However, at 4000 L$_{\odot}$ these low $\chi^{2}_{\rm{core}}$  values are only obtained for three specific setups of model 2 (which contains only a disk and ambient cloud around the central heating source), while at 500 L$_{\odot}$ low values are obtained for all three models. Interestingly, we can see that while Model 1 and 3 runs are doing a better job at low luminosity and a worse job at high luminosity, Model 2 runs do not show this tendency. This is a direct consequence of the reprocessing of the radiation from the central protostar by dust. In Model 2, photons can escape the system almost freely and therefore contribute significantly less to heat up the dust within the disc, leading to far-infrared fluxes that are compatible with an infrared quiet core. However, for Model 3 and 1, the dense central Plummer core absorbs and re-emits much of the source radiation. At high luminosity this translates into a central core that is too bright compared to the observed fluxes. As a result, Model 3 and 1 runs only manage to reproduce the 250 $\mu$m, 350 $\mu$m and 500 $\mu$m emission properties of H1 at low luminosity. 

\begin{figure}
    \centering
    \includegraphics[width=\hsize]{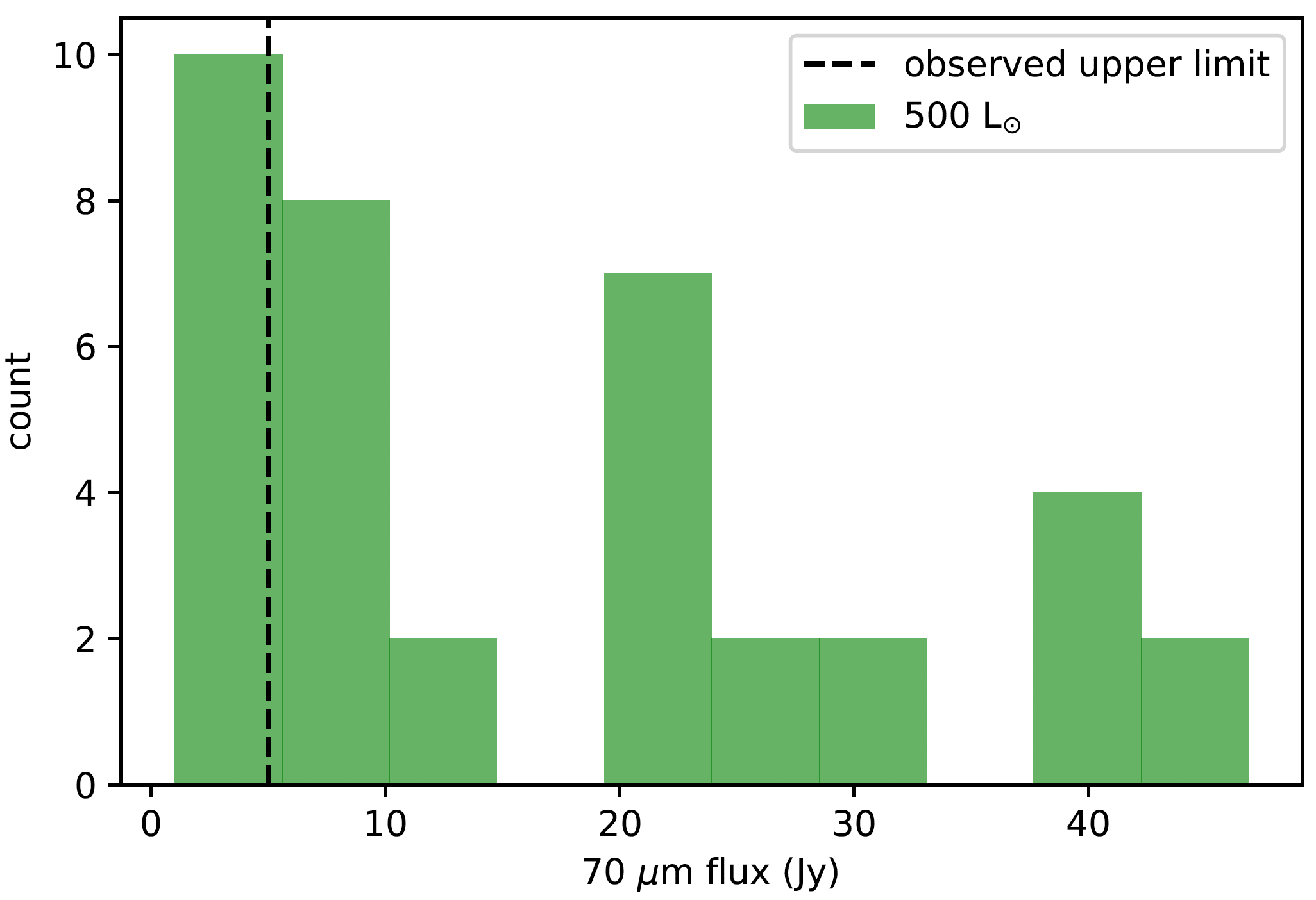}
    \caption{70 $\mu$m flux distributions for the simulated cores with $\chi^{2}_{\rm{core}} \le 2$ for an internal luminosity of 500 L$_{\odot}$. The vertical dashed line indicates the upper limit of 5 Jy observed towards G345.88-1.10.
    }
    \label{fig:core70distribution}
\end{figure}

At 70 $\mu$m, the complexity of the nebulosity emission makes it hard to determine a 70 $\mu$m flux for the H1 fragment. However, we can still derive an upper limit for the H1 70 $\mu$m flux of 5 Jy.  
We can now check whether the radiative transfer calculations that provide the best fits to the emission properties of the core at long wavelengths are also compatible with this 70 $\mu$m flux upper limit. To select the radiative transfer calculations that fit reasonably well the observations of the H1 fragment in the 250 $\mu$m - 500 $\mu$m range, we consider models for which $\chi^{2}_{\rm{core}} \le 2$. 
This provides a set of 54 simulations to work with, 37 of which have a central luminosity of 500 L$_{\odot}$, 9 have a central luminosity of 2000 L$_{\odot}$ and 8 have a central luminosity of 4000 L$_{\odot}$. Within this subset of models, the 2000 L$_{\odot}$ and 4000 L$_{\odot}$ runs have a minimum 70 $\mu$m flux of 81 Jy and 29 Jy, respectively. This is more than a factor of 6 higher than the observed upper limit. For the 500 L$_{\odot}$ runs, the distribution of obtained 70 $\mu$m fluxes are plotted in Fig. \ref{fig:core70distribution}. These radiative transfer calculations produce 70 $\mu$m fluxes that are low enough to be compatible with the observed flux upper limit. This shows that the central protostellar source embedded within H1 must have a luminosity $\le$ 500 L$_{\odot}$. However, for such low luminosity models, the discrepancy at 70 $\mu$m between the observed and modeled cavities increases to several orders of magnitude (see Appendix \ref{sec:combCavCorRT}). This implies that our radiative transfer calculations cannot reconcile the presence of an infrared quiet core with a luminous mid-infrared bright cavity.  

 Lastly, it should be noted that we used the same thin ice mantle dust grain model for all calculations. Changing the dust distribution to include larger grain mantles can increase the emissivity up to a factor three at 70 $\mu$m \citep{Ossenkopf1994}. However, this is insufficient to resolve the orders of magnitude discrepancy with the observations of G345.88-1.10.

\section{Discussion}

In the following discussion, we will consider that the major axis of the G345.88-1.10 bipolar cavity and its associated outflow have a large inclination angle with respect to the line of sight. As discussed in previous sections, this is mostly supported by the similar morphology and brightness of the two lobes of the nebulosity. 

\subsection{A very young protostar generating powerful mass ejection?}

All four fragments identified within G345.88-1.10 are massive and quiet in the mid-infrared. This includes the H1 fragment for which radiative transfer calculations show that the luminosity of the embedded protostar has to be $\le$ 500 L$_{\odot}$. The same radiative transfer calculations also show that radiative heating from such a low luminosity source appears to be unable to account for the order of magnitude brighter cavities. Additionally, with the H30$\alpha$ recombination line and the 5 GHz radio continuum we do not detect a thermal HII region. Therefore, we consider here the possibility that the heating in the cavity is mostly mechanical, originating from shocks of a powerful protostellar outflow. In the following we investigate the plausibility of such a scenario. 

In the case where the cavity brightness is powered by the outflow kinetic energy, then the cavity luminosity could provide an estimate of the outflow mechanical power. This is however complicated by the fact that shock heating is generally not radiated away by dust continuum emission. It could be considered that emission from shock cooling lines like [OI] contributes to the total observed continuum emission (e.g. in the \textit{Herschel} 70 $\mu$m band). This is however very unlikely as this would require excessively bright cooling lines. Recently developed shock models did however show that up to 30\% of shock kinetic flux can escape as FUV Ly$\alpha$ photons \citep{Lehmann2020,Lehmann2021}. These photons could then heat the dust and excite cooling lines within the cavities.\\\\ 

In a scenario where the observed cavity luminosity is powered by the mechanical heating of the protostellar outflow, one may link luminosity and outflow momentum rate via
\begin{equation}
\text{L} = \frac{1}{2}\frac{f_{irr}}{f_{ent}}\text{F}_{CO}\text{v}_{\text{jet}}
\end{equation} 
where v$_{\rm{jet}}$ is the jet velocity.
To correct for the fact that only a fraction of the outflow mechanical energy produces Ly$\alpha$ photons \citep{Lehmann2020,Lehmann2021} that can heat the dust, we include the factor f$_{irr}$. Here, we assume that 20\% of the outflow energy is being radiated away. Further, we use a CO entrainment efficiency of 50\% (i.e. f$_{ent}$; \citet{DuarteCabral2013}) and the observed CO momentum rate at an 80$^{\circ}$ inclination angle  (F$_{CO}$). 
This gives a jet velocity of 3.8$\times$10$^{3}$ km s$^{-1}$. 
This is significantly above the typical velocity range of 100 to 1000 km s$^{-1}$ that is reported for protostellar jets \citep[e.g.][]{Torrelles2011,Gusdorf2017,Anglada2018}. It thus appears that mechanical heating by a protostellar outflow does not provide a straightforward explanation for the high luminosity of the cavities.\\
One should keep in mind that the different parameters (i.e. distance, inclination angle, and CO entrainment efficiency) that enter the jet velocity estimate are highly uncertain. Towards some high-mass star forming regions, it was found that part of the outflow is also detected in [CII] \citep[e.g.][]{Schneider2018,Sandell2020}. This implies that a fraction of the outflow mechanical power might be completely missed by CO observations. If the CO entrainment efficiency is lower than 0.5, this could reduce the required jet velocity towards a more compatible value compared to other protostellar jet velocities. This argument is particularly important if Ly$\alpha$ emission is produced by the shocks as these photons can dissociate CO.


\begin{figure}
    \centering
    \includegraphics[width=\hsize]{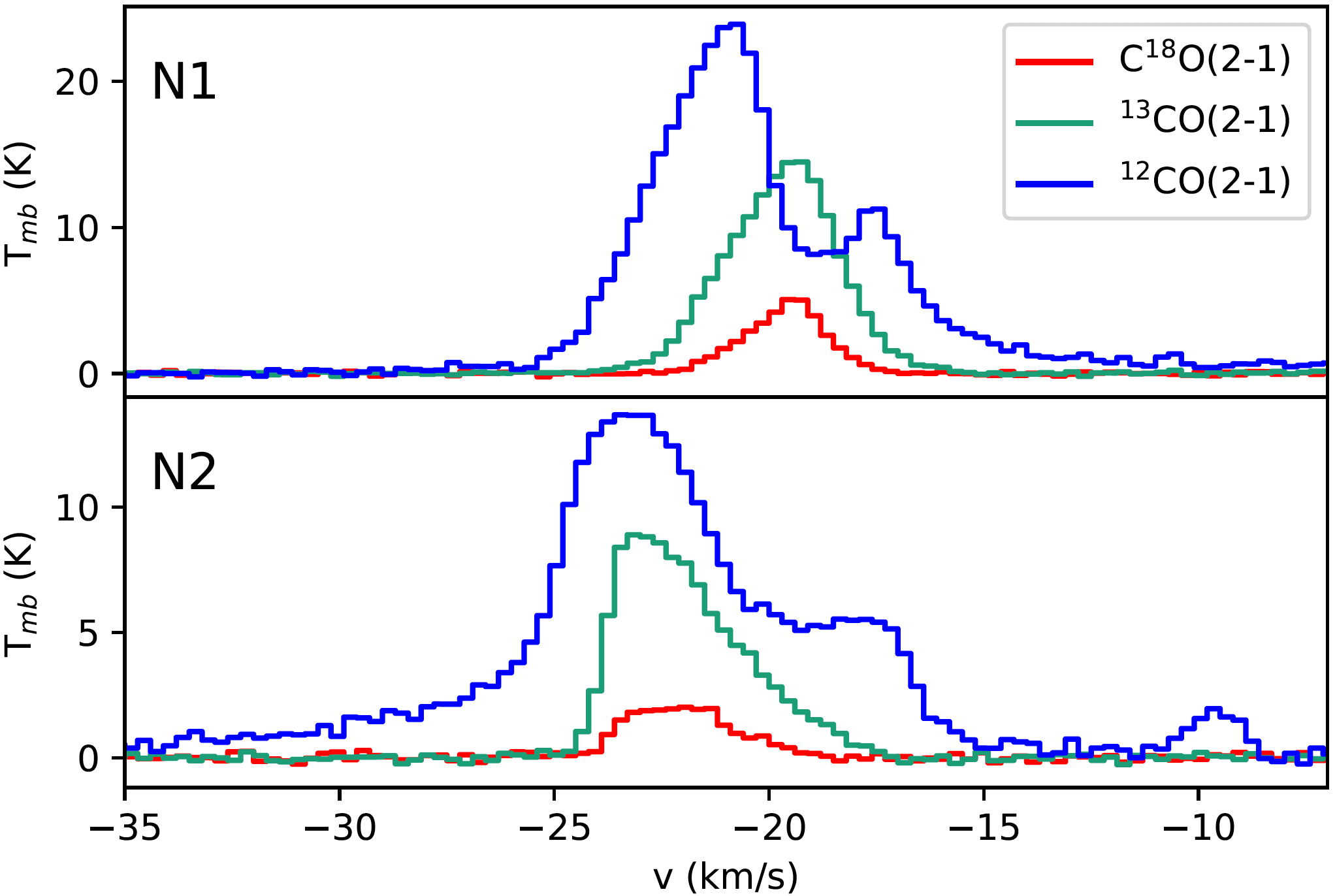}
    \caption{$^{12}$CO(2-1), $^{13}$CO(2-1) and C$^{18}$O(2-1) spectra extracted in the N1 (top) and N2 (bottom) cavities. It is observed that the high-velocity outflow wing, seen in $^{12}$CO(2-1), are at the opposite side of the spectrum with respect to the lower-velocity tails that are also seen in $^{13}$CO(2-1) and C$^{18}$O(2-1) on top of $^{12}$CO(2-1). 
    }
    \label{fig:oppositeWings}
\end{figure}


Finally, if mechanical heating were to be the origin of the source's luminosity, then one might expect evidence of ongoing shocks on the cavity walls. Unfortunately, we do not have shock tracer observations at the moment, but the C$^{18}$O(2-1) and $^{13}$CO(2-1) spectra observed towards G345.88-1.10  present several skewed line shapes (see Fig. \ref{fig:oppositeWings} and App. \ref{sec:coEmissionApp}). A closer inspection of these skewed spectra shows that they are not randomly positioned, but instead are prominent in, and at the edges of, the two lobes of the bipolar nebulosity, suggesting a direct link between skewed line profiles and the cavities. These skewed spectra, especially $^{13}$CO(2-1), show a notable resemblance in shape to the spectra observed with the shock tracer SiO(2-1) around IR quiet massive dense cores in Cygnus-X  with both a narrow and broader components \citep{Motte2007,DuarteCabral2014}. In G345.88-1.10, the velocity range for C$^{18}$O(2-1) and $^{13}$CO(2-1) cover up to 8-10 km s$^{-1}$. This is generally smaller than in \cite{Motte2007} where the tail covers a velocity interval up to 20-30 km s$^{-1}$. However, such a difference could easily be explained by the high inclination angle of the outflow axis with respect to the line-of-sight. Even though the shape of the $^{13}$CO(2-1) and C$^{18}$O(2-1) spectra towards the cavity are suggestive of the presence of shocked gas, it should be noted that with current observations it is not possible to exclude that these spectra might also be associated with strong line-of-sight velocity gradients. High-angular resolution observations of well-known shock tracers will be necessary to resolve this issue. 

\subsection{Global collapse of a hub}\label{sec:discScenario}

\begin{figure}[t]
    \centering
    \includegraphics[width=\hsize]{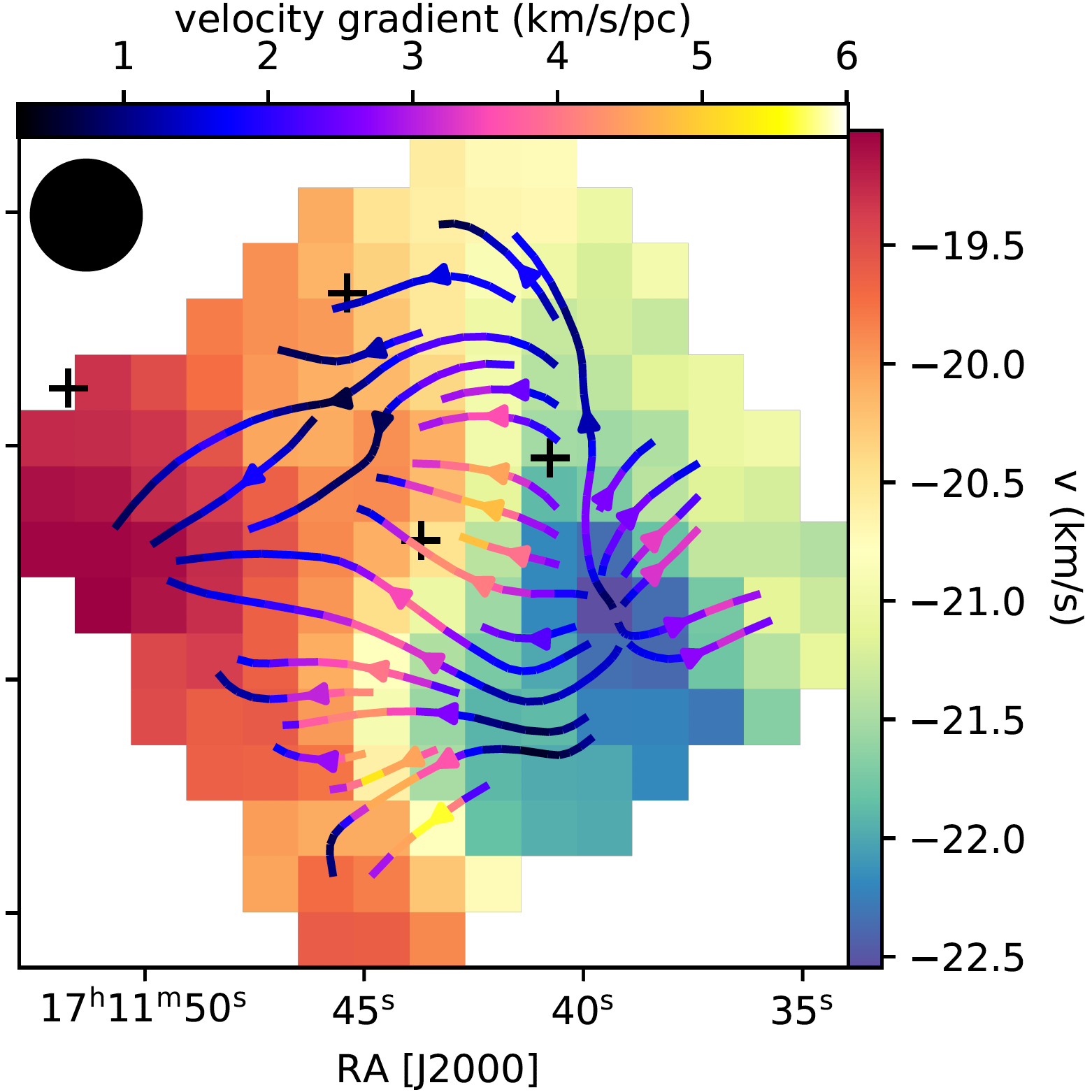}
    \caption{Velocity centroid map obtained by fitting a Gaussian profile to the C$^{18}$O(2-1) spectra. The calculated velocity gradient is overplotted on top of it, using streamplot\textsuperscript{a}, showing organised motion. The colorbar of this streamplot, located on top of the figure, indicates the magnitude of the velocity gradient at each location. The black circle indicates the beam size of the presented APEX observations and the black crosses indicate the position of the 4 massive fragments.\\
    \textsuperscript{a} https://matplotlib.org/stable/api/\_as\_gen/matplotlib.pyplot.streamplot.html}
    \label{fig:velFieldMap}
\end{figure}

Mass rates from infall and protostellar outflows are connected to each other.  While in the low-mass case, it is believed that protostellar outflows are powered by the collapse of protostellar cores, in the high-mass case there are indications that it is the collapse of the entire parsec-scale parent clump \citep[e.g.][]{Avison2021}. In a scenario where the observed mid-infrared cavities might be shaped by the H1 outflow, a link between clump collapse and outflows suggests a link between clump collapse and the presence of cavities. In this section, we investigate the kinematics of the G345.88-1.10 hub in an attempt to characterise its collapse.   

Figure \ref{fig:invPCygni} shows the $^{12}$CO(2-1), $^{13}$CO(2-1), and C$^{18}$O(2-1) spectra averaged over the extent of the G345.88-1.10 hub. While the former two spectra show a double peaked spectrum with predominant blue-shifted spectral emission, the latter shows a single-peaked line centred near the dip of the other two. This is the well-known blueshifted self-absorbed spectral line shape which is often interpreted as a signature of collapse \citep[e.g.][]{Mardones1997}. CO isotopologues are not the most commonly used tracer of gravitational inflow as they tend to trace more external layers of the clouds. However, in cases where clouds are self-gravitating and collapsing on large scales, these tracers can be used to probe infall velocities \citep{Schneider2015a}. 
Using the simple analytical model of \cite{Myers1996}, a first estimate of the line-of-sight infall velocity can be made from the $^{13}$CO(2-1) spectrum, 
which gives a derived infall velocity of 0.5 km s$^{-1}$ (using $^{12}$CO(2-1) gives a similar value of 0.6 km s$^{-1}$). However, \citet{DeVries2005} noted that Myers' model tends to underestimate the infall velocity by a factor 2, leading to a corrected infall velocity $v_{\rm{inf}}\sim 1$ km s$^{-1}$. As a result of the large opacity of CO lines and the depletion of CO isotopologues at moderate densities \citep[e.g.][]{Tafalla2004}, the derived infall velocity is most likely representative of the dynamics of the hub outer layers only.

Figure \ref{fig:velFieldMap} presents a map of the C$^{18}$O(2-1) hub centroid velocity obtained in C$^{18}$O(2-1). On that plot it can be clearly seen that velocity gradients show a rapid increase around the H1 and H2 fragments, in the densest region of the hub. This behaviour has been proposed to be the result of gravitational collapse by several groups \citep[e.g.][]{Hartmann2007,Peretto2007,Peretto2013,Smith2013,Gomez2014,Hacar2017,Watkins2019}. The velocity gradients near the H1 and H2 fragments range from 4 to 6 km s$^{-1}$ pc$^{-1}$, while the average value over the hub is 1.9$^{+0.3}_{-0.2}$ km s$^{-1}$ pc$^{-1}$. Velocity gradients of this magnitude and larger have been observed at significantly smaller scales \citep[e.g.][]{Beuther2015b,Dhabal2018}. However, at the spatial resolution of the APEX data (i.e. $\sim0.3$ pc) these gradients are large when comparing to other star forming regions, whether they are low-mass or high-mass \citep[e.g.][]{Schneider2010,Henshaw2013,Henshaw2014,Peretto2014,Bonne2020}. This indicates that the  G345.88-1.10 hub is particularly dynamic overall, and gas might be accelerated towards its centre, as expected in the case of the global collapse of centrally condensed clouds \citep[e.g.][]{Peretto2007}. However, another origin for organised gas motion is rotation. We therefore checked whether the observed velocity field could be explained by rotation alone. But assuming both spherical and flattened geometries, the rotational energy for the hub is less than 15\% of the hub gravitational energy. Thus, even though we cannot rule out its presence, it is clear that rotation on its own cannot prevent the collapse of the G345.88-1.10 hub. It could be that the hub velocity field is the result of both rotation and kinematics driven by a strong magnetic fields that provides support against collapse \citep[e.g.][]{Inoue2018,Bonne2020,Arzoumanian2022}. However, a simpler explanation would be that gravity is the main driver of the observed velocity field, as suggested by the infall spectral signatures discussed above.

To further investigate the role of gravity one can compare the observationally-derived inflow timescale to the theoretical free-fall time. The inflow timescale is given by:

\begin{equation}
\rm{t}_{\rm{inf}} = \frac{R}{\rm{v}_{\rm{inf}}}
\end{equation}

\noindent where $R$ is the hub radius. Taking $R=1$~pc and  v$_{\rm{inf}}=1$~km s$^{-1}$, we obtain t$_{\rm{inf}}=1$~Myr. On the other hand, the free fall time scale is given by:

\begin{equation}
\rm{t}_{\rm{ff}} = \sqrt{\frac{3\pi}{32G\rho}}
\label{eq:eqToala2012}
\end{equation}

\noindent where $\rho$ is the average mass density within the hub. With a mass of 2900~M$_{\odot}$ within a radius of 1~pc (see Sec. \ref{sec:colDensMap}), we obtain $\rho=4.6\times10^{-20}$ g cm$^{-3}$ which corresponds to a molecule number density of n$_{H_{2}}=1.2\times10^4$~cm$^{-3}$. Plugging this density in the equation of the free-fall time we get $t_{\rm{ff}}=0.3$~Myr. We can conclude that the collapse time of the G345.88-1.10 hub is about 3 times longer than its free-fall time. This implies that the collapse of G345.88-1.10 is not pressure-free, but it is still significantly faster than what we expect for the quasi-static evolution of clumps \citep[e.g.][]{Krumholz2007}. 
Now, assuming that the mass inflow is spherically symmetric we can compute the hub mass infall rate using: 
\begin{equation}
\dot{M}_{\rm{inf}}^{\rm{hub}} = 4\pi R^{2} v_{\rm{inf}}\rho 
\end{equation}
Here we use the same values as those used for the timescales derivation, except for the radius where we use $R=0.5$~pc. The reason for this is that the average hub density $\rho$ is reached at a radius that is about half the external hub radius of 1~pc. With these values in hand we obtain  $\dot{M}_{\rm{inf}}^{\rm{hub}} =2.2\times10^{-3}$~M$_{\odot}$ yr$^{-1}$. This is very similar to what has been derived for the prototypical SDC335 hub filament system \citep{Peretto2013}.\\\\

The gravitational collapse of the G345.88-1.10 hub provides the means for rapid mass provision to the center, along with non-steady accretion flows which might open the cavity on short timescales through an accretion burst. This can be the result of e.g. jet precession \citep{Rosen2020} and opening of the jet itself \citep{Cesaroni2018}. The modeling by \citet{Cesaroni2018} indicates that the jet opening angle can increase to values as high as 40-50$^{o}$, but it should also be noted that this is model dependent. The simulations by \citet{Rosen2020} indicate that a jet precession can be very rapid at the early stages of high-mass star formation with rapid angle changes that on average reach 40$^{o}$ kyr$^{-1}$ with apparent peaks as high as 250$^{o}$ kyr$^{-1}$. However, it should be noted that the opening of the cavities by an accretion burst also faces the issue that this is generally associated with a strong increase of the core brightness in the far-infrared \citep[e.g.][]{Stecklum2021,Hunter2021}, which is not seen in G345.88-1.10. It could be that, after an accretion burst has occurred, there is a time delay between the dimming of the core and the dimming of the cavity. Such a time delay would be of the order of the jet's crossing time of the cavity, i.e. $\lesssim$1000 yr. We could therefore be catching G345.88-1.10 right at that moment where the cavity is still bright but the core has dimmed. Although this would be a noteworthy coincidence, it can fit with the cooling timescale we estimate for the massive H1 fragment which should be lower than 20 years and can be as low as a couple of days or months, see App. \ref{sec:appCoolTimes}. These timescales also seem to be in agreement with surveys of continuum variability, at least in low-mass star forming cores that typically find very rapid changes on the timescales of weeks to several years \citep[e.g.][]{Johnstone2013,Mairs2017,Lee2021}. Because of these very short cooling timescales, it is even possible that multiple accretion bursts spread over few 10 to 100 years are at the origin of the heated cavities. In App.  \ref{sec:appCoolTimes}, we also estimated the cooling timescale in the cavities for which we find values between 1.7$\times$10$^{3}$ to 2.7$\times$10$^{4}$ yr which is significantly longer than for the core. Furthermore, the lower end of this estimated timescale is of the same order of magnitude as the crossing timescale for the jet in the cavities. This might explain why the full cavity is bright instead of only several subregions in the cavity.\\

\section{Summary and future prospects}
We have presented the first study of the G345.88-1.10 hub filament system ($d=2.26^{+0.30}_{-0.21}$ kpc, M=2.9$\times$10$^{3}$ M$_{\odot}$ within a radius of 1 pc). The analysis in this paper focuses on the massive 70 $\mu$m-quiet H1 fragment, located at the hub centre, and the associated bright bipolar cavity. We combined archival \textit{Herschel}, WISE, Spitzer and 2MASS data with APEX observations of $^{12}$CO(2-1), $^{13}$CO(2-1), C$^{18}$O(2-1) and H30$\alpha$ towards the hub. The main results are:\\\\
- The H1 fragment has an estimated mass of 2.1$^{+1.2}_{-1.1}\times10^{2}$ M$_{\odot}$ within a radius $R_{\rm{eff}}=0.14$~pc, and a luminosity upper limit of 10$^{2}$ L$_{_\odot}$ from direct integration of the SED.\\\\
- Bipolar infrared nebulosities are centered on the H1 fragment. The two lobes of the nebulosity have similar overall shapes and luminosities, with a combined luminosity  $\sim$4.4$\times$10$^{3}$ L$_{\odot}$. It is believed that the main axis of the bipolar nebulosity is lying close to the plane of the sky. \\\\
- $^{12}$CO(2-1) high velocity wings,  centred on the H1 fragment, are observed in the bipolar nebulosities with an estimated momentum rate of 3.1$\times$10$^{-2}$ M$_{\odot}$ km s$^{-1}$ yr$^{-1}$. 
The nebulosities thus trace the outflow cavity of a powerful protostellar outflow emanating from the H1 fragment with an opening angle of $\sim$ 90$\pm$15$^{\circ}$.\\\\
- Weak 843 MHz radio continuum emission is detected towards the cavities in the SUMSS survey. 
However, H30$\alpha$ and 5 GHz emission is not detected. This puts upper limits on the thermal pressure of potential \HII\ gas. As a result, ionising radiation appears unable to open the cavities.\\\\ 
- Radiative transfer calculations of a disc/core/cavity system confirms that the luminosity of the embedded source in the massive H1 fragment has to be $\le$ 500 L$_{\odot}$. Most importantly, these calculations also show that the bright 70 $\mu$m nebulosities at the centre of G345.88-1.10 cannot be explained by the radiative heating from an embedded object.\\\\
- Potential heating of the cavities by shocks from the powerful protostellar outflow also appears unable to explain the observations.\\\\
- CO isotologue emission lines show evidence for a global collapse of the hub with a corresponding large mass infall rate (i.e. $\sim10^{-3}$~M$_{\odot}$ yr$^{-1}$). \\\\

In conclusion, G345.88-1.10 is an intriguing source where the combined presence of an infrared dark fragment (H1) and associated  mid-infrared bright bipolar cavities remains unexplained. The uniqueness of G345.88-1.10 strongly suggests that it traces a short and currently unknown phase of the high-mass star formation process.\\\\

The two main limitations of the study presented in this paper are the angular resolution of the available data ($\sim$ 20$^{\prime\prime}$-30$^{\prime\prime}$, i.e. 0.2 to 0.3~pc at the distance of G345.88-1.10) and the lack of shock tracers.\\
ALMA observations at (sub)arcsecond resolution of the molecular outflow, fragments, cavity walls, and hub filaments would resolve a number of unsolved issues and place G345.88-1.10 in the larger context of high-mass star formation. Such observations would allow to: i. determine the nature of the molecular outflow, i.e. whether it is made of a single outflow or multiple ones; ii. measure the mass of the core(s) that are driving the outflow(s) and e.g. place them in a luminosity/mass evolutionary diagram \citep{DuarteCabral2013}; iii. map the spatial distribution of shocked gas via the detection of SiO emission; iv. map the velocity field of the hub at higher resolution, showing in more detail how the large-scale inflow of matter is connected to the smaller-scale core/disc.\\
Additionally, SOFIA observations of \CII\, [OI], H$_2$, and high-J CO transitions would allow to specifically look at the physics at play within the cavities, directly tackling the question of the heating mechanism. Lastly, sensitive arcsecond resolution radio continuum and RRL observations of the cavities with ATCA or a SKA precursor could allow the detection of an eventual radio jet at the origin of the cavities.

\begin{acknowledgements}
      We thank the referee for insightful comments that significantly improved the content of this paper. We thank P. Clark and S. Mairs for helpful discussions on the cooling time and dust continuum variability of dense cores. This publication is based on data acquired with the Atacama Pathfinder Experiment (APEX) under program ID 097.F-9301(A). L.B. acknowledges support from the Erasmus+ program for his stay at Cardiff University. L.B. and N.S. acknowledge support by the French ANR and the German DFG through the GENESIS project (ANR-16-CE92-0035-01/DFG1591/2-1). N.P. acknowledges the support from  STFC under the consolidated grant number ST/R005028/1. A.S. acknowledges the support of the Max Planck Society. This research made use of Astropy \citep[https://www.astropy.org/][]{Astropy2013,Astropy2018}, a community-developed core Python package for Astronomy. This research made use of APLpy, an open-source plotting package for Python \citep{Robitaille2012}.
\end{acknowledgements}

%
   \bibliographystyle{aa} 
   \bibliography{citations1.bib} 
\bibliography{template.bib} 
%

%
%
%
%
%
%
%
%
%

\begin{appendix}

\section{Flux determination}
\label{app: fluxCalc}
\begin{figure}
\begin{center}
\includegraphics[width=\hsize]{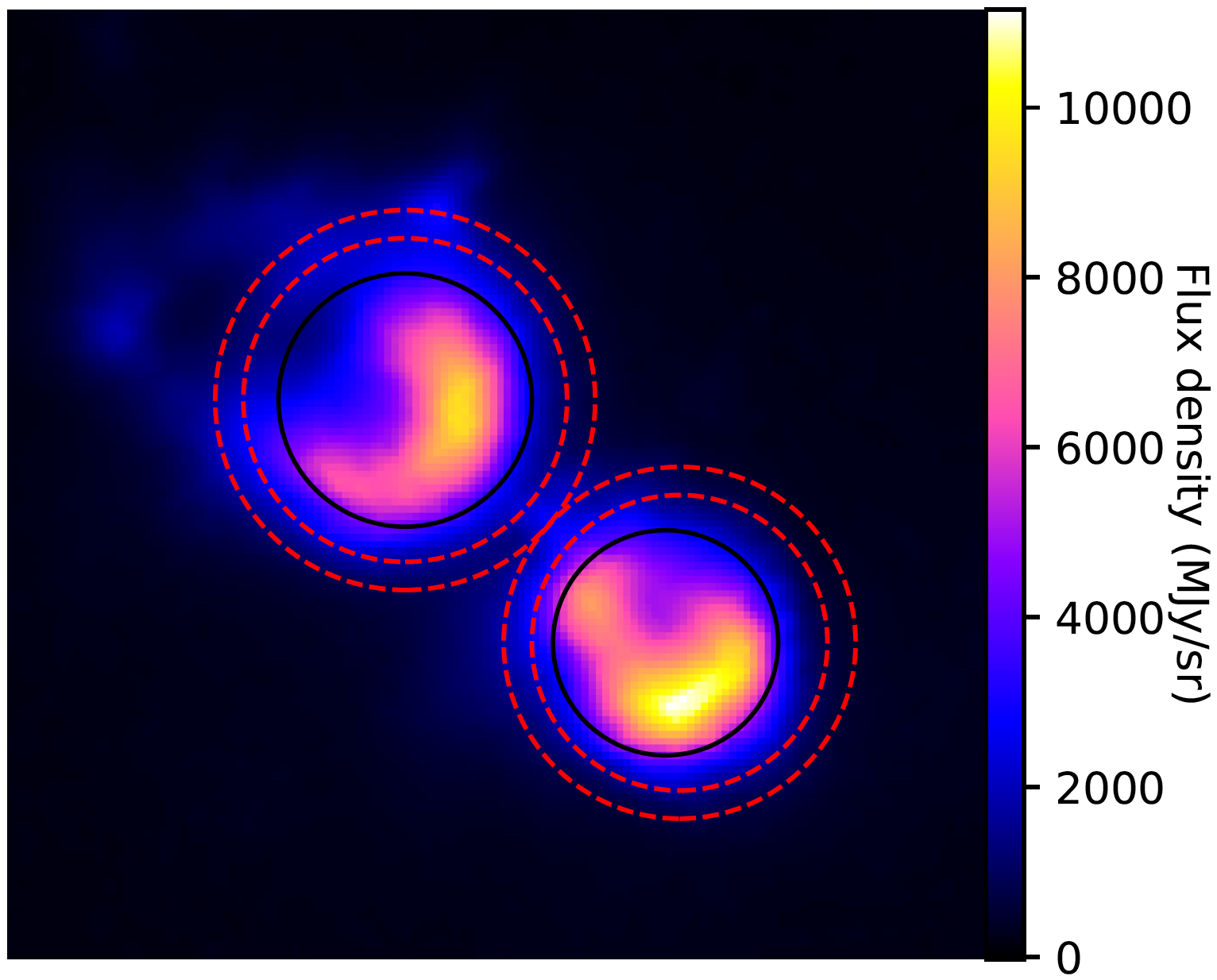}
\caption{The apertures used to calculate the flux of the nebulosities in G345.88-1.10 (black). The annuli to correct for background emission are indicated with the dashed red circles.}
\label{apertureCavities}
\end{center}
\end{figure}

In order to calculate the nebulosity and core fluxes, we performed aperture photometry. For the nebulosity, we used two apertures of $25''$ radius centred on each of the nebulosity lobes, encompassing most of the 70 $\mu$m emission . To correct for background emission, annuli were constructed around these apertures (see Fig. \ref{apertureCavities}). For the H1 core flux estimates, different aperture sizes were used depending on the wavelength and whether the core was unresolved. The aperture radius used for 160 $\mu$m and 250 $\mu$m, which resolve the fragment, was taken to be 13$^{\prime\prime}$ and centered on the H1 fragment as this corresponds to the fragment size. For the 350 $\mu$m and 500 $\mu$m data, the area was taken that corresponds to a single beam size (i.e. $\frac{\pi~\rm{FWHM}^{2}}{4\rm{ln(2)}}$) centred on the H1 fragment. As for the nebulosity, annuli were constructed around the different apertures to correct for background emission, the annuli were taken to cover emission between 2 and 3 times the radius of the beamsize.

\section{$^{12}$CO(2-1) channel maps}\label{sec:chanMapSec}
\begin{figure*}
    \centering
    \includegraphics[width=\hsize]{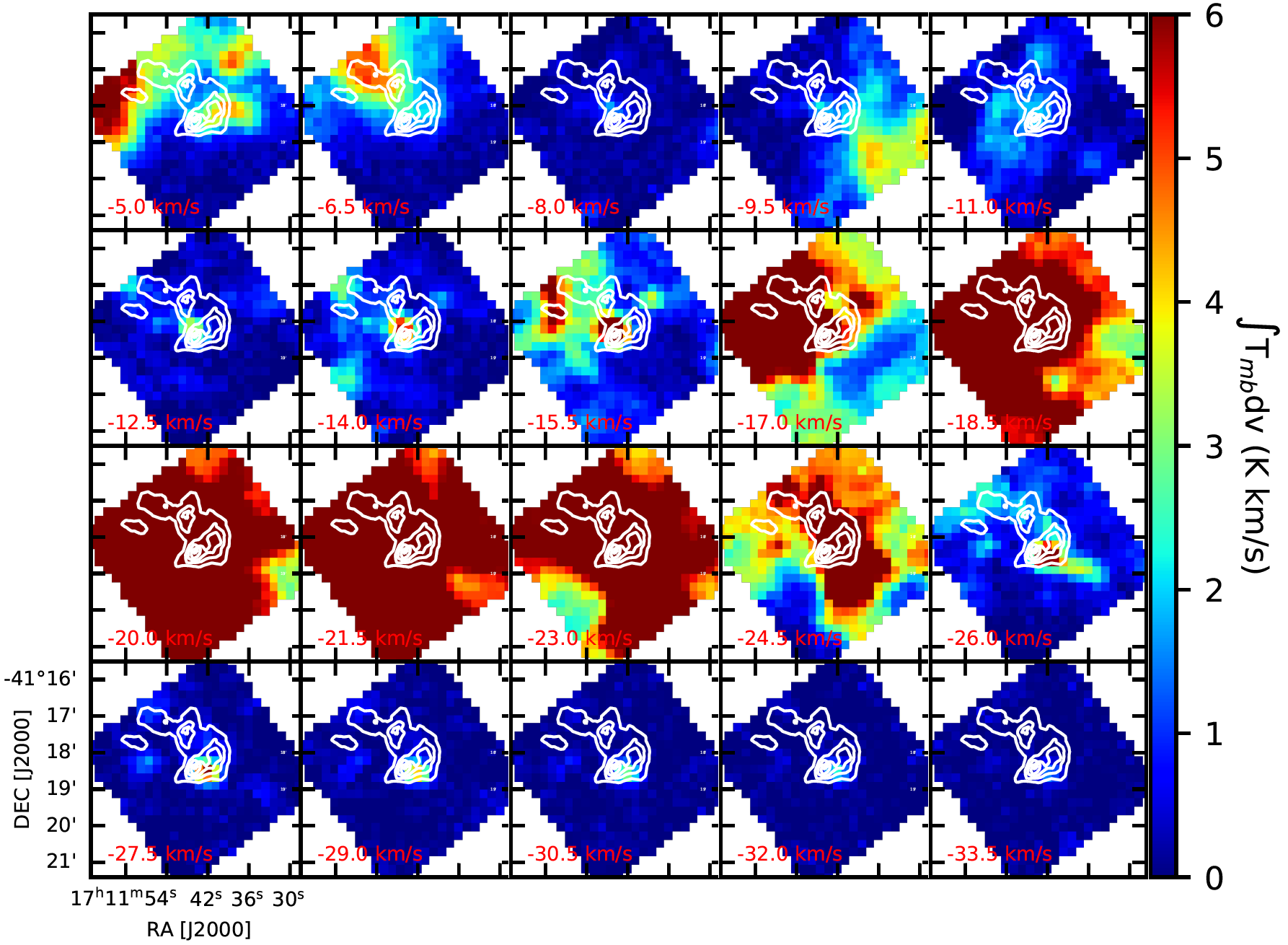}
    \caption{$^{12}$CO(2-1) channel maps over the full velocity range of the spectra with white Herschel column density contours highlighting the G345.88-1.10 hub. The colourbar was chosen to highlight the emission associated with the $^{12}$CO(2-1) wings, which explain why the figure is saturated in the figure between -25 km s$^{-1}$ and -15 km s$^{-1}$.}
    \label{fig:chanMapsCO}
\end{figure*}

Fig. \ref{fig:chanMapsCO} presents the channel maps of the $^{12}$CO(2-1) emission line. Focusing on the velocity ranges associated with the blueshifted wing (v$_{LSR} <$ -25 km s$^{-1}$) and the redshifted wing (v$_{LSR} >$ -15 km s$^{-1}$), it appears that there is weak high-velocity CO emission over the full cavities and no evident velocity substructure in the outflow. This suggests that the observed protostellar outflow might be dominated by a single powerful outflow. Because of the limited resolution and sensitivity it is however challenging to be conclusive about this emission. Sensitive high-resolution observations will be able to shed more light on this.

\section{The C$^{18}$O(2-1) and $^{13}$CO(2-1) emission}\label{sec:coEmissionApp}
\begin{figure*}
\centering
\includegraphics[width=0.445\hsize]{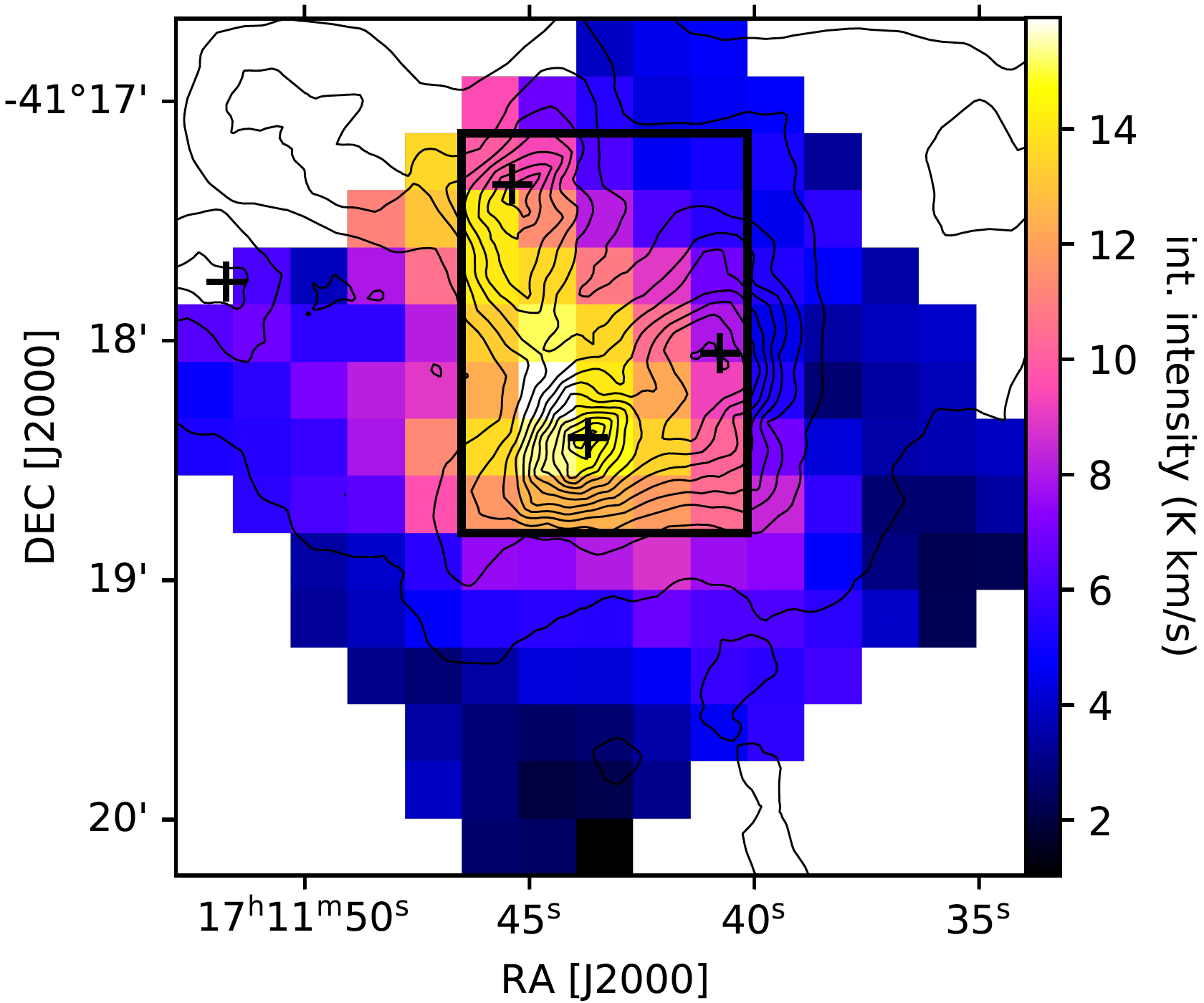}
\includegraphics[width=0.41\hsize]{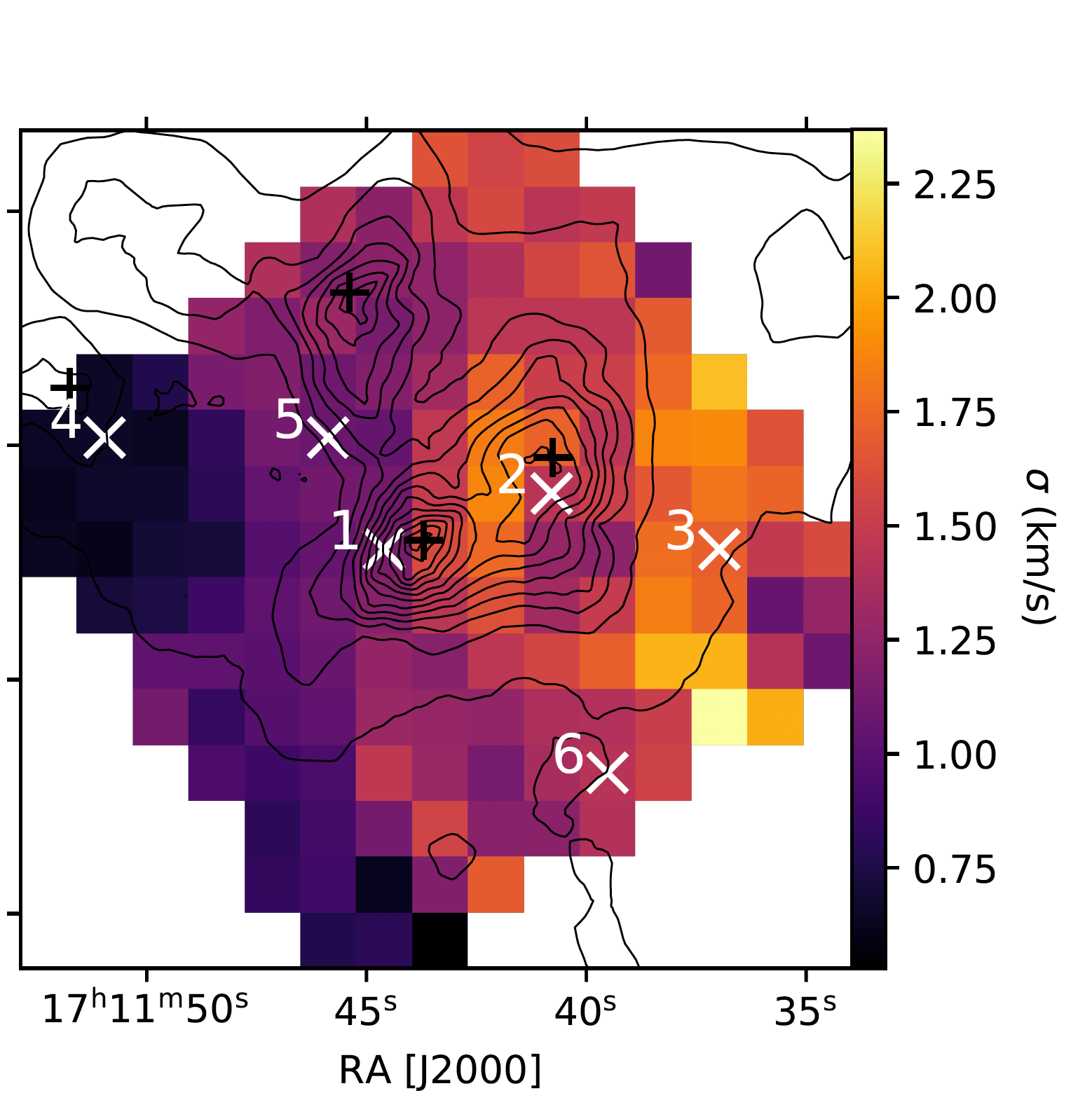}
\includegraphics[width=0.315\hsize]{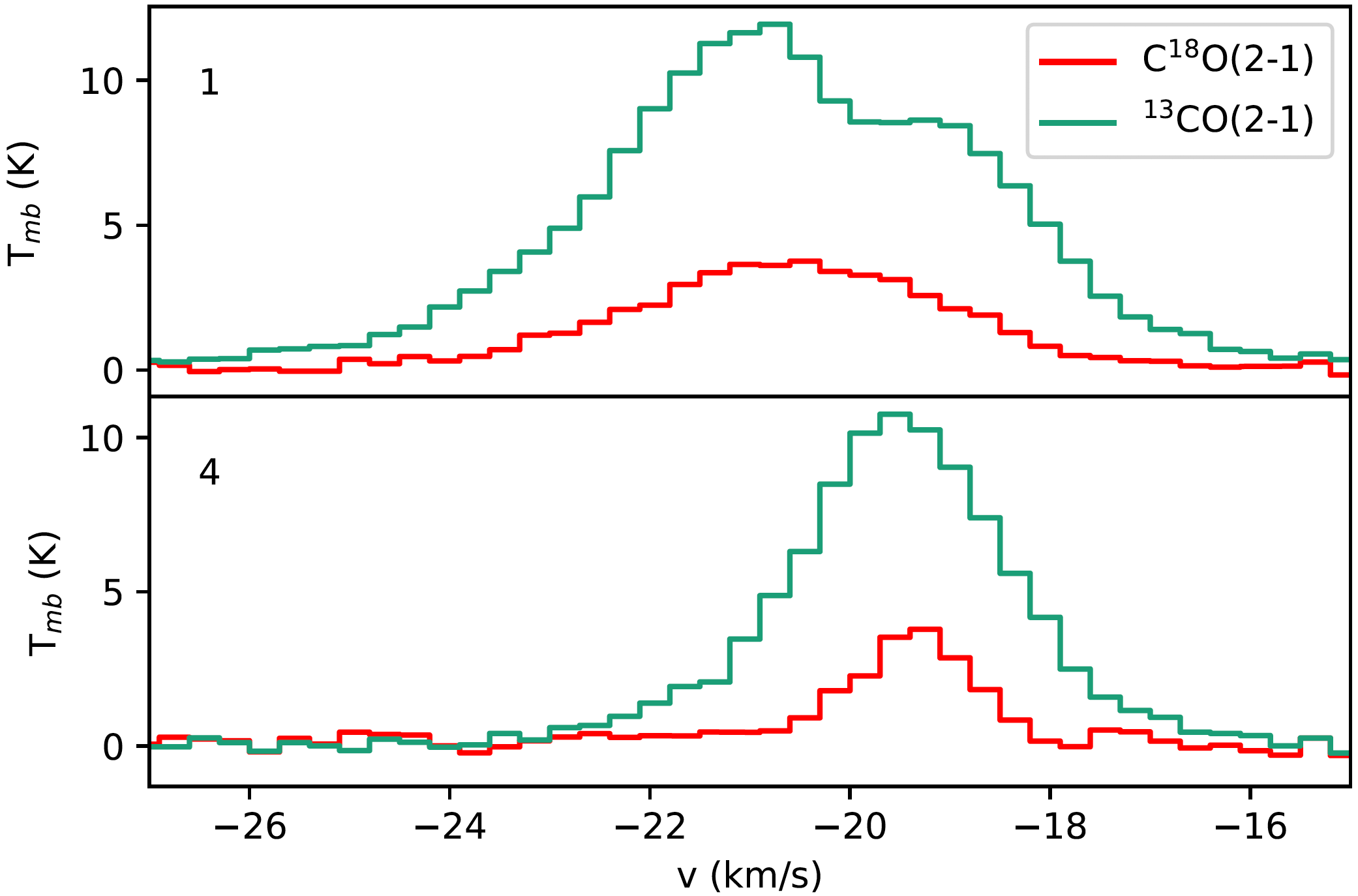}
\includegraphics[width=0.315\hsize]{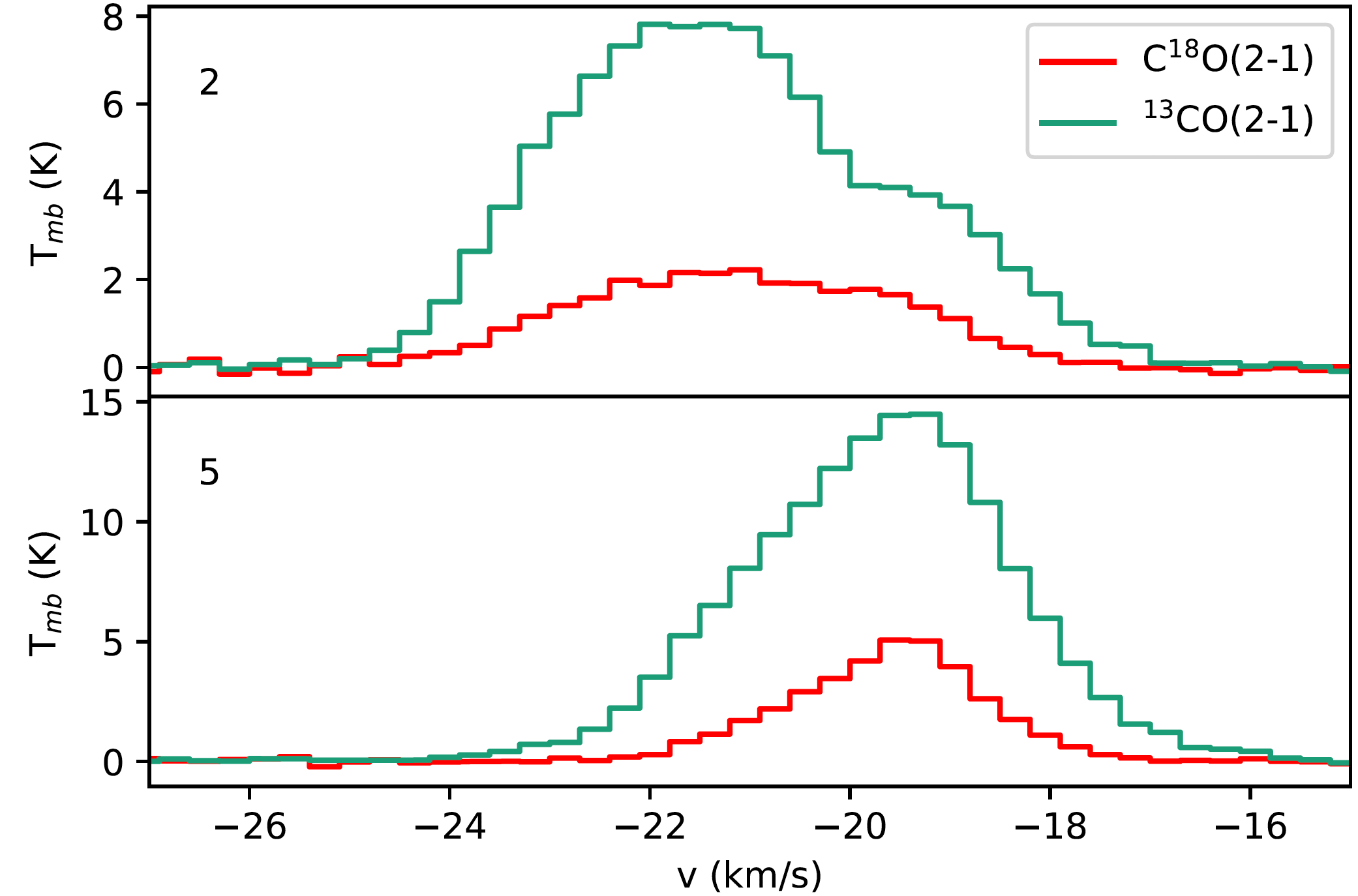}
\includegraphics[width=0.33\hsize]{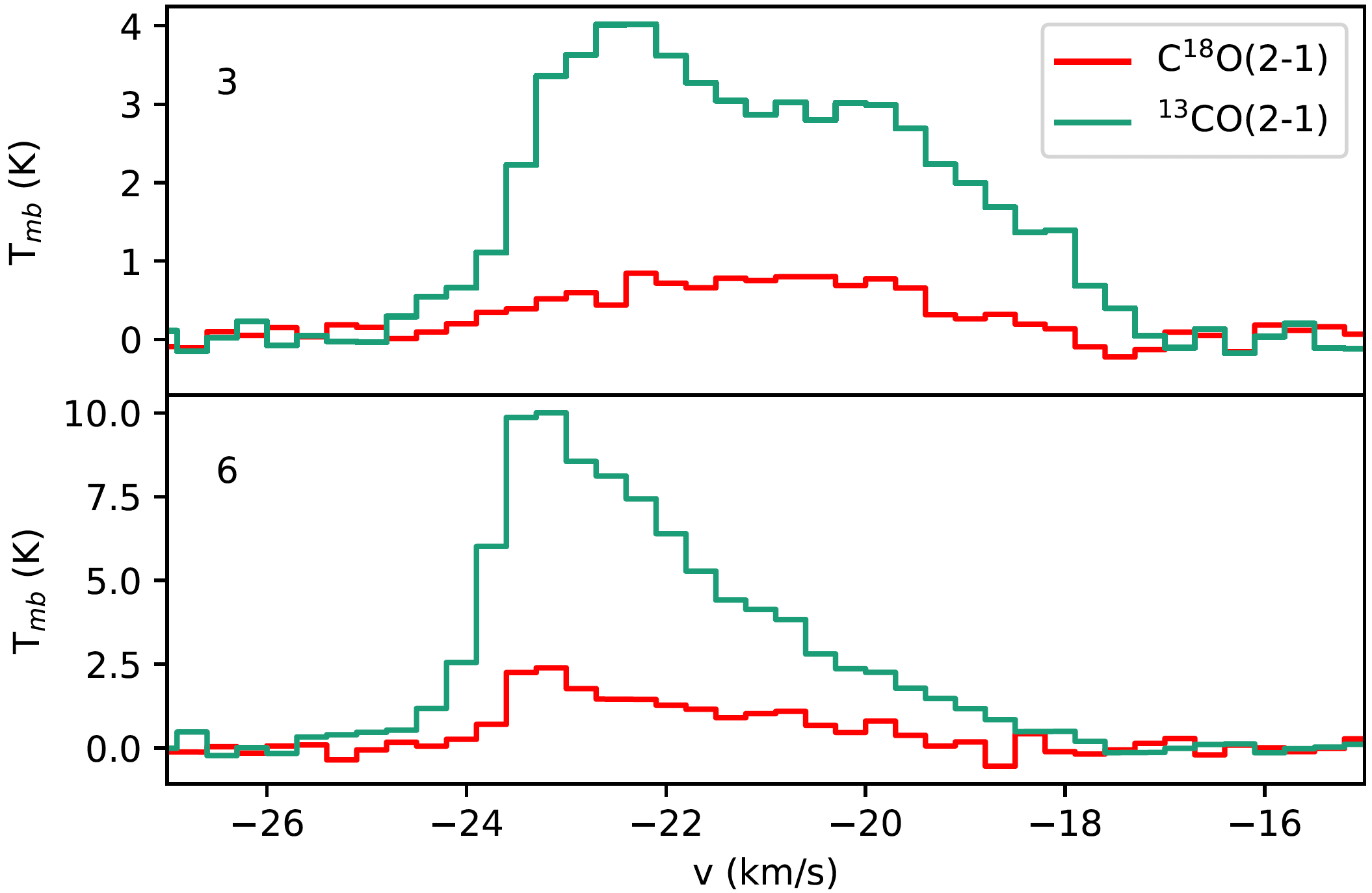}
\caption{(left) C$^{18}$O(2-1) integrated intensity map from -30 km s$^{-1}$ to -10 km s$^{-1}$. Overplotted in black are the \textit{Herschel} column density contours starting at N$_{H_{2}}$ = 2$\times$10$^{22}$ cm$^{-2}$ with increments of 2$\times$10$^{22}$ cm$^{-2}$. The black crosses indicate the 4 extracted fragments from the column density map and the black rectangle indicates the region used to extract the spectra presented in Fig. \ref{fig:invPCygni}. 
(right) The velocity dispersion map from fitting the Gaussian to the C$^{18}$O(2-1) spectra, with overplotted in black the \textit{Herschel} column density contours. The white crosses indicate the locations of the spectra presented below. (bottom) The $^{13}$CO(2-1) and C$^{18}$O(2-1) spectra at the locations in the hub indicated in the velocity dispersion map. The spectra at the location of the cavities (spectra 5 \& 6) show skewed spectra in both $^{13}$CO(2-1) and C$^{18}$O(2-1).}
\label{C18O13CO}
\end{figure*}

Fig. \ref{C18O13CO} presents the velocity integrated C$^{18}$O emission between -30 km s$^{-1}$ and -10 km s$^{-1}$ over the G345.88-1.10 hub. It is found that the C$^{18}$O(2-1) emission broadly follows the morphology of the hub. However, the dust column density peaks of the fragments do not coincide perfectly with the C$^{18}$O(2-1) peak emission. This could be due to several factors. First, CO could be depleted in the densest fragments. Second, strong temperature gradients could contribute to the line strength variations on its own. Further C$^{18}$O observations of different transitions, which we currently lack, would help to constrain this question.\\\\
Some C$^{18}$O(2-1) and $^{13}$CO(2-1) spectra from the hub are presented in Fig. \ref{C18O13CO}. In particular towards the cavities (spectra 5 \& 6 in Fig.~\ref{C18O13CO}), both the C$^{18}$O(2-1) and $^{13}$CO(2-1) spectra show skewed line profiles. This presence of skewed line profiles in the cavities was also highlighted in Fig. \ref{fig:oppositeWings}. Towards the fragments, we observe blueshifted self-absorbed $^{13}$CO spectra towards fragment H1 and H2,  a well-established signature of collapse signature which is widespread across the hub (see Sec. \ref{sec:discScenario}).\\
To obtain a centroid velocity and velocity dispersion maps, a single gaussian profile was fitted to the C$^{18}$O(2-1) spectra. The resulting velocity field is presented in Fig. \ref{fig:velFieldMap}, showing organised kinematics with a velocity gradient peak centered on the H1 and H2 fragments. This strong velocity gradient towards the presumed centre of collapse is expected in the context of gravitational acceleration. In Fig. \ref{C18O13CO}, the velocity dispersion map is presented. It shows large spectral linewidths to the west of the hub, whereas the smallest linewidths are found to the east/north-east where the 10 pc long filament connects to the hub. Lastly, it can be noted that there is no observed increase in velocity dispersion towards the identified massive fragments which suggests they do not experience increased turbulent support compared to the rest of the hub.

\section{Outflow momentum rate calculation}
\label{app:calculationMomentumFlux}
\begin{figure}
\begin{center}
\includegraphics[width=\hsize]{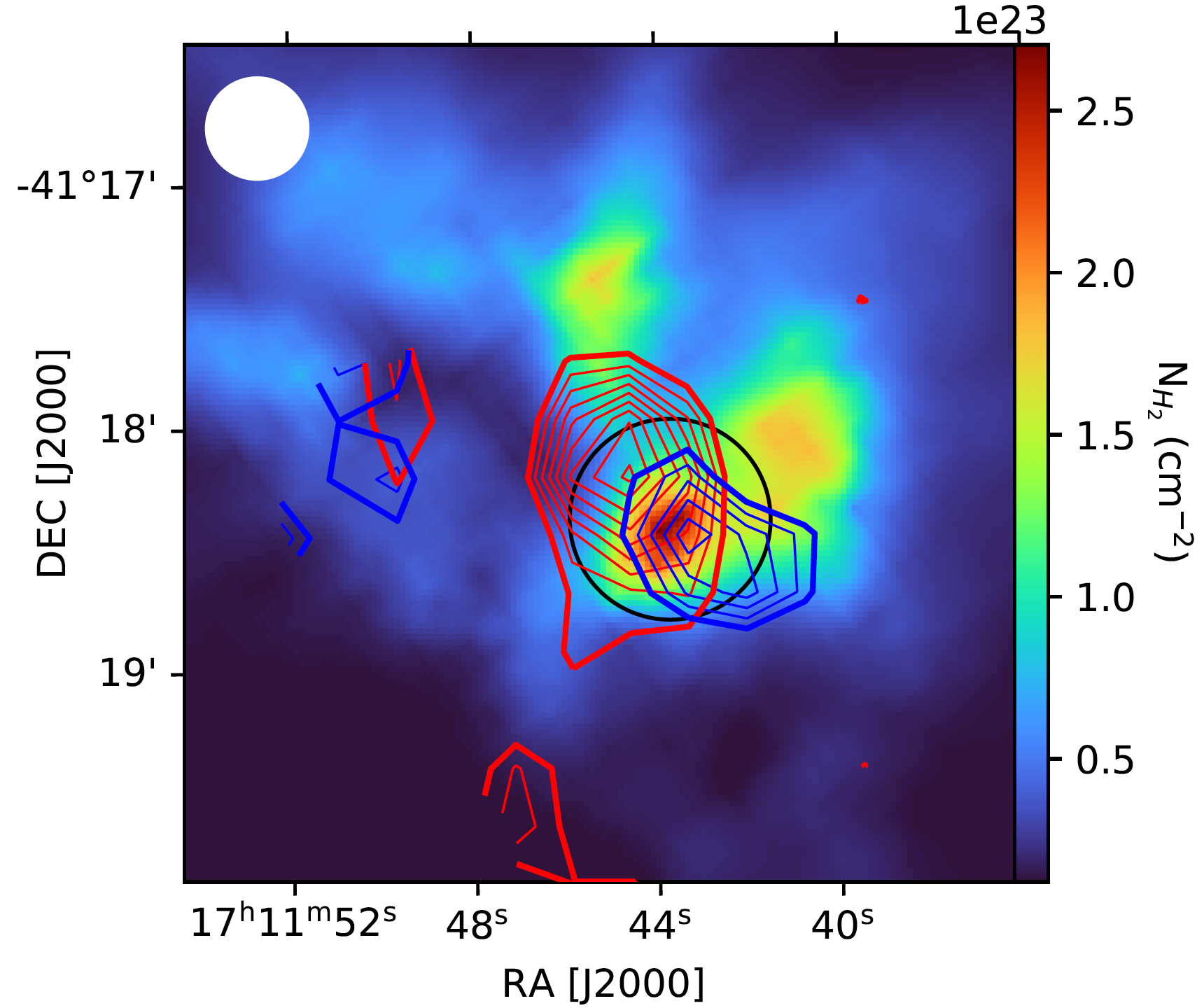}
\caption{The circle (in black) with a FWHM radius encompasses the emission peak of the blue- and redshifted outflow. To reduce noise, the outer shape (fat line) of the outflow contours is then used to cut out the outflow inside the circle. This allows to focus only on the actual outflow located inside the circle and reduces the noise that is not part of the outflow. The white circle in the top left indicates the beamsize. 
}
\label{outflowCalc}
\end{center}
\end{figure}

To determine the outflow momentum rate, we make use of the conservation of momentum rate along the outflow axis. For this exercise, an annulus with size $\Delta$r allows to focus on a specific location along the outflow axis. The construction of the annulus thus allows to define a region along the outflow axis over which we work with conservation of momentum rate. 
Here, we chose $\Delta r=28''$ ($\sim0.32$~pc at the distance of G345.88-1.10), which is the APEX beam FWHM in $^{12}$CO(2-1). In a second step, a more complex shape was cut out from this annulus that is encompassed by the outflow contours, see Fig. \ref{outflowCalc}, to reduce noise from non-outflow material that is found in the annulus \citep{DuarteCabral2013}. 
From Fig. \ref{outflowCalc}, it is observed that the peak intensity is very close to the central source. The inner radius of the annulus around the outflow peak is thus unresolved. As a result, the annulus becomes effectively a circle with radius $\Delta$r. Higher angular resolution observations will allow us to resolve the outflow structure, and provide a more accurate determination of its corresponding momentum rate.\\\\
The two annulus/circle for the outflow wings was constructed around the maximum of the outflow intensities to maximize the outflow rate since this should best trace the impact on the ambient cloud \citep{DuarteCabral2013}. Within the obtained areas, the momentum rate can be estimated for the outflows using:
\begin{equation}
\label{eq:outflowIntegral}
F_{CO} \propto \int \frac{T(\rm{v})(\rm{v}-\rm{v}_{0})^{2}d\rm{v}}{\Delta \rm{r}}
\end{equation}
In this equation, T is the main beam temperature in Kelvin and v$_{0}$ the velocity of the powering object which is assumed to be at -21 km s$^{-1}$. Further, a CO to H$_{2}$ conversion factor of 10$^{4}$, a temperature of 20 K and a molecular weight of 2.33 g mole$^{-1}$ were used.\\
We include a correction factor for the opacity of the CO wings, given by $\frac{\tau_{CO}}{1-e^{-\tau_{CO}}}=3.5$ \citep{CabritBertout1992}. Outflow momentum rates for both the red- and blueshifted sides of the two outflows are given in Tab. ~\ref{sourceTable}.\\
Lastly, we also include a correction for the inclination of the outflow. The actual $\Delta$r is given by $\Delta$r/sin(i), and the real outflow velocity is given by v/cos(i). The correction for the inclination angle in the outflow momentum rate calculation is therefore given by $\frac{\text{sin(i)}}{\text{cos}^{2}\text{(i)}}=32.7$, which corresponds to the assumed inclination angle with the line of sight of 80$^{\circ}$.

\section{Outflow opening angle}
\label{app:appendixOutflowAngle}
\begin{figure}
\begin{center}
\includegraphics[width=\hsize]{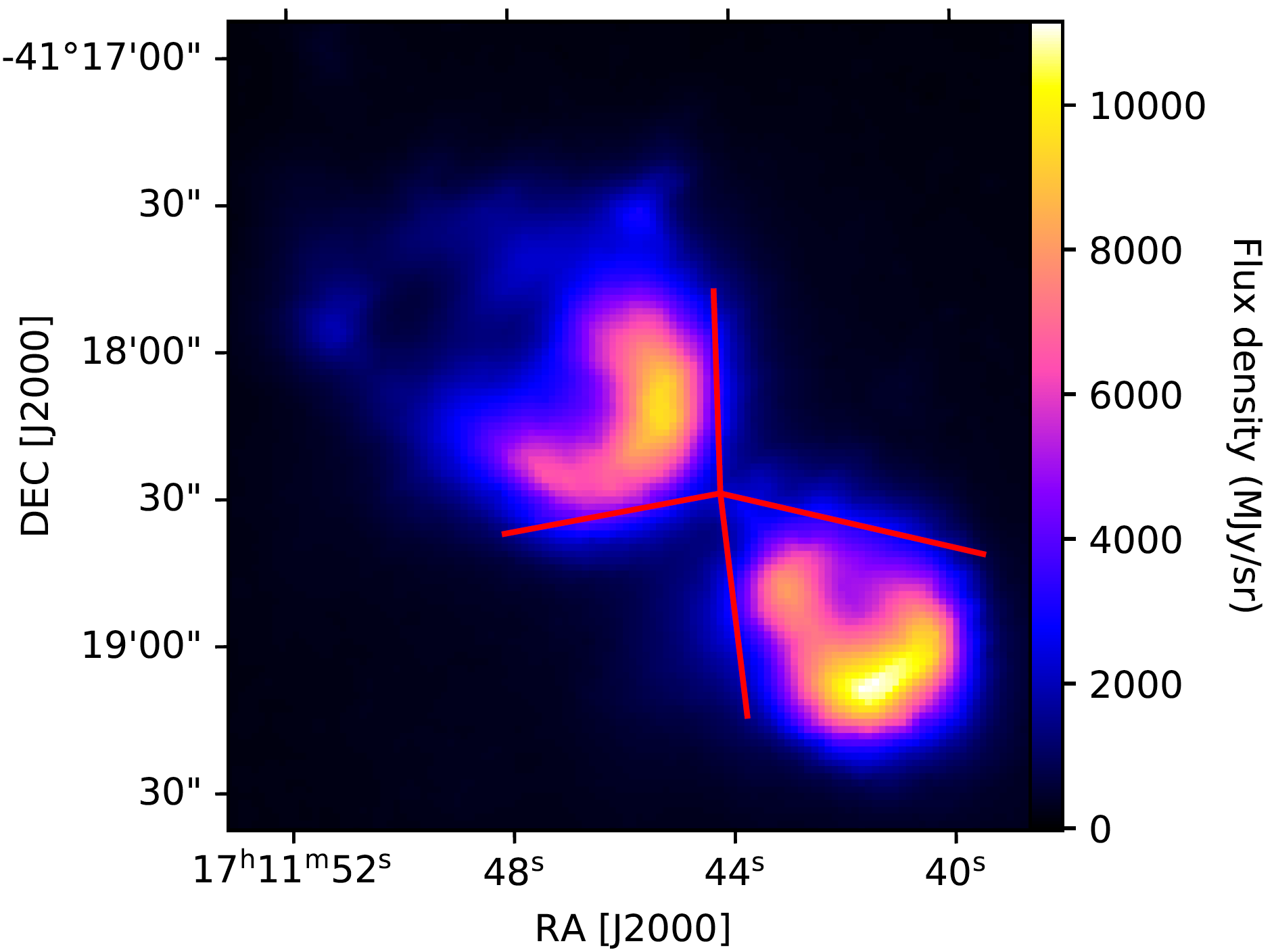}
\caption{
Illustration of the cones used to determine the opening angle of the 70 $\mu$m nebulosities.}
\label{outflowOpeningAngle}
\end{center}
\end{figure}

To determine the opening angle for the bipolar outflow cavities, a cone was constructed that connects the center of the H1 fragment with the outer edge of the cavity. This is illustrated in Fig. \ref{outflowOpeningAngle} for the 70 $\mu$m emission. This estimate of the outflow opening angle assumes axial symmetry which is uncertain seeing the proposed precession in the paper.

\section{H30$\alpha$ non-detection}
\label{app:h30alpha}
\begin{figure}
    \centering
    \includegraphics[width=\hsize]{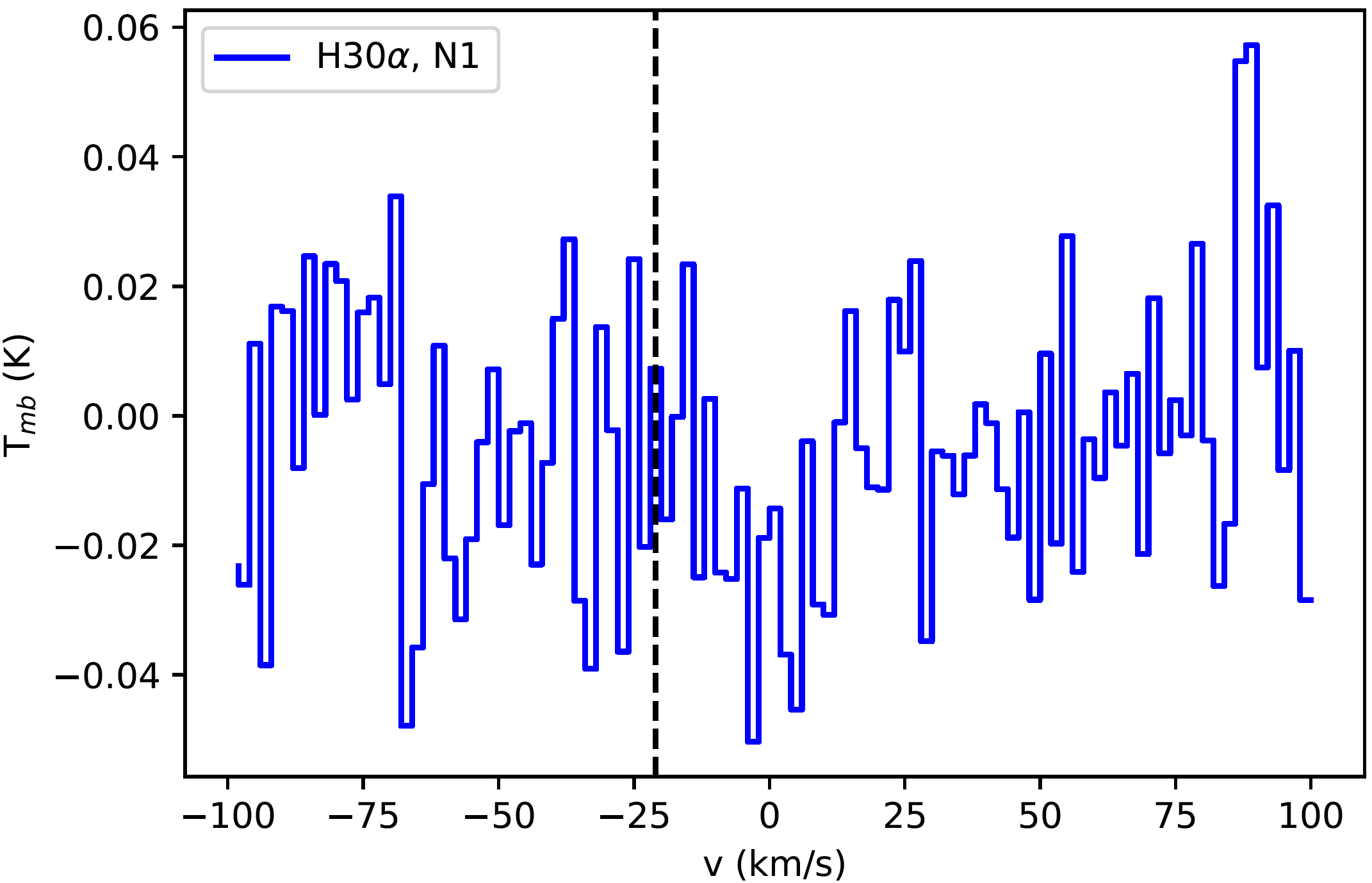}
    \includegraphics[width=\hsize]{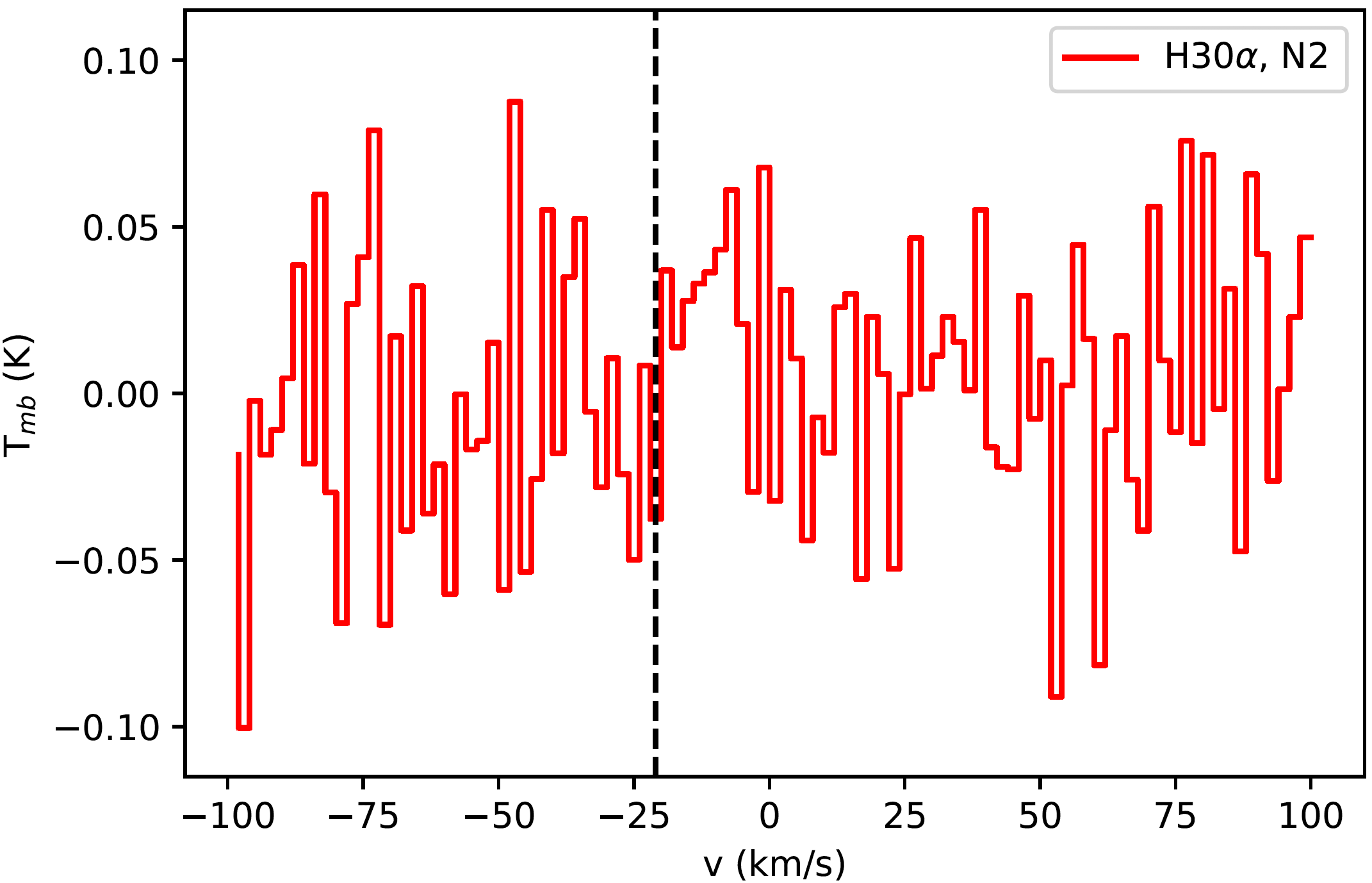}
    \caption{Average H30$\alpha$ spectra, sampled at 2 km s$^{-1}$ towards towards the N1 (top) and N2 (bottom) cavity. The vertical dashed lines indicates the expected central velocity of the line which remains undetected in both cavities.}
    \label{fig:h30specs}
\end{figure}

The spectra in Fig. \ref{fig:h30specs} present the H30$\alpha$ non-detection in both cavities. This provides 3$\sigma$ upper limits of 63 mK and 110 mK for N1 and N2, respectively. Using the optically thin assumption for the H30$\alpha$ line, a size of 0.6 pc for the cavity and a typical electron temperature (T$_{e}$) of 7500 K for the Galactocentric distance of G345.88-1.10 \citep{Quireza2006}, it is then possible to calculate an upper limit for the average electron density. This can be done from the formula below \citep{Gordon2002}:
\begin{equation}
\left( \frac{T_{L}}{K} \right) \simeq 3076\left( \frac{T_{e}}{K} \right)^{-3/2} \left( \frac{EM}{cm^{-6} pc} \right) \left( \frac{\nu_{0}}{GHz} \right)^{-1} \left( \frac{\Delta V}{km s^{-1}} \right)^{-1} \Delta n M_{\Delta n}
\end{equation}
with T$_{L}$ the line peak intensity, EM the emission measure, $\nu_{0}$ the rest frequency of the line, $\Delta$n = 1 and M$_{\Delta n}$ = 0.19. The emission measure is approximated by n$_{e}^{2}$L, with n$_{e}$ the electron density and L the size of the cavity in the line of sight. Plugging all values in this equation gives an electron density upper limit of n$_{e} <$ 3.4$\times$10$^{2}$ cm$^{-3}$ and n$_{e} <$ 4.6$\times$10$^{2}$ cm$^{-3}$ for the N1 and N2 cavities, respectively.

\section{The combination of a 70 $\mu$m bright cavity and 70 $\mu$m quiet core in the radiative transfer simulations}\label{sec:combCavCorRT}

\begin{figure}
\includegraphics[width=\hsize]{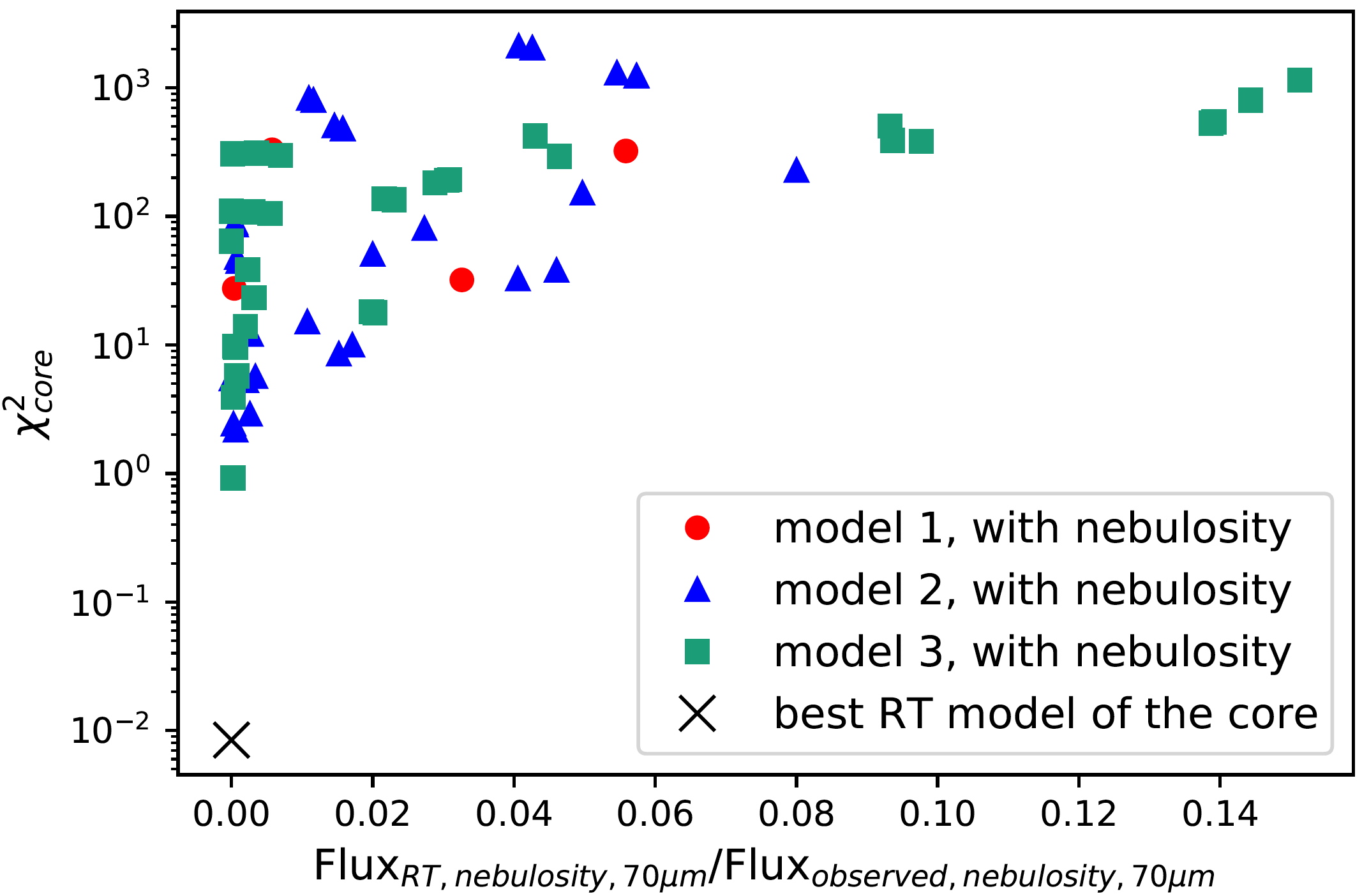}
\caption{$\chi^{2}_{core}$ for the radiative transfer models that have nebulosities as a function of the 70 $\mu$m flux for the simulated nebulosities (normalised to the observed nebulosities). This shows that none of the radiative transfer simulations with 70 $\mu$m nebulosities fit with the core at longer wavelengths.}
\label{chi2vsNeb}
\end{figure}

We combine the gathered information on both the emission properties of the nebulosity and that of the core to check whether the radiative transfer simulations with nebulosities also correspond to the best core runs. To do this we take all the radiative transfer calculations for which a nebulosity has formed, and plot the corresponding 70 $\mu$m nebulosity fluxes (normalised to the observed nebulosity flux) as a function of their associated $\chi^{2}_{\rm{core}}$ (see Fig. \ref{chi2vsNeb}). First, we find that only a small number of runs with nebulosity have a $\chi^{2}_{\rm{core}} <3$. Second, those good core runs are also the worst at reproducing the 70 $\mu$m nebulosity flux. In fact, these two conditions (i.e. a quiet mid-infrared core and a bright mid-infrared cavity) are conflicting, with the best models for each sitting at both ends of the point distribution in Fig.~\ref{chi2vsNeb}. This important discrepancy and the fact that it appears impossible to even create radiative transfer nebulosities with more than 15\% of the observed 70 $\mu$m flux in G345.88-1.10 strongly suggests that radiative heating from an embedded object is not responsible for the the bright cavities in G345.88-1.10. This also justifies our approach not to apply a model optimizer and explore the parameter space by hand as it will search for a non existing solution.

\begin{figure}
\includegraphics[width=\hsize]{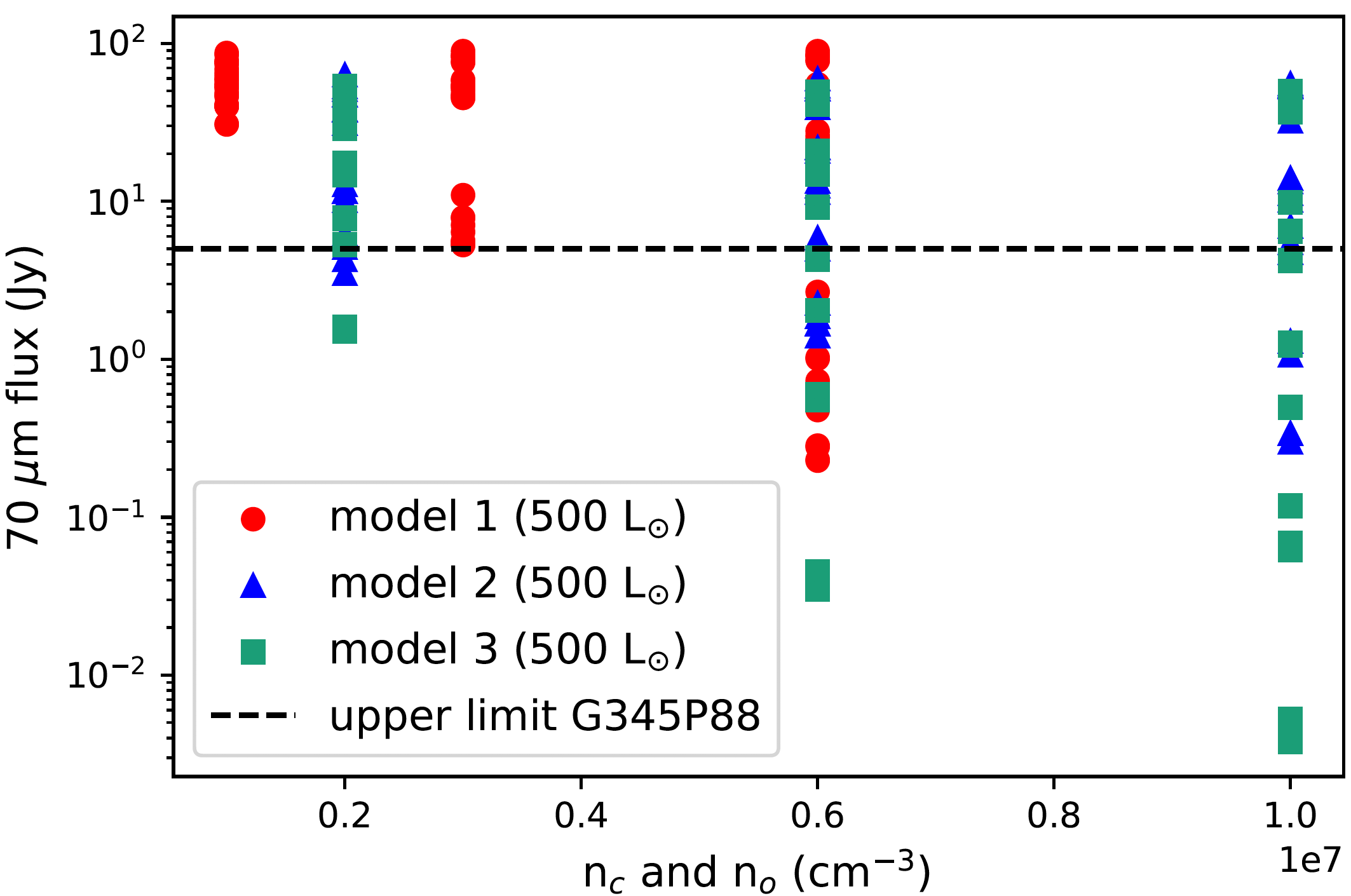}
\caption{The 70 $\mu$m flux predicted by the radiative transfer calculations as a function of the central (n$_{c}$) and characteristic (n$_{0}$) density for all models with a luminosity of 500 L$_{\odot}$. The horizontal line indicates the upper limit for the 70 $\mu$m flux towards the H1 fragment.}
\label{70micronDark}
\end{figure}

As already pointed out in Sect. \ref{sec:70quietRT}, radiative transfer calculations with a central source with  $L\geq$ 4000 $L_{\odot}$ are not compatible with the observations of H1. Investigating the relation between the 70 $\mu$m flux of the core and the density of the models (see Fig.~\ref{70micronDark}), it is also found that a high density is necessary to reproduce a weak 70 $\mu$m core emission. At 500 L$_{\odot}$, the central density $n_c$ of the Plummer profile, or the characteristic density $n_0$ of the passive flared disc, needs to be larger than n$_{H_{2}}$ = 5$\times$10$^{6}$ cm$^{-3}$  to sufficiently suppress the 70 $\mu$m emission towards the core. For an embedded object with a luminosity of 2000 L$_{\odot}$ and more, the required density increases to values n$_{H_{2}}$ $\ge$ 10$^{7}$ cm$^{-3}$. However, as pointed out in Sec. \ref{sec:70quietRT}, at 2000 L$_{\odot}$ there is still the problem that emission at longer wavelengths is not in agreement with what is observed towards the H1 fragment.


\section{Calculating the cooling timescale for the H1 fragment and the cavities}\label{sec:appCoolTimes}
The cooling time scale is given by
\begin{equation}
    \tau_{cool} = \frac{\rho \text{u}}{\lambda_{cool}}
\end{equation}
with $\rho$ the gas density, u the specific internal energy and $\lambda_{cool}$ the cooling rate per unit volume. The specific internal energy is given by
\begin{equation}
    \text{u} = \frac{3}{2}\frac{\text{k}_{b}\text{T}_{gas}}{\mu \text{m}_{p}}
\end{equation}
with k$_{b}$ the Boltzmann constant, T$_{gas}$ the kinetic gas temperature, $\mu$ the mean molecular mass (=2.33) and m$_{p}$ the mass of a proton. The last term that has to be constrained is the cooling rate per unit volume ($\lambda_{cool}$). If the gas is dense enough, typically n$_{H_{2}} >$ 10$^{5}$ cm$^{-3}$ \citep{Glover2012}, then it can be assumed that the cooling is driven by dust through gas-grain energy transfer. Given that the estimated average density in the H1 fragment is n$_{H_{2}}$ = 3.2$\times$10$^{5}$ cm$^{-3}$, we can use this condition \citep{Glover2012}. When the cooling is dominated by gas-grain energy transfer the cooling rate per unit volume is given by
\begin{equation}
    \lambda_{cool} = 4\pi \rho \text{D} \int_{0}^{\infty}\kappa_{\nu}\text{B}_{\nu}\text{(T)}\text{d}\nu
\end{equation}
where $\kappa_{\nu}$ is the dust opacity, D the dust to gas mass fraction and B$_{\nu}$ the Planck function. Using \citet{Ossenkopf1994} this results in 
\begin{equation}
    \lambda_{cool} = 4.68\times 10^{-31} \text{n} \text{T}_{dust}^{6} \text{( in erg s}^{-1} \text{ cm}^{-3}\text{)}
\end{equation}
with T$_{dust}$ the dust temperature and n the number density of the gas for which we assume it is fully molecular in the H1 fragment.\\
We considered several input parameters for the heated gas in the fragment ranging from 20 K to 100 K and for T$_{dust}$ we considered that the heated dust could reach 20 K, 30 K and 40 K. The resulting cooling timescales are given in Tab. \ref{tab:coolingTimes} and are typically of the order of a few days to a few years for these conditions. Assuming even higher dust temperature values would only further decrease the cooling timescale.\\
For the cavities we get typical densities n$_{H_{2}}$ = 1.0-2.5$\times$10$^{4}$ cm$^{-3}$ based on the Herschel column density and cavity size. From \citet{Goldsmith1978} we find that for a typical density n$_{H_{2}}$ = 10$^{4}$ cm$^{-3}$, $\lambda_{cool}$ is given by
\begin{equation}
    \lambda_{cool} = 1.5\times10^{-26} \text{T}^{2.7} \text{( in erg s}^{-1} \text{ cm}^{-3}\text{)}
\end{equation}
where T is the kinetic temperature in the region. Using the same average kinetic temperatures of 20-100 K in the cavities, we obtain $\lambda_{cool}$ between 4.9$\times$10$^{-23}$ - 3.8$\times$10$^{-21}$ erg cm$^{-3}$ s$^{-1}$ which is consistent with cooling rates found in the simulations by \citet{Glover2012} for these densities. Using these values, we obtain estimated cooling timescales between 1.7$\times$10$^{3}$ and 2.7$\times$10$^{4}$ yr for the cavities. Note that the lower end of these estimated cooling timescales is of the same order of magnitude as the propagation timescale for the jet through the cavity. This could thus be consistent with the observation that the full cavity is bright and not a specific subregion, Because of the lower densities, the cavities thus have significantly longer cooling timescales than the dense fragment.
\begin{table*}[]
    \centering
    \caption{The cooling timescales for the H1 fragment for a grid of dust and gas temperatures.}
    \begin{tabular}{cccccc}
    \hline
    \hline
         & T$_{gas}$ = 20 K & T$_{gas}$ = 40 K & T$_{gas}$ = 60 K & T$_{gas}$ = 80 K & T$_{gas}$ = 100 K \\
         \hline
        T$_{dust}$ = 20 K & 4.4 yr & 8.8 yr & 13 yr & 18 yr & 22 yr\\
        T$_{dust}$ = 30 K & 3.9$\times$10$^{-1}$ yr & 7.7$\times$10$^{-1}$ yr & 1.2 yr & 1.5 yr & 1.9 yr\\
        T$_{dust}$ = 40 K & 6.9$\times$10$^{-2}$ yr & 1.4$\times$10$^{-1}$ yr & 2.1$\times$10$^{-1}$ yr & 2.7$\times$10$^{-1}$ yr & 3.4$\times$10$^{-1}$ yr\\
        \hline
    \end{tabular}
    \label{tab:coolingTimes}
\end{table*}

\end{appendix}

\end{document}